\documentstyle[12pt]{article}
\renewcommand{\theequation}{\arabic{equation}}
\newcommand{\EQ}{\begin{equation}}
\newcommand{\EN}{\end{equation}}

\newcommand{\ket}[1]{\left|#1\right\rangle}      

\newcommand{\bear}{\begin{eqnarray}}
\newcommand{\ear}{\end{eqnarray}}
\newcommand{\bt} { \begin{tabular} }
\newcommand{\et}{ \end{tabular} }
\newcommand{\bc} { \begin{center} }
\newcommand{\ec}{ \end{center} }

\newcommand{\btb} { \begin{table} }
\newcommand{\etb}{ \end{table} }

\begin{document}

\topmargin 0pt
\oddsidemargin 5mm
\newcommand{\NP}[1]{Nucl.\ Phys.\ {\bf #1}}
\newcommand{\PL}[1]{Phys.\ Lett.\ {\bf #1}}
\newcommand{\NC}[1]{Nuovo Cimento {\bf #1}}
\newcommand{\CMP}[1]{Comm.\ Math.\ Phys.\ {\bf #1}}
\newcommand{\PR}[1]{Phys.\ Rev.\ {\bf #1}}
\newcommand{\PRL}[1]{Phys.\ Rev.\ Lett.\ {\bf #1}}
\newcommand{\MPL}[1]{Mod.\ Phys.\ Lett.\ {\bf #1}}
\newcommand{\JETP}[1]{Sov.\ Phys.\ JETP {\bf #1}}
\newcommand{\TMP}[1]{Teor.\ Mat.\ Fiz.\ {\bf #1}}
     
\renewcommand{\thefootnote}{\fnsymbol{footnote}}
     
\newpage
\setcounter{page}{0}
\begin{titlepage}     
\begin{flushright}
IFTA-97-35/UFSCARTH-97-19
\end{flushright}
\vspace{0.5cm}
\begin{center}
\large{ The Quantum Inverse Scattering Method for Hubbard-like Models }
\vspace{1cm}
\vspace{1cm}
{\large $M.J. Martins^{1,2}$   and $P.B. Ramos^{2}$ } \\
\vspace{1cm}
{\em 1. Instituut voor Theoretische Fysica, Universiteit van Amsterdam\\
Valcknierstraat 65, 1018 XE Amsterdam, The Netherlands}\\
\vspace{.5cm}
{\em 2. Universidade Federal de S\~ao Carlos\\
Departamento de F\'isica \\
C.P. 676, 13560~~S\~ao Carlos, Brazil}\\
\end{center}
\vspace{0.5cm}
     
\begin{abstract}
This work is concerned with various aspects of the formulation of the
quantum inverse scattering method for the one-dimensional 
Hubbard model. We first 
establish the essential tools to solve the  
eigenvalue problem  for the transfer matrix 
of the classical ``covering'' Hubbard model within the
algebraic Bethe Ansatz framework. The fundamental 
commutation rules exhibit 
a hidden $6$-vertex symmetry which plays a  crucial role in the whole
algebraic construction. Next we apply this formalism  to study the
$SU(2)$ highest weights properties of the eigenvectors and 
the solution of a related coupled spin model with twisted boundary conditions.
The machinery developed in this paper is applicable to many other
models, and as an example we present  
the algebraic solution of the Bariev $XY$ coupled model.
\end{abstract}
\vspace{.15cm}
\vspace{.1cm}
\vspace{.15cm}
\end{titlepage}

\renewcommand{\thefootnote}{\arabic{footnote}}
\section{Introduction}

The discovery of the quantum version of the inverse scattering method 
in the late seventies was undoubtedly a remarkable contribution to the 
development of the field of exactly solvable models in $(1+1)$ dimensions
\cite{STF}. This method provides a means for integrating models in 
two-dimensional classical 
statistical mechanics and $(1+1)$ quantum field theory, 
unifying major achievements such as the transfer matrix ideas, the Bethe 
Ansatz and the Yang-Baxter equation. Nowadays detailed reviews on this 
subject are available in the literature, for 
instance see refs. \cite{FA,FA1,TAK,KO}.

We shall start this paper illustrating the essential features of this 
method in the context of lattice models of statistical mechanics.  For example,
consider  
a vertex model on the square lattice and suppose that its  row-to-row transfer
matrix can be constructed from an elementary local vertex operator
${\cal L}_{{\cal A}i}(\lambda)$. This operator, known as the Lax operator,
contains all information about the  
structure of the Boltzmann weights which are parametrized through the spectral
parameter $\lambda$. The operator 
${\cal L}_{{\cal A}i}(\lambda)$ is frequently viewed as 
a matrix on the auxiliary space 
${\cal A}$, corresponding in the vertex model to the space of states of 
the horizontal degrees of freedom. Its 
matrix elements are operators on the Hilbert space 
$\displaystyle \prod_{i=1}^{L} \otimes V_{i}$, where $V_{i}$ corresponds to 
the space of vertical degrees of freedom and $i$ 
denotes the sites of a one-dimensional 
lattice of size $L$. In this paper  we shall consider
the situation in which the auxiliary space 
${\cal A}$  and the quantum space $V_{i}$  are equivalent. A
sufficient condition for integrability  of ultralocal models,
i.e. those in which the matrices elements of  the operator
${\cal L}_{{\cal A}i}(\lambda)$ commute for distinct values of index $i$,
is the existence of an invertible matrix 
$R(\lambda,\mu)$ satisfying the following property
\EQ
R(\lambda,\mu) {\cal L}_{{\cal A}i}(\lambda) \otimes {\cal L}_{{\cal A}i}(\mu) =
{\cal L}_{{\cal A}i}(\mu) \otimes {\cal L}_{{\cal A}i}(\lambda) R(\lambda,\mu)
\EN
where the tensor product is taken only with respect to the auxiliary space 
${\cal A}$. The matrix $R(\lambda,\mu)$ is defined on the tensor product 
${\cal A} \otimes {\cal A}$ and its matrix elements are c-numbers. A 
ordered product of Lax operators gives rise to the  monodromy operator
${\cal T}(\lambda)$ 
\EQ
{\cal T}(\lambda) = {\cal L}_{{\cal A}L}
(\lambda){\cal L}_{{\cal A}L-1}(\lambda) \ldots {\cal L}_{{\cal A}1}(\lambda)
\EN

It is possible to extend property (1) to the monodromy matrix, and such
global  intertwining relation
reads
\EQ
R(\lambda,\mu) {\cal T}(\lambda) \otimes {\cal T}(\mu) =
{\cal T}(\mu) \otimes {\cal T}(\lambda) R(\lambda,\mu)
\EN

The transfer matrix of the vertex model, for periodic boundary conditions,
can be written as the trace of 
the monodromy matrix on the auxiliary space ${\cal A}$
\EQ
T(\lambda) = Tr_{{\cal A}} {\cal T}(\lambda)
\EN

From the above  definition and property (3) we can derive that the transfer 
matrix is the generating function of the conserved currents. Indeed, taking 
the trace of equation (3) on the tensor 
${\cal A} \otimes {\cal A}$ space and using the trace cyclic property
we find
\EQ
\left[ T(\lambda), T(\mu) \right] = 0
\EN

Consequently, the expansion of  the transfer matrix in the spectral 
parameter yields an infinite number of conserved charges. We recall that
local charges are in general obtained as logarithm derivatives of $T(\lambda)$
\cite{LU,TAR}. Furthermore,  
the compatibility condition of ordering three Lax operators 
${\cal L}_{{\cal A}1}(\lambda_{1})$, ${\cal L}_{{\cal A}2}(\lambda_{2})$ 
and ${\cal L}_{{\cal A}3}(\lambda_{3})$ through
the intertwining relation (1) implies the famous Yang-Baxter equation
\EQ
R_{23}(\lambda_{1},\lambda_{2})R_{12}(\lambda_{1},\lambda_{3})R_{23}(\lambda_{2},\lambda_{3}) = 
R_{12}(\lambda_{2},\lambda_{3})R_{23}(\lambda_{1},\lambda_{3})R_{12}(\lambda_{1},\lambda_{2})
\EN
where $R_{ab}(\lambda,\mu)$ denotes the action of matrix $R(\lambda,\mu)$ on the
 spaces $V_{a} \otimes V_{b}$. 

Equation (3) is the starting point of solving two-dimensional 
classical statistical models by an exact operator formalism. This equation 
contains all possible commutation relation between the matrix elements of the 
monodromy operator ${\cal T}(\lambda)$. The diagonal terms of 
${\cal T}(\lambda)$ define the  transfer matrix eigenvalue problem 
and the off-diagonal ones play the role of creation and 
annihilation fields. The eigenvectors of the transfer matrix are constructed by 
applying the creation operators on a previously chosen reference state, 
providing us with an elegant formulation of the Bethe states. For this 
reason this framework is often denominated in the literature as the algebraic 
Bethe Ansatz approach. This situation resembles much the matrix
formulation 
of $(0+1)$ quantum mechanics. It is well known that the 
harmonic oscillator can 
either be solved by the Schr\"odinger formalism or 
by the Heisenberg algebra of 
creation and annihilation operators.  The later approach, however,
is conceptually 
much simpler provided the relevant dynamical symmetry has been
identified for a given quantum system.
One successful example is the solution 
of the hydrogen atom  through
the $SO(4)$ algebra \cite{SH}.

In this paper we are primarily interested in applying the quantum inverse 
scattering method for the one-dimensional Hubbard model. We recall that, after 
the Heisenberg model, the second one-dimensional lattice paradigm in the 
theory of magnetism solved by Bethe Ansatz method was the Hubbard model. 
The solution was found by Lieb and Wu in 1968 \cite{LW} using the extension of 
the coordinate Bethe Ansatz to the problem of 
fermions interacting via $\delta$-functions \cite{YY}. Considering the success 
of the solution of the Heisenberg model by the inverse method \cite{FA1}, the 
next natural target for this program would then be the Hubbard model. However, 
it turns out that the solution of this problem followed  a more 
arduous path than one could imagine from the very beginning. Indeed, nearly 18 
years were to pass before it was found the classical statistical vertex model 
whose transfer matrix generates the conserved charges commuting with the 
Hubbard Hamiltonian. This remarkable step was done by Shastry \cite{SA,SA1,SA2} 
who also found the $R$-matrix solution and thus proved the 
integrability of the Hubbard model from the quantum inverse method point of 
view. Shastry himself attempted to complete the inverse scattering program, but 
he was only able to conjecture the eigenvalues of the transfer matrix guided by 
a phenomenological approach which goes by the name of analytical 
Bethe Ansatz \cite{SA2}. Subsequently Bariev presented a 
coordinate Bethe Ansatz solution for the classical Shastry's model,
however on the basis of the 
$diagonal$-$to$-$diagonal$ transfer matrix method \cite{BA}.

One of the main results of this paper is the solution
of the one-dimensional 
Hubbard model by a first principle method, namely via the algebraic
Bethe Ansatz approach 
\footnote{A brief summary of some 
of our results has appeared in ref. \cite{PM}. }.
For this purpose we will use Shastry's $R$-matrix 
as well as the  
modifications introduced by Wadati and co-workers \cite{WA}. 
Apart from the  
fact that the solution of the one-dimensional
Hubbard model by the algebraic Bethe Ansatz framework 
remains an important unsolved theoretical 
challenge in the field of integrable models, there are
also other motivations to pursue this program.
Recent developments of new 
powerful methods to deal with finite temperature properties of 
integrable models (see for e.g. refs. \cite{KOM,KUP,DEV})
show clearly that the central object to be 
diagonalized is the quantum transfer matrix rather the underlying 
one-dimensional Hamiltonian. The transfer matrix eigenvalues provide us 
with the 
spectrum of all conserved charges, a fact which could be helpful in the 
study of transport properties \cite{ZOT} and level statistics behaviour
\cite{POI}. Lastly, there is a hope that this program is the first step 
towards the formulation of a general approach for computing lattice correlation 
functions \cite{KO}. 

We would like to remark that the ideas developed in this 
paper transcend the solution of the one-dimensional Hubbard model. In fact, the
original basis of our approach might be traced back to the solution of the
supersymmetric $spl(2|1)$ vertex model \cite{PM2}. Very recently, we have 
shown that this
method provide us with a unified way of solving a wider class of 
integrable models based on the braid monoid algebra \cite{MP1}. Here we also
will see that the lattice analog of the coupled $XY$ Bariev chain \cite{BAR}
can be solved
by this technique. The unusual feature of the Hubbard and
Bariev models is that they both have a non-additive R-matrix solution.

We have organized this paper as follows. To make our presentation 
self-contained,
in next section we briefly 
review the basic properties of the embedding of 
the one-dimensional Hubbard model into a classical vertex model, 
originally due to Shastry \cite{SA,SA1,SA2}. In section 3 we discuss the 
commutation rules coming from the Yang-Baxter algebra. In particular, a hidden 
symmetry of $6$-vertex type, which is 
crucial for integrability, is found. We use 
these properties in section 4 in order to construct the eigenvectors and the 
eigenvalues of the transfer matrix of the classical statistical model. The 
Lieb's and Wu's  results as well as the spectrum of higher conserved charges can be 
obtained from our expression for the transfer matrix eigenvalues. In 
section 5 we present complementary results such as extra comments on 
systems with twisted boundary conditions and a discussion on the $SU(2)$
highest weights properties  of the eigenvectors.
Section 6 is dedicated to the solution of the classical analog of the
coupled $XY$ Bariev model. Our conclusions are presented in section 7.
Finally, five appendices
summarize Boltzmann weights, extra commutation rules
and  technical details we omitted in the main
text.

\section{The classical covering Hubbard model}

We begin this section reviewing the work of Shastry 
\cite{SA,SA1,SA2} on the identification of the classical statistical model 
whose row-to-row transfer matrix commutes with the one-dimensional 
Hubbard Hamiltonian. Originally, Shastry  
looked at this problem considering the 
coupled spin version of the Hubbard model, since in one-dimension fermions 
and spin-$\frac{1}{2}$ Pauli operators are related to each other via 
Jordan-Wigner transformation. In the context of 
statistical mechanics, however, 
the later representation  is sometimes more appealing. Here we will 
consider the coupled spin model introduced by Shastry with general twisted 
boundary conditions. Its Hamiltonian is 
\bear
H & = &
\sum_{i=1}^{L-1} \sigma^{+}_{i}\sigma^{-}_{i+1} 
+ \sigma^{-}_{i}\sigma^{+}_{i+1} 
+ \tau^{+}_{i}\tau^{-}_{i+1} + \tau^{-}_{i}\tau^{+}_{i+1} + 
\frac{U}{4}\sigma^{z}_{i}\tau^{z}_{i} 
\nonumber \\
&& + e^{-i\phi_{1}}\sigma^{+}_{L}\sigma^{-}_{1} +
e^{i\phi_1}\sigma^{-}_{L}\sigma^{+}_{1} +
e^{-i\phi_{2}}\tau^{+}_{L}\tau^{-}_{1} +
e^{i\phi_2}\tau^{-}_{L}\tau^{+}_{1} +
\frac{U}{4}\sigma^{z}_{L}\tau^{z}_{L}
\ear
where $\{\sigma^{\pm}_{i},\sigma^{z}_{i}\}$ and $\{\tau^{\pm}_{i},\tau^{z}_{i}\}$ are two commuting sets of Pauli matrices acting on the site i of a lattice of size L. The second term in (7) stands for the boundary conditions 
$\sigma^{\pm}_{L+1}=e^{\pm i\phi_1} \sigma^{\pm}_{1}$, 
$\tau^{\pm}_{L+1}=e^{\pm i\phi_2} \tau^{\pm}_{1}$,  
$\sigma^{z}_{L+1} = \sigma^{z}_{1}$, and $\tau^{z}_{L+1} = \tau^{z}_{1}$ where 
$\phi_{1}$ and $\phi_{2}$ are arbitrary angles $0 \leq \phi_{1},
\phi_{2} < 2\pi$. The coupling constant $U$ represents the 
Hubbard on-site Coulomb interaction.

In order to relate the coupled spin model to the Hubbard model 
we have to perform the following Jordan-Wigner transformation \cite{SA}
\EQ
c_{i\uparrow} = \prod_{k=1}^{i-1} \sigma^{z}_{k} \sigma^{-}_{i}~,~ 
c_{i\downarrow} = \prod_{k=1}^{L} \sigma^{z}_{k} \prod_{k=1}^{i-1} 
\tau^{z}_{k} \tau^{-}_{i}
\EN
where $c_{i\sigma}$ are canonical Fermi operators of spins $\sigma=\uparrow, 
\downarrow$ on site i, with anti-commutation relations given by 
$\{c^{\dagger}_{i\sigma},c_{j\sigma'}\} = \delta_{i,j} \delta_{\sigma,\sigma'}$.
Defining the number operator $n_{i\sigma} = c^{\dagger}_{i\sigma} c_{i\sigma}$ 
for electrons with spin $\sigma$ on site i and performing transformation (8) we 
find that 
\bear
H & = & 
-\sum_{i=1}^{L-1} \sum_{\sigma=\uparrow,\downarrow} 
[ c^{\dagger}_{i\sigma}c_{i+1\sigma} + c^{\dagger}_{i+1\sigma}c_{i\sigma} ]
+U \sum_{i=1}^{L-1}(n_{i\uparrow}-\frac{1}{2})(n_{i\downarrow}-\frac{1}{2})  
\nonumber \\
&& -e^{-i\phi_{\uparrow}}c^{\dagger}_{L\uparrow}c_{1\uparrow} 
-e^{i\phi_{\uparrow}}c^{\dagger}_{1\uparrow}c_{L\uparrow} 
-e^{-i\phi_{\downarrow}}c^{\dagger}_{L\downarrow}c_{1\downarrow} 
-e^{i\phi_{\downarrow}}c^{\dagger}_{1\uparrow}c_{L\downarrow} +
U (n_{L\uparrow}-\frac{1}{2})(n_{L\downarrow}-\frac{1}{2}) 
\ear
where the angles $\phi_{\uparrow}$ and $\phi_{\downarrow}$ are given by
\EQ
\phi_{\uparrow} = \phi_{1} + \pi + \pi N^{h}_{\uparrow}~,~ 
\phi_{\downarrow} = \phi_{2} + \pi + \pi N^{h}_{\downarrow} 
\EN
and $N^{h}_{\sigma}$ is the number of holes  
(eigenvalues of the operator $\displaystyle \sum_{i=1}^{L} 
c_{i\sigma} c^{\dagger}_{i\sigma}$) of spin 
$\sigma$ of a given sector of the Hubbard model. Therefore, the Hubbard model 
with periodic boundary conditions $(\phi_{\uparrow}=\phi_{\downarrow})$ is 
related to the coupled spin model with dynamically (sector dependent) twisted 
boundary conditions imposed. This was the reason why we started with a more 
general coupled spin model, since the two representations are fully equivalent 
only for free boundary conditions.

From the point of view of a 
vertex model, twisted boundary conditions correspond 
to the introduction of a seam of different Boltzmann weights along the infinite 
direction on the cylinder. In practice this is accomplished by multiplying 
one of the elementary vertex operator, 
${\cal L}_{{\cal A}L}(\lambda)$ say, by a 
``gauge'' matrix $G_{{\cal A}}$ (see section 5). Such matrix is usually
related to
additional hidden
invariances of the $R$-matrix \cite{DEVI}. Hence, although twisted boundary 
conditions may affect eigenvalues and Bethe Ansatz equations in a significative 
way, the relevant features of the integrability still remain intact. Since 
this section is concerned with the later point, we can 
assume periodic boundary conditions 
without losing generality. As Shastry \cite{SA,SA1,SA2} pointed out, 
the mapping of the Hubbard model (modulo above subtlety) into a coupled 
spin system is quite illuminating in searching for a ``covering''
vertex model. It 
is known that the decoupled spin model $(U=0)$ can be derived in terms of a 
pair of uncoupled free-fermion $6$-vertex models. This suggests that, for 
the interacting model, we have to look for a copy of 
two free-fermion $6$-vertex models
coupled in an appropriate way. Shastry \cite{SA,SA1,SA2} 
determined the nature of this coupling by demanding that it should reproduce the higher conserved charges \cite{SA} \footnote{For further discussion on 
Hubbard's conserved charges see refs. \cite{INV}.} when the corresponding 
transfer matrix $T(\lambda)$ was expanded in powers of the spectral 
parameter $\lambda$. The solution found by Shastry for the Lax operator is 
given by \cite{SA1,SA2}
\EQ
{\cal L}_{{\cal A}i}(\lambda) = 
{\cal L}^{\sigma}_{{\cal A}i}(\lambda) {\cal L}^{\tau}_{{\cal A}i} (\lambda) 
e^{ h(\lambda) \sigma^{z}_{{\cal A}} \tau^{z}_{{\cal A}} \otimes I_{i}}
\EN

The form of operators ${\cal L}^{\sigma}_{{\cal A}i}(\lambda)$ and 
 ${\cal L}^{\tau}_{{\cal A}i}(\lambda)$ obey the $6$-vertex structure
\EQ
{\cal L}^{\sigma}_{{\cal A}i}(\lambda) =
\frac{a(\lambda)+b(\lambda)}{2} + 
\frac{a(\lambda)-b(\lambda)}{2} \sigma^{z}_{{\cal A}}\sigma^{z}_{i} +
(\sigma^{+}_{{\cal A}}\sigma^{-}_{i} + \sigma^{-}_{{\cal A}}\sigma^{+}_{i})
\EN
and
\EQ
{\cal L}^{\tau}_{{\cal A}i}(\lambda) =
\frac{a(\lambda)+b(\lambda)}{2} + 
\frac{a(\lambda)-b(\lambda)}{2} \tau^{z}_{{\cal A}}\tau^{z}_{i} +
(\tau^{+}_{{\cal A}}\tau^{-}_{i} + \tau^{-}_{{\cal A}}\tau^{+}_{i})
\EN
where the weights $a(\lambda)$ and $b(\lambda)$ satisfy the free-fermion 
condition $a^{2}(\lambda)+b^{2}(\lambda)=1$. Furthermore, 
the constraint $h(\lambda)$ is determined in terms 
of the weights and the coupling 
$U$ by 
\EQ
\sinh[2h(\lambda)] = \frac{U}{2} a(\lambda)b(\lambda)
\EN

A second important result due to Shastry \cite{SA1,SA2} was the 
solution of the Yang-Baxter algebra for the Lax operator (11), and thus 
determinating the
form of the $R$-matrix. 
The matrix 
$R(\lambda,\mu)$ is a $16 \times 16$ matrix whose non-null elements are given 
in terms of $10$ distinct Boltzmann weights $\alpha_{i}(\lambda,\mu)$, 
$i=1, \dots, 10$. For practical calculations it is helpful to display 
its matrix form 
{\scriptsize
\bear
 R(\lambda,\mu) = 
\pmatrix{
 \alpha_{2} &0 &0 &0 &0  &0  &0  &0  &0  &0  &0  &0  &0  &0  &0  &0    \cr
 0 &\alpha_{5} &0 &0 &-\alpha_{9}  &0  &0  &0  &0  &0  &0  &0  &0  &0  &0  &0    \cr
 0 &0 &\alpha_{5} &0 &0  &0  &0  &0  &-\alpha_{9}  &0  &0  &0  &0  &0  &0  &0    \cr
 0 &0 &0 &\alpha_{4} &0  &0  &\alpha_{10}  &0  &0  &\alpha_{10}  &0  &0  &-\alpha_{7}  &0  &0  &0    \cr
 0 &\alpha_{8} &0 &0 &\alpha_{5}  &0  
&0  &0  &0  &0  &0  &0  &0  &0  &0  &0    \cr
 0 &0 &0 &0 &0  &\alpha_{1}  &0  &0  &0  &0  &0  &0  &0  &0  &0  &0    \cr
 0 &0 &0 &\alpha_{10} &0  &0  &\alpha_{3}  &0  &0  &\alpha_{6}  &0  &0  &\alpha_{10}  &0  &0  &0    \cr
 0 &0 &0 &0 &0  &0  &0  &\alpha_{5}  &0  &0  &0  &0  &0  &\alpha_{8}  &0  &0    \cr
 0 &0 &\alpha_{8} &0 &0  &0  &0  &0  &\alpha_{5}  &0  &0  &0  &0  &0  &0  &0    \cr
 0 &0 &0 &\alpha_{10} &0  &0  &\alpha_{6}  &0  &0  &\alpha_{3}  &0  &0  &\alpha_{10}  &0  &0  &0    \cr
 0 &0 &0 &0 &0  &0  &0  &0  &0  &0  &\alpha_{1}  &0  &0  &0  &0  &0    \cr
 0 &0 &0 &0 &0  &0  &0  &0  &0  &0  &0  &\alpha_{5}  &0  &0  &\alpha_{8}  &0    \cr
 0 &0 &0 &-\alpha_{7} &0  &0  &\alpha_{10}  
&0  &0  &\alpha_{10}  &0  &0  
&\alpha_{4} &0  &0  &0    \cr
 0 &0 &0 &0 &0  &0  &0  &-\alpha_{9}  &0  &0  &0  &0  &0  &\alpha_{5}  &0  &0    \cr
 0 &0 &0 &0 &0  &0  &0  &0  &0  &0  &0  &-\alpha_{9}  &0  &0  &\alpha_{5}  &0    \cr
 0 &0 &0 &0 &0  &0  &0  &0  &0  &0  &0  &0  &0  &0  &0  &\alpha_{2}    \cr}
\nonumber \\
\ear
}
where the expressions for the weights $\alpha_{i}(\lambda,\mu)$ in terms of the 
free-fermion weights $a(\lambda)$, $b(\lambda)$ and the constraint $h(\lambda)$ 
can be found in appendix $A$. The striking feature of this solution is that 
$R$-matrix (15) is non-additive with respect the spectral parameters. In fact, 
after an unitary transformation, $R(\lambda,\mu)$ can be written in a more 
compact form \cite{SA2} which shows that it depends  on both  
the difference and the sum  of the spectral 
parameters. As 
far we know, it is still an open question whether 
or not there exists an embedding for the Hubbard model  
satisfying the standard difference property. As a final remark we mention that
an 
analytical proof that  $R(\lambda,\mu)$ indeed 
satisfies the Yang-Baxter equation (6) has been recently presented in 
ref. \cite{SHWA}.

We close this section presenting the graded Yang-Baxter formalism 
\cite{SKU} for the Hubbard model. This interesting approach was pursued by 
Wadati and co-workers \cite{WA} and it has the 
advantage of making real distinction 
between bosonic and fermionic degrees of freedom. In the 
Hubbard model, the empty and doubly occupied sites play the role of 
bosonic states while the spin up and down states are the fermionic ones. This 
formalism is an elegant mathematical procedure\footnote{This scheme accommodates
a particular class of models 
having ``nonultralocal'' Yang-Baxter relations. 
For more general implications of  nonultralocality
see the recent review  \cite{KUN}.} of avoiding the subtlety 
on boundary condition raised in 
the beginning of this section. In other words, the graded version of the 
inverse scattering method guarantees that the ``non-local'' anticommutation 
rules of fermionic degrees of freedom is satisfied for any lattice sites. In 
general, the basic changes we need to perform is to consider the analogs of the 
trace and the tensor product properties on the graded space. For example, the 
graded Yang-Baxter for the monodromy matrix now reads \cite{SKU}
\EQ
R_{g}(\lambda,\mu) {\cal T}(\lambda) \stackrel{s}{\otimes} {\cal T}(\mu) =
{\cal T}(\mu) \stackrel{s}{\otimes} {\cal T}(\lambda) R_{g}(\lambda,\mu)
\EN
where the symbol $\stackrel{s}{\otimes}$ stands for the supertensor product 
$(A \stackrel{s}{\otimes} B)_{ab}^{cd} = (-1)^{p(b)[p(a)+p(c)]} A_{ac}B_{bd}$. 
The index $p(a)$ is the Grassmann parity of the $a$-th degree of freedom, 
assuming values $p(a)=0$ for bosonic specie and $p(a)=1$ for fermionic ones. 
Other important change is on the transfer matrix definition, which is now given 
in terms of the supertrace of the monodromy matrix
\EQ
T(\lambda) = Str_{\cal A} {\cal T}(\lambda) = 
\sum_{a \epsilon {\cal A}} (-1)^{p(a)} {\cal T}_{aa}(\lambda)
\EN

There is no extra effort to obtain the matrix $R_{g}(\lambda,\mu)$ from
the original solution found by Shastry.  One just have to 
perform a Jordan-Wigner transformation on the Lax operator (11), taking into 
account the gradation of the space of states \cite{WA}. It turns out that 
the graded $R$-matrix is related to  Shastry's solution (15) by a
unitary transformation, and its explicit form is given by \cite{WA}
{\scriptsize
\bear
 R_g(\lambda,\mu) = 
\pmatrix{
 \alpha_{2} &0 &0 &0 &0  &0  &0  &0  &0  &0  &0  &0  &0  &0  &0  &0    \cr
 0 &\alpha_{5} &0 &0 &-i\alpha_{9}  &0  &0  &0  &0  &0  &0  &0  &0  &0  &0  &0    \cr
 0 &0 &\alpha_{5} &0 &0  &0  &0  &0  &-i\alpha_{9}  &0  &0  &0  &0  &0  &0  &0    \cr
 0 &0 &0 &\alpha_{4} &0  &0  &-i\alpha_{10}  &0  &0  &i\alpha_{10}  &0  &0  &\alpha_{7}  &0  &0  &0    \cr
 0 &-i\alpha_{8} &0 &0 &\alpha_{5}  &0  
&0  &0  &0  &0  &0  &0  &0  &0  &0  &0    \cr
 0 &0 &0 &0 &0  &\alpha_{1}  &0  &0  &0  &0  &0  &0  &0  &0  &0  &0    \cr
 0 &0 &0 &i\alpha_{10} &0  &0  &\alpha_{3}  &0  &0  &-\alpha_{6}  &0  &0  &-i\alpha_{10}  &0  &0  &0    \cr
 0 &0 &0 &0 &0  &0  &0  &\alpha_{5}  &0  &0  &0  &0  &0  &-i\alpha_{8}  &0  &0    \cr
 0 &0 &-i\alpha_{8} &0 &0  &0  &0  &0  &\alpha_{5}  &0  &0  &0  &0  &0  &0  &0    \cr
 0 &0 &0 &-i\alpha_{10} &0  &0  &-\alpha_{6}  &0  &0  &\alpha_{3}  &0  &0  &i\alpha_{10}  &0  &0  &0    \cr
 0 &0 &0 &0 &0  &0  &0  &0  &0  &0  &\alpha_{1}  &0  &0  &0  &0  &0    \cr
 0 &0 &0 &0 &0  &0  &0  &0  &0  &0  &0  &\alpha_{5}  &0  &0  &-i\alpha_{8}  &0    \cr
 0 &0 &0 &\alpha_{7} &0  &0  &i\alpha_{10}  
&0  &0  &-i\alpha_{10}  &0  &0  
&\alpha_{4} &0  &0  &0    \cr
 0 &0 &0 &0 &0  &0  &0  &-i\alpha_{9}  &0  &0  &0  &0  &0  &\alpha_{5}  &0  &0    \cr
 0 &0 &0 &0 &0  &0  &0  &0  &0  &0  &0  &-i\alpha_{9}  &0  &0  &\alpha_{5}  &0    \cr
 0 &0 &0 &0 &0  &0  &0  &0  &0  &0  &0  &0  &0  &0  &0  &\alpha_{2}    \cr}
\nonumber \\
\ear
}
where here we assumed that the first and the fourth degrees of freedom 
are bosonic  $(p(1)=p(4)=0)$ while the 
remaining ones are fermionic $(p(2)=p(3)=1)$.

In the next sections we are going to use the graded formalism in order to find the appropriate commutation rules, the eigenvectors and the 
eigenvalues of the transfer matrix (17). Afterwards, we will get back to the 
standard quantum inverse formalism, but now with 
twisted boundary conditions.

\section{The fundamental commutation rules}

In addition to the Lax operator and the $R$-matrix the existence of a local 
reference state 
is another important object in the quantum inverse scattering program. 
This is a vector $\ket{0}_{i}$ such that the result of the action of the 
Lax operator on it is a matrix having a triangular form. We choose 
$\ket{0}_{i}= {\scriptsize 
\left( \renewcommand{\arraystretch}{0.5} \begin{array}{@{}c@{}} 1 \\ 0 \end{array} \right)_i \otimes 
\left( \renewcommand{\arraystretch}{0.5} \begin{array}{@{}c@{}} 1 \\ 0 \end{array} \right)_i } $
as the standard spin up ``ferromagnetic state'', which in the 
fermionic language corresponds to the doubly occupied state.  
The action of the vertex operator in this state satisfies the following
property
\EQ 
{\cal L}_{{\cal A}i}(\lambda) \ket{0}_{i} = 
\pmatrix{
\omega_1(\lambda) \ket{0}_i &  \ddagger  & \ddagger & \ddagger  \cr
0  &  \omega_2(\lambda) \ket{0}_{i}  &  0 & \ddagger  \cr
0  &  0  &  \omega_2(\lambda) \ket{0}_i  & \ddagger  \cr
0  &  0  &  0  & \omega_3(\lambda) \ket{0}_i  \cr}
\EN
where the symbol $\ddagger$ represents arbitrary non-null values  and  
the functions $\omega_1(\lambda)$, 
$\omega_2(\lambda)$ and $\omega_3(\lambda)$ are given by
\EQ
\omega_1(\lambda) = [a(\lambda)]^2 e^{h(\lambda)}~~,~~
\omega_2(\lambda) = a(\lambda)b(\lambda)e^{-h(\lambda)}~~,~~
\omega_3(\lambda) = [b(\lambda)]^2 e^{h(\lambda)} 
\EN

The global reference state  $\ket{0}$  
is then defined by the tensor product 
$\ket{0} = \displaystyle  \prod_{i=1}^{L} \otimes \ket{0}_{i}$. This 
state is
an eigenstate of the transfer matrix since the triangular property 
is easily  extended
to the monodromy matrix.
In order to construct other eigenstates it is necessary  
to seek for an appropriate  representation of 
the monodromy matrix. By this we mean a structure which is able to 
distinguish creation and annihilation fields as well as possible hidden 
symmetries. The triangular property of the Lax operator suggests us the 
following form
\EQ
{\cal T}(\lambda) = 
\pmatrix{
B(\lambda)       &   \vec{B}(\lambda)   &   F(\lambda)   \cr
\vec{C}(\lambda)  &  \hat{A}(\lambda)   &  \vec{B^{*}}(\lambda)   \cr
C(\lambda)  & \vec{C^{*}}(\lambda)  &  D(\lambda)  \cr}_{4 \times 4}
\EN
where $\vec{B}(\lambda)$,  $\vec{C^{*}}(\lambda)$ and
$\vec{B^{*}}(\lambda)$, 
$\vec{C}(\lambda)$  are two component vectors 
with dimensions $1 \times 2$ and $2 \times 1$, respectively. The operator 
$\hat{A}(\lambda)$ is a $2 \times 2$ matrix and we shall denote its 
elements by $\hat{A}_{ab}(\lambda)$. The
remaining operators  $B(\lambda)$, $C(\lambda)$, $D(\lambda)$ and $F(\lambda)$
are 
scalars.  In this paper we will  use the
symbol $ABCDF$ to refer to the above way  
of representing the elements of the monodromy matrix.
We recall that such Ansatz is quite
distinct from the traditional $ABCD$ form proposed originally by 
Faddeev  and co-workers
\cite{STF,FA,FA1}. 

In the $ABCDF$ 
representation the eigenvalue problem for the graded transfer matrix
becomes
\EQ
[B(\lambda)-\sum_{a=1}^{2}A_{aa}(\lambda)+D(\lambda)] \ket{\Phi} = 
\Lambda(\lambda) \ket{\Phi}
\EN
where $\Lambda(\lambda)$ and $\ket{\Phi}$ correspond to 
the eigenvalues and  to the
eigenvectors, respectively. As a consequence of the triangular property we
can derive important relations for the monodromy matrix elements. For the
diagonal part of
$\cal{T}(\lambda)$ we have
\EQ
B(\lambda)\ket{0} = [\omega_1(\lambda)]^{L}\ket{0}~~,~~ D(\lambda)\ket{0} = 
[\omega_3(\lambda)]^{L}\ket{0}~~,~~
\hat{A}_{aa}(\lambda)\ket{0} = [\omega_2(\lambda)]^{L}\ket{0}~\mathrm{for}~a=1,2
\EN

Also one expects that the operators
$\vec{B}(\lambda)$, $\vec{B^{*}}(\lambda)$ and $F(\lambda)$ 
play the role  of creation fields over the reference state
$\ket{0}$. It also follows from the triangular property  
the  annihilation properties 
\EQ
\vec{C}(\lambda)\ket{0} = 0~~,~~ \vec{C}^{*}(\lambda)\ket{0} = 0~~,~~
C(\lambda) \ket{0} = 0~~,~~  
\hat{A}_{ab}(\lambda)\ket{0} = 0~ \mathrm{for}~  a \neq b
\nonumber \\
\EN

To make further progress we have to recast the graded Yang-Baxter algebra
in the form of commutation relations for the creation and annihilation 
fields. In general it is not known how and when such job can be performed 
for a particular representation, and one could surely say that the
``artistic'' part of the algebraic Bethe Ansatz construction begins here.
Within the $ABCDF$ formalism, the solution of this problem
turns out to be  more complicated than a
similar situation occurring for the $6$-vertex model 
\cite{STF,FA,FA1} and its multi-state generalizations  
\cite{RES,BAB}. The new feature present 
here is that we have a mixture of two classes 
of creation fields, the non-commutative vectors $\vec{B}(\lambda)$ or 
$\vec{B}^{*}(\lambda)$ and one commutative operator 
represented by $F(\lambda)$. We shall start 
our discussion by the
commutation 
rule between the fields  $\vec{B}(\lambda)$
and  $\vec{B}(\mu)$. In this case the relation that comes out from the
Yang-Baxter algebra is not the convenient one for further computations.
It turns out to be necessary to perform a second step which consists in
substituting the exchange relation for the scalar operators 
$B(\lambda)$ and $F(\mu)$ (see equation (38)) back on the original
commutation rule we just derived for the 
fields  $\vec{B}(\lambda)$
and  $\vec{B}(\mu)$. The basic trick is to keep the diagonal operator
$B(\lambda)$ always in the 
right-hand side position in the commutation rule \cite{PM2}. After
performing this
two step procedure we are able to get the 
appropriate commutation rule,  which is
\EQ
\vec{B}(\lambda) \otimes \vec{B}(\mu) = \frac{\alpha_{1}(\lambda,\mu)}{\alpha_{2}(\lambda,\mu)}
[ \vec{B}(\mu) \otimes \vec{B}(\lambda) ] .\hat{r}(\lambda,\mu)
\nonumber \\
-i\frac{\alpha_{10}(\lambda,\mu)}{\alpha_{7}(\lambda,\mu)}  
\{ F(\lambda)B(\mu) - F(\mu)B(\lambda) \} \vec{\xi}
\EN
where $\vec{\xi}$ is a $1\times 4$ vector and $\hat{r}(\lambda,\mu)$ is an
auxiliary $4 \times 4 $ matrix given by
\EQ
{\vec \xi} = 
\matrix{(
0  &1  &-1  &0 )  \cr}~~,~~
\hat{r}(\lambda,\mu) = 
\pmatrix{
1  &0  &0  &0  \cr
0  &\bar{a}(\lambda,\mu)  &\bar{b}(\lambda,\mu)  &0  \cr
0  &\bar{b}(\lambda,\mu)  &\bar{a}(\lambda,\mu)  &0  \cr
0  &0  &0  &1  \cr}
\EN
and the 
functions $\bar{a}(\lambda,\mu)$ and 
$\bar{b}(\lambda,\mu)$ are  given in terms
of the Boltzmann weights by
\EQ
\bar{a}(\lambda,\mu) = \frac{\alpha_{3}(\lambda,\mu)\alpha_{7}(\lambda,\mu)+\alpha^{2}_{10}(\lambda,\mu)}{\alpha_{1}(\lambda,\mu)\alpha_{7}(\lambda,\mu)}
~,~~ \bar{b}(\lambda,\mu) = -\frac{\alpha_{6}(\lambda,\mu)\alpha_{7}(\lambda,\mu)+\alpha^{2}_{10}(\lambda,\mu)}{\alpha_{1}(\lambda,\mu)\alpha_{7}(\lambda,\mu)}
\EN

It turns out that the auxiliary matrix
$\hat{r}(\lambda,\mu)$ is precisely the rational $R$-matrix 
of the isotropic $6$-vertex  model or the $XXX$ spin chain. In order to
see that, we first 
simplify a bit more the auxiliary weights 
$\bar{a}(\lambda,\mu)$
and $\bar{b}(\lambda,\mu)$ with the help of identities (A.10-A.12).
We find that they satisfy the following relations 
\EQ
\bar{a}(\lambda,\mu)=1- \bar{b}(\lambda,\mu)~~,~~ \bar{b}(\lambda,\mu)= 
\frac{\alpha_{8}(\lambda,\mu)\alpha_{9}
(\lambda,\mu)}{\alpha_{1}(\lambda,\mu)\alpha_{7}(\lambda,\mu)}
\EN

Next we simplify as much as possible the ratios 
$\frac{\alpha_9(\lambda,\mu)}{\alpha_1(\lambda,\mu)}$ and
$\frac{\alpha_8(\lambda,\mu)}{\alpha_7(\lambda,\mu)}$ in terms of
the free-fermion Boltzmann weights and the constraint $h(\lambda)$. After
some algebra we write these ratios as
\EQ
\frac{\alpha_9(\lambda,\mu)}{\alpha_1(\lambda,\mu)}  = 
\frac{ a(\mu) b(\lambda) e^{2[h(\mu)-h(\lambda)]}- a(\lambda) b(\mu)}
{ b(\lambda)b(\mu) +a(\lambda) a(\mu) e^{2[h(\mu)-h(\lambda)]} }
\EN
\EQ
\frac{\alpha_8(\lambda,\mu)}{\alpha_7(\lambda,\mu)}   = 
\frac{ b(\lambda)b(\mu) +a(\lambda) a(\mu) e^{2[h(\mu)+h(\lambda)]} }
{ a(\mu) b(\lambda) e^{2[h(\mu)+h(\lambda)]} -  a(\lambda)b(\mu) }
\EN

Now if we take into account the identity
\EQ
\frac{a(\lambda)}{b(\lambda)}e^{2h(\lambda)} -\frac{b(\lambda)}{a(\lambda)}
e^{-2h(\lambda)} =
\frac{a(\lambda)}{b(\lambda)}e^{-2h(\lambda)} -\frac{b(\lambda)}{a(\lambda)}
e^{2h(\lambda)} +U
\EN
and perform the following reparametrization
\EQ
\tilde{\lambda} = \frac{a(\lambda)}{b(\lambda)}e^{2h(\lambda)}
- \frac{b(\lambda)}{a(\lambda)}
e^{-2h(\lambda)} - \frac{U}{2}
\EN
we finally can rewrite  the auxiliary weights as
\EQ
\bar{a}(\tilde{\lambda},\tilde{\mu}) = \frac{U}{\tilde{\mu}-\tilde{\lambda} + U}~,~
\bar{b}(\tilde{\lambda},\tilde{\mu}) = \frac{\tilde{\mu}-\tilde{\lambda}}{\tilde{\mu}-\tilde{\lambda} + U}
\EN

Clearly, these are  the non-trivial Boltzmann weights of 
the isotropic $6$-vertex model. This is an important 
hidden symmetry, which is known to play a decisive role on 
the exact solution of the Hubbard model since the work of 
Lieb and Wu \cite{LW}. The derivation of this
symmetry in the context of the quantum inverse scattering program
is however a rather non-trivial result. One of the virtues of this result
is that it becomes 
valid for the generator of the
commuting conserved charges and not only for the Hubbard
Hamiltonian. Moreover, we also  recall that this  
symmetry is of relevance 
to the Yangian invariance of the Hubbard model which emerges in the
thermodynamic limit \cite{UK,GOMU1}.

To solve the eigenvalue problem (22) we still need the help of 
several other commutation relations. For instance, the commutation rules 
between the diagonal and creation operators play an important role in the 
eigenvalue construction. It turns out that in some cases we have 
to take into account similar trick discussed above. 
This is specially important for the field $\hat{A}(\lambda)$, where we 
have to use an auxiliary exchange relation between 
the operator $B(\mu)$ and $\vec{B^{*}}(\lambda)$, in order to 
obtain  a more appropriate commutation rule 
with the creation operator $\vec{B}(\lambda)$. In general, the task is quite 
cumbersome and here we limit ourselves to list the final results. The 
commutation relations between the diagonal fields and the creation operator 
$\vec{B}(\lambda)$ are 
\bear
\hat{A}(\lambda) \otimes \vec{B}(\mu) & = &
-i\frac{\alpha_{1}(\lambda,\mu)}{\alpha_{9}(\lambda,\mu)}[\vec{B}(\mu) \otimes 
\hat{A}(\lambda) ]. \hat{r}(\lambda,\mu)
+i \frac{\alpha_{5}(\lambda,\mu)}{\alpha_{9}(\lambda,\mu)} \vec{B}(\lambda) \otimes \hat{A}(\mu)   
\nonumber \\
&& -i\frac{\alpha_{10}(\lambda,\mu)}{\alpha_{7}(\lambda,\mu)} \left[ \vec{B^{*}}(\lambda)B(\mu) 
+i\frac{\alpha_{5}(\lambda,\mu)}{\alpha_{9}(\lambda,\mu)}F(\lambda)\vec{C}(\mu)  
-i \frac{\alpha_{2}(\lambda,\mu)}{\alpha_{9}(\lambda,\mu)}F(\mu)\vec{C}(\lambda) \right ] 
\otimes \vec{\xi}
\nonumber \\
\ear
\EQ
B(\lambda)\vec{B}(\mu) = 
i \frac{\alpha_{2}(\mu,\lambda)}{\alpha_{9}(\lambda,\mu)} \vec{B}(\mu)B(\lambda) -i 
\frac{\alpha_{5}(\mu,\lambda)}{\alpha_{9}(\lambda,\mu)} \vec{B}(\lambda)B(\mu)
\EN
\bear
D(\lambda)\vec{B}(\mu) & = &
-i \frac{\alpha_{8}(\lambda,\mu)}{\alpha_{7}(\lambda,\mu)} \vec{B}(\mu)D(\lambda) 
+ \frac{\alpha_{5}(\lambda,\mu)}{\alpha_{7}(\lambda,\mu)} F(u)\vec{C^{*}}(\lambda)
 \nonumber \\ 
 && - \frac{\alpha_{4}(\lambda,\mu)}{\alpha_{7}(\lambda,\mu)} F(\lambda)\vec{C^{*}}(\mu)
- i \frac{\alpha_{10}(\lambda,\mu)}{\alpha_{7}(\lambda,\mu)} 
\vec{\xi}. [ \vec{B^{*}}(\lambda) \otimes \hat{A}(\mu)] 
\ear
while those for the scalar field $F(\lambda)$ are
\bear
\hat{A}_{ab}(\lambda)F(\mu) & = &
[1 + \frac{\alpha^{2}_{5}(\lambda,\mu)}{\alpha_{9}(\lambda,\mu)\alpha_{8}(\lambda,\mu)}] F(\mu)\hat{A}_{ab}(\lambda)
- \frac{\alpha^{2}_{5}(\lambda,\mu)}{\alpha_{9}(\lambda,\mu)\alpha_{8}(\lambda,\mu)} F(\lambda) \hat{A}_{ab}(\mu)
\nonumber \\
 && + i\frac{\alpha_{5}(\lambda,\mu)}{\alpha_{9}(\lambda,\mu)} 
 [\vec{B}(\lambda) \otimes \vec{B^{*}}(\mu)]_{ba} 
+i\frac{\alpha_{5}(\lambda,\mu)}{\alpha_{8}(\lambda,\mu)} 
 [\vec{B^{*}}(\lambda) \otimes \vec{B}(\mu)]_{ab} 
\ear
\EQ
B(\lambda)F(\mu) = 
\frac{\alpha_{2}(\mu,\lambda)}{\alpha_{7}(\mu,\lambda)} F(\mu)B(\lambda) - 
\frac{\alpha_{4}(\mu,\lambda)}{\alpha_{7}(\mu,\lambda)} F(\lambda)B(\mu)
+i\frac{\alpha_{10}(\mu,\lambda)}{\alpha_{7}(\mu,\lambda)} 
\{ \vec{B}(\lambda) \otimes \vec{B}(\mu) \}. \vec{\xi}^{t}
\EN
\EQ
D(\lambda)F(\mu) = 
\frac{\alpha_{2}(\lambda,\mu)}{\alpha_{7}(\lambda,\mu)} F(\mu)D(\lambda) - 
\frac{\alpha_{4}(\lambda,\mu)}{\alpha_{7}(\lambda,\mu)} F(\lambda)D(\mu)
- i\frac{\alpha_{10}(\lambda,\mu)}{\alpha_{7}(\lambda,\mu)} \vec{\xi} . 
\{ \vec{B^{*}}(\lambda) \otimes \vec{B^{*}}(\mu) \}
\EN
where $\vec{\xi}^{t}$ stands for the transpose of $\vec{\xi}$.
Furthermore, the relations closing the commutation rules between the 
creation operators $\vec{B}(\lambda)$ and $F(\lambda)$ are
\EQ
\left[ F(\lambda), F(\mu) \right] = 0
\EN
\EQ
F(\lambda) \vec{B}(\mu) = \frac{\alpha_{5}(\lambda,\mu)}{\alpha_{2}(\lambda,\mu)} F(\mu) \vec{B}(\lambda) -i\frac{\alpha_{8}(\lambda,\mu)}{\alpha_{2}(\lambda,\mu)} \vec{B}(\mu) F(\lambda)
\EN
\EQ
\vec{B}(\lambda) F(\mu) = \frac{\alpha_{5}(\lambda,\mu)}{\alpha_{2}(\lambda,\mu)} \vec{B}(\mu) F(\lambda)  -i \frac{\alpha_{9}(\lambda,\mu)}{\alpha_{2}(\lambda,\mu)} F(\mu) \vec{B}(\lambda)
\EN

Finally, it  remains to consider the commutation rules for 
the creation field $\vec{B^{*}}(\lambda)$. To avoid overcrowding this
section with more heavier formulae we have collected them in appendix $B$. We
see that they are quite similar to those we just derived for the field
$\vec{B}(\lambda)$. In fact, it is possible
to establish an  equivalence between these two sets 
of commutation rules if we formally interpret the symbol
$*$ as a mathematical operation acting on the elements of the
monodromy matrix. For  lack of a better name we call it ``dual'' 
transformation and we impose that it satisfies the following properties:
$\left ( O^{*}(\lambda) \right )^{*}  
\equiv O(\lambda)$, $ A^{*}(\lambda) \equiv -A^{t}(\lambda) $, $B^{*}(\lambda)
\equiv D(\lambda) $,  $F^{*}(\lambda)=F(\lambda)$ and $C^{*}(\lambda)=C(\lambda)$.
Applying the ``dual'' transformation on
the commutation rules of field $\vec{B}(\lambda)$ we obtain those for the
field $\vec{B^*}(\lambda)$ with new Boltzmann weights $\alpha^{*}_j(\lambda,
\mu;h) \equiv \alpha_j(\lambda,\mu;-h)$, where, for sake of clarity, we
stressed the dependence on the constraint $h(\lambda)$. This means that
the functional
form of the weights  remains unchanged but now we have to perform the 
transformation 
$ h(\lambda) \rightarrow - h(\lambda)$ ($ U \rightarrow -U$). We recall
that in this last
step we used  the following identities for 
the Boltzmann weights:
$\alpha_j(\lambda,\mu;h) = \alpha_j(\mu,\lambda;-h)~ j=1,\ldots 7$,
$\alpha_8(\lambda,\mu;h) = \alpha_9(\mu,\lambda;-h)$ and 
$\alpha_{10}(\lambda,\mu;h) = -\alpha_{10}(\mu,\lambda;-h)$. 
Therefore, we expect that the construction of the eigenvectors
will be based either on the pair of fields 
 $ \vec{B}(\lambda)$ and $F(\lambda) $ or on the ``dual'' ones
$ \vec{B^{*}}(\lambda)$ and $ F(\lambda) $ rather 
than on a general combination 
of the three  creation fields. This redundance
is in accordance to what one would expect from the space of states of the 
Hubbard Hamiltonian, since at a given site we can either  create a single 
electron (spin up or down) or a pair of electrons with opposite spins. 
We remark that such ``duality'' property  is not particular
to the Hubbard model but  it is rather a general feature present in
our framework
\cite{MP1}. 

At this point we have set up the basic tools to start the construction 
of the eigenvectors of the eigenvalue problem (22). In next section we will 
show how this problem can be solved with the help of the commutations rules 
$(25-26;34-42)$ and few other relations presented in appendix $B$.

\section{The eigenvectors and the eigenvalue construction}

The purpose of this section is to solve the eigenvalue problem for the
graded transfer matrix.
We shall 
begin by considering the construction of 
an Ansatz for the corresponding eigenvectors. The multi-particle
state are going to satisfy an important recurrence
relation. We will see that the eigenvalue 
problem (22) has a nested structure, i.e. it will depend on the 
solution of an 
inhomogeneous auxiliary problem related to the $6$-vertex hidden
symmetry.

\subsection{The eigenvalue problem}

The eigenvectors of the transfer matrix are in principle built up in 
terms of a linear combination of products of the many creation fields 
acting on the reference state. These Bethe states are often thought as 
multi-particle states, characterized by a set of rapidities parametrizing the 
creation fields. Before embarking on the technicalities of the construction of
an arbitrary $n$-particle state we first define it by the following
scalar product
\EQ
\ket{\Phi_{n}(\lambda_{1}, \ldots ,\lambda_{n})} =  
\vec {\Phi}_{n}(\lambda_{1}, \ldots ,\lambda_{n}).\vec{\cal{F}} \ket{0}
\EN
where the mathematical structure of vector 
$\vec {\Phi}_{n}(\lambda_{1}, \ldots ,\lambda_{n})$ will be described
in terms of the creation fields. At this stage the components of vector
$\vec{\cal{F}}$ are simply thought as coefficients of an arbitrary 
linear combination which would be determined later on. This reflects
the ``spin'' degrees of freedom of the space of states and we shall
denote such coefficients by
${\cal{F}}^{a_n \ldots a_1}$ where the index $a_{i}$ run 
over two possible values $a_{i}=1,2$. 

Let us now turn our attention to the construction of vector  
$\vec {\Phi}_{n}(\lambda_{1}, \ldots ,\lambda_{n})$. As mentioned at
the end of the previous section, it is   
sufficient 
to look for 
combinations between the fields $\vec{B}(\lambda)$ and $F(\lambda)$. 
In general, there is no known recipe which is able to provide us with an 
educated Ansatz for this vector and as it is 
customary we shall start the construction considering few particle 
excitations over the reference state. A single particle excitation 
is made by creating a hole of spin up or down 
on the full band pseudovacuum $\ket{0}$. 
From the point of view of the inverse scattering method this excitation is 
represented by $\vec{\Phi}_{1}(\lambda_{1})=\vec{B}(\lambda_{1})$ and 
consequently the one-particle state is
\EQ
\ket{\Phi_{1}(\lambda_{1})} = \vec{B}(\lambda_1). \vec{\cal{F}} \ket{0} = B_{a}(\lambda_{1}){\cal{F}}^{a}\ket{0}
\EN
where from now on we assume sum over repeated index.

It is not difficult to solve the eigenvalue problem (22) for such 
one-particle state. If we use the commutation relations (34-36), and 
the pseudovacuum properties (23-24) we find  that the one-particle state 
satisfies the
following  relations
\EQ
B(\lambda) \ket{\Phi_{1}(\lambda_{1})} = 
i\frac{\alpha_{2}(\lambda_{1},\lambda)}{\alpha_{9}(\lambda_{1},\lambda)} 
[\omega_1(\lambda)]^{L}
\ket{\Phi_{1}(\lambda_{1})}
-i\frac{\alpha_{5}(\lambda_{1},\lambda)}
{\alpha_{9}(\lambda_{1},\lambda)}  
[\omega_1(\lambda_1)]^{L} \vec{B}(\lambda).\vec{{\cal{F}}} \ket{0}
\EN
\EQ
D(\lambda) \ket{\Phi_{1}(\lambda_{1})} =
-i\frac{\alpha_{8}(\lambda,\lambda_{1})}{\alpha_{7}(\lambda,\lambda_{1})} 
[\omega_3(\lambda)]^{L}
\ket{\Phi_{1}(\lambda^{(1)}_{1})}
-i\frac{\alpha_{10}(\lambda,\lambda_{1})}
{\alpha_{7}(\lambda,\lambda_{1})}
[\omega_2(\lambda_{1})]^{L}
[\vec{\xi}.(\vec{B^{*}}(\lambda) \otimes \hat{I})]. \vec{{\cal{F}}} \ket{0}
\EN
\bear
\sum_{a=1}^{2}  A_{aa}(\lambda) \ket{\Phi_{1}(\lambda_{1})} & = &
-i\frac{\alpha_{1}(\lambda,\lambda_{1})}{\alpha_{9}(\lambda,\lambda_{1})} 
\hat{r}_{c_{1}a_{1}}^{a_{1}b_{1}}(\lambda,\lambda_{1}) 
[\omega_2(\lambda)]^{L}
B_{c_{1}}(\lambda_{1}) {\cal F}^{b_{1}} \ket{0}
\nonumber \\
&& +i\frac{\alpha_{5}(\lambda,\lambda_{1})}{\alpha_{9}(\lambda,\lambda_{1})} 
[\omega_2(\lambda_1)]^{L} \vec{B}(\lambda). \vec{{\cal F}} \ket{0} 
\nonumber \\
&& -i\frac{\alpha_{10}(\lambda,\lambda_{1})}
{\alpha_{7}(\lambda,\lambda_{1})} 
[\omega_1(\lambda_{1})]^{L}
[\vec{\xi}.(\vec{B^{*}}(\lambda) \otimes \hat{I})]. \vec{{\cal{F}}} \ket{0}
\ear
where $\hat{I}$ is the $2 \times 2$ identity matrix.
The terms proportional to the eigenvector $ \ket{\Phi_{1}(\lambda_{1})}$ 
are denominated wanted terms because they contribute directly to the 
eigenvalue. The remaining ones are  called unwanted terms and  
they  can be eliminated 
by imposing further restriction on the rapidity $\lambda_{1}$. 
This constraint, known as the Bethe Ansatz equation, is given by
\EQ
\left[ \frac{\omega_1(\lambda_{1})}{\omega_2(\lambda_1)} \right]^{L} = 1
\EN

It is now straightforward to go ahead and to determine the one-particle
eigenvalue. However, it is convenient to start introducing suitable
notation which can be extended to accommodate multi-particle states.
With this in mind,  we define 
the following auxiliary eigenvalue
problem 
\EQ
T^{(1)}(\lambda,\lambda_{1})_{b_{1}}^{a_{1}} {\cal F}^{a_{1}} = 
\hat{r}_{b_{1}\alpha}^{\alpha a_{1}}(\lambda,\lambda_{1}) {\cal F}^{a_{1}} =
\Lambda^{(1)}(\lambda,\lambda_{1}) {\cal F}^{b_{1}}
\EN
and we see that, in terms of equation (49),  the one-particle eigenvalue can be 
expressed by
\EQ
\Lambda(\lambda,\lambda_{1}) = 
i\frac{\alpha_{2}(\lambda_{1},\lambda)}{\alpha_{9}(\lambda_{1},\lambda)} 
[\omega_1(\lambda)]^{L} 
-i\frac{\alpha_{8}(\lambda,\lambda_{1})}{\alpha_{7}(\lambda,\lambda_{1})} 
[\omega_3(\lambda)]^{L}
+i\frac{\alpha_{1}(\lambda,\lambda_{1})}{\alpha_{9}(\lambda,\lambda_{1})}
\Lambda^{(1)}(\lambda,\lambda_{1})
[\omega_2(\lambda)]^{L}
\EN

Up to the level of the one-particle state there is no extra effort 
to solve the corresponding auxiliary problem.  
Considering 
the $6$-vertex 
structure of matrix $\hat{r}(\lambda,\mu)$ it is easily seen that the solution
is
\EQ
\Lambda^{(1)}(\lambda,\lambda_{1}) = 1+\bar{b}(\lambda,\lambda_{1})
\EN

We next turn to the analysis of the two-particle state. We expect that 
such state will be a composition between two single hole excitations of 
arbitrary spins and a local hole pair with opposite spins. The former is 
made by tensoring two fields of $\vec{B}(\lambda)$ type while the later should
be  
represented by $F(\lambda)$. The vector $\vec{\xi}$ has also a physical meaning.
It plays the role of 
an ``exclusion'' principle, forbidding  two spin up or 
two spin down at the same site. Thus, an educate  Ansatz for the
two-particle vector should be the linear combination
\bear
\vec{\Phi}_{2}(\lambda_{1},\lambda_{2}) =
\vec{B}(\lambda_{1}) \otimes \vec{B}(\lambda_{2}) + 
\vec{\xi} F(\lambda_{1}) 
B(\lambda_{2})\hat{g}^{(2)}_{0}(\lambda_{1},\lambda_{2})
\ear
where $\hat{g}^{(2)}_{0}(\lambda_{1},\lambda_{2})$ 
is an arbitrary function to be 
determined. We found also convenient to 
add the 
diagonal field $B(\lambda_2)$  on the right-hand side of the 
two-particle vector
Ansatz. We see that when
the Ansatz (52) is projected out 
on the subspace of equal spins, no contribution coming from $F(\lambda)$ 
appears, which is in perfect accordance to what one would expect from the 
Pauli principle. In other words, using the definition (43) we have 
\EQ
\ket{\Phi_{2}(\lambda_{1},\lambda_{2})} =
B_{i}(\lambda_{1}) B_{j}(\lambda_{2}) {\cal F}^{ji} \ket{0} + 
[\omega_1(\lambda_{2})]^{L} 
F(\lambda_{1}) \hat{g}^{(2)}_{0}(\lambda_{1},\lambda_{2})
({\cal F}^{21}-{\cal F}^{12})
\EN

In order to tackle the eigenvalue problem for the 
two-particle state,  
besides the commutation 
rules of the last section, we have to use extra relations between the fields 
$\vec{B}(\lambda)$, $\vec{B}^{*}(\lambda)$, $\vec{C}(\lambda)$ and 
$\vec{C}^{*}(\lambda)$. These relations have been summarized in
the beginning of appendix $B$. After turning
the diagonal fields over the two-particle state, we find that there are
two classes of unwanted terms. The first class we call
``easy'' unwanted terms because they are only produced  by the same
diagonal operator ( $\hat{A}(\lambda)$ or $D(\lambda)$ ) and they can
be eliminated by an appropriate choice of 
function $\hat{g}^{(2)}_{0}(\lambda_{1},\lambda_{2})$. 
There are   three terms of this sort
\EQ
F(\lambda) D(\lambda_{1}) B(\lambda_{2})~~,~~
\vec{B}(\lambda).\vec{B^{*}}(\lambda_{1}) B(\lambda_{2})~~,~~
\vec{\xi}.[\vec{B^{*}}(\lambda) \otimes \vec{B^{*}}(\lambda_{1})] B(\lambda_{2}) \EN
and all of them are cancelled out provided we chose   
function $\hat{g}^{(2)}_{0}(\lambda_{1},\lambda_{2})$ as
\EQ
\hat{g}^{(2)}_{0}(\lambda_{1},\lambda_{2}) = i\frac{\alpha_{10}(\lambda_{1},\lambda_{2})}
{\alpha_{7}(\lambda_{1},\lambda_{2})}
\EN

Now, besides the wanted terms, we are only left with 
standard unwanted terms, i.e.
those that require further restriction on the rapidities. We shall see below
that
these terms can be simplified in rather closed forms with the help of the
two-particle auxiliary problem. Similar to the one-particle analysis, the
auxiliary eigenvalue problem is figured out by looking at the wanted terms
coming from the operator $ \displaystyle \sum_{a=1}^{2} \hat{A}_{aa}(\lambda)$. Considering the
commutation rule (34) we soon realize that the  two-particle
auxiliary problem is
\EQ
T^{(1)}(\lambda,\{\lambda_{l}\})_{b_{1}b_{2}}^{a_{1}a_{2}} {\cal F}^{a_{2}a_{1}} = 
\hat{r}_{b_{1}d_{1}}^{c_{1}a_{1}}(\lambda,\lambda_{1}) 
\hat{r}_{b_{2}c_{1}}^{d_{1}a_{2}}(\lambda,\lambda_{2}) {\cal F}^{a_{2}a_{1}} =
\Lambda^{(1)}(\lambda,\{\lambda_{l}\}) {\cal F}^{b_{2}b_{1}}
\EN

With the above  information we move on simplifying as much as possible the
action of the diagonal fields on the two-particle state. We keep in mind that
we want to present the results in a way that would be amenable to multi-particle
states generalization. After a cumbersome algebra we find that
\bear
B(\lambda) \ket{\Phi_{2}(\lambda_{1},\lambda_2)} & = &
[\omega_1(\lambda)]^{L} \prod_{j=1}^{2} 
i\frac{\alpha_{2}(\lambda_{j},\lambda)}{\alpha_{9}(\lambda_{j},\lambda)} 
\ket{\Phi_{2}(\lambda_{1},\lambda_2)}
\nonumber \\
&& -\sum_{j=1}^{2}  
[\omega_1(\lambda_{j})]^{L}  \ket{\Psi^{(1)}_{1}(\lambda,\lambda_{j}; \{ \lambda_{l} \})}  
\nonumber \\
&& + H_1(\lambda,\lambda_1,\lambda_2) 
[\omega_1(\lambda_1) \omega_1(\lambda_2)]^{L} 
\ket{\Psi^{(3)}_{0}(\lambda,\lambda_j,\lambda_l;\{ \lambda_k \} ) }
\ear
\bear
D(\lambda) \ket{\Phi_{2}(\lambda_{1},\lambda_2)} & = &
[\omega_3(\lambda)]^{L} \prod_{j=1}^{2} 
-i\frac{\alpha_{8}(\lambda,\lambda_j)}{\alpha_{7}(\lambda,\lambda_j)} 
\ket{\Phi_{2}(\lambda_{1},\lambda_2)}
\nonumber \\
&& -\sum_{j=1}^{2} 
[\omega_2(\lambda_{j})]^{L} 
\Lambda^{(1)}(\lambda=\lambda_j,\{ \lambda_l \})
  \ket{\Psi^{(2)}_{1}(\lambda,\lambda_{j}; \{ \lambda_{l} \})}  
\nonumber \\
&& + H_{2}(\lambda,\lambda_1,\lambda_2) 
[\omega_2(\lambda_1) \omega_2(\lambda_2)]^{L} 
\ket{\Psi^{(3)}_{0}(\lambda,\lambda_j,\lambda_l;\{ \lambda_k \} ) }
\ear
\bear
\sum_{a=1}^{2} {\hat{A}}_{aa}(\lambda) 
\ket{\Phi_{2}(\lambda_{1},\lambda_2)} & = &
[\omega_2(\lambda)]^{L} \prod_{j=1}^{2} 
-i\frac{\alpha_{1}(\lambda,\lambda_j)}{\alpha_{9}(\lambda,\lambda_j)} 
\Lambda^{(1)}(\lambda,\{ \lambda_l \})
\ket{\Phi_{2}(\lambda_{1},\lambda_2)}
\nonumber \\
&& -\sum_{j=1}^{2}  
[\omega_2(\lambda_{j})]^{L} 
\Lambda^{(1)}(\lambda=\lambda_j,\{ \lambda_l \})
\ket{\Psi^{(1)}_{1}(\lambda,\lambda_{j};\{ \lambda_{l}\} )}
\nonumber \\
&& -\sum_{j=1}^{2} [\omega_1(\lambda_{j})]^{L} 
\ket{\Psi^{(2)}_{1}(\lambda,\lambda_{j};\{ \lambda_{l} \} )}
\nonumber \\
&& + H_3(\lambda,\lambda_1,\lambda_2) 
[\omega_1(\lambda_1) \omega_2(\lambda_2)]^{L} 
\ket{\Psi^{(3)}_{0}(\lambda,\lambda_j,\lambda_l;\{ \lambda_k \} ) }
\nonumber \\
&& +H_4(\lambda,\lambda_1,\lambda_2)
[\omega_1(\lambda_2) \omega_2(\lambda_1)]^{L} 
\ket{\Psi^{(3)}_{0}(\lambda,\lambda_j,\lambda_l;\{ \lambda_k \} ) }
\ear

For sake of clarity we have shortened the notation for the unwanted terms
and represented them by the eigenfunctions 
$ \ket{\Psi^{(j)}_{1}(\lambda,\lambda_{j} ;\{ \lambda_{k}\} )} $
and $ \ket{\Psi^{(3)}_{0}(\lambda,\lambda_{j} ;\{ \lambda_{k}\} )} $. We
see that there are three classes of unwanted terms and their explicit 
expressions in terms of the creation fields are
\EQ
\ket{\Psi^{(1)}_{1}(\lambda,\lambda_{j};\{ \lambda_{l}\} )}  = 
i\frac{\alpha_{5}(\lambda_{j},\lambda)}
{\alpha_{9}(\lambda_{j},\lambda)} \prod^2_{ \stackrel {k=1}{ k \neq j}} 
i\frac{\alpha_{2}(\lambda_{k},\lambda_j)}{\alpha_{9}(\lambda_{k},\lambda_j)} 
\vec{B}(\lambda) \otimes \vec{\Phi}_1(\lambda_k) 
\hat{O}^{(1)}_j(\lambda_j; \{ \lambda_k \} )
.\vec{\cal{F}}
\ket{0}
\EN
\EQ
\ket{\Psi^{(2)}_{1}(\lambda,\lambda_{j};\{ \lambda_{l}\} )}  = 
i\frac{\alpha_{10}(\lambda,\lambda_j)}
{\alpha_{7}(\lambda,\lambda_j)} \prod^2_{ \stackrel {k=1}{ k \neq j}} 
i\frac{\alpha_{2}(\lambda_{k},\lambda_j)}{\alpha_{9}(\lambda_{k},\lambda_j)} 
[\vec{\xi}.(\vec{B^{*}}(\lambda)\otimes \hat{I})]  
\otimes \vec{\Phi}_1(\lambda_k) 
\hat{O}^{(1)}_j(\lambda_j; \{ \lambda_k \} )
.\vec{\cal{F}}
\ket{0}
\EN
\EQ
\ket{\Psi^{(3)}_{0}(\lambda,\lambda_{j}, \lambda_l;\{ \lambda_{k}\} )}  = 
F(\lambda) \vec{\xi} . \vec{\cal{F}}
\ket{0}
\EN
where the operator
$\hat{O}^{(1)}_j(\lambda_j; \{ \lambda_k \} )$ is a sort of ``ordering'' 
factor for the unwanted terms and it is given by the formula
\EQ
\hat{O}^{(1)}_j(\lambda_j; \{ \lambda_k \} ) =
\prod_{k=1}^{j-1} \frac{\alpha_1(\lambda_k,\lambda_j)}
{\alpha_2(\lambda_k,\lambda_j)} 
\hat{r}_{k,k+1}(\lambda_k,\lambda_j) 
\EN

Before proceeding with a discussion of the results,
we should pause to comment on the ``brute-force'' analysis
we performed so far for the two-particle state problem. Roughly speaking,
one can estimate the wanted terms by keeping the first term of the right-hand
side of the commutation rules (34-36) when we turn the diagonal
fields over the creation operators $\vec{B}(\lambda_j)$. This 
procedure gives us the
coefficients proportional to the first part of the eigenvector and to
show that this is also true for the second part we need 
to use  some identities
between the Boltzmann weights. The situation for the unwanted terms is even
worse due to the proliferation of many different terms, common in a
Bethe Ansatz ``brute-force'' analysis. The ``ordering'' factor just accounts
for these many different contributions to the unwanted terms.
Later on it will become clear that the origin
of this factor is due to a permutation property satisfied by the two-particle
eigenvector.
In appendix $C$ we provide the details
about the less straightforward simplifications carried 
out for the two-particle state, since some of them will be also useful to
multi-particle states as well.
Finally, within a ``brute force'' computation, we have found nine 
contributions to the third unwanted term which come from many different 
sources. It is possible to recast them in terms of four functions 
$H_i(x,y,z)~i=1, \ldots, 4$ whose expressions  
are 
\bear
H_1(x,y,z) & = &i\frac{\alpha_2(y,x) \alpha_5(z,x) \alpha_{10}(y,x)}
{\alpha_9(y,x) \alpha_9(z,x) \alpha_7(y,x)}
-i\frac{\alpha_4(y,x) \alpha_{10}(y,z)}
{\alpha_7(y,x) \alpha_7(y,z)}
\nonumber \\
H_2(x,y,z) & =& i\frac{\alpha_5(x,y) \alpha_{10}(x,z)}{\alpha_7(x,y) \alpha_7(x,z)}
-i \frac{\alpha_4(x,y) \alpha_{10}(y,z)}{\alpha_7(x,y) \alpha_7(y,z)}
\nonumber \\
H_3(x,y,z) & =&  i\frac{\alpha_{10}(x,y) \alpha_5(x,y) \alpha_5(y,z)}{\alpha_7(x,y)
\alpha_9(x,y) \alpha_9(y,z)} 
-i \frac{\alpha_2(x,y) \alpha_5(x,z) \alpha_{10}(x,y)}{\alpha_9(x,y) \alpha_9(x,z)
\alpha_7(x,y)}
\nonumber \\
H_4(x,y,z) &  = & -i\frac{\alpha_{10}(x,y) \alpha_5(x,y) \alpha_5(y,z) }{\alpha_7(x,y)
\alpha_9(x,y) \alpha_9(y,z)} 
+i \frac{\alpha_1(x,y) \alpha_{10}(x,z) \alpha_5(x,y)[1+\bar{a}(x,y)]}
{\alpha_9(x,y) \alpha_7(x,z)
\alpha_8(x,y)}
\nonumber \\
&& -2i \frac{\alpha_5^2(x,y) \alpha_{10}(y,z)}
{\alpha_8(x,y) \alpha_9(x,y) \alpha_7(y,z)}
\ear

Now we return to the discussion of the two-particle state results. For 
the first two classes of unwanted terms we only have two main 
contributions and from equations (57-59) it is direct to see that they vanish
provided that 
the rapidities satisfy the following Bethe Ansatz
equations
\EQ
\left[ \frac{\omega_1(\lambda_{i})}{\omega_2(\lambda_{i})} \right]^{L} = 
\Lambda^{(1)}(\lambda=\lambda_{i},\{\lambda_{j}\}), ~~ i=1,2
\EN

Furthermore, the above  Bethe Ansatz equations are also
sufficient to cancel out altogether the four contributions 
proportional  to the unwanted term  
$F(\lambda) \vec{\xi} . \vec{\cal{F}}$.
A simple way of seeing that
is first to factorize a common factor $[\omega_2(\lambda_1) \omega_2(\lambda_2)]^{L}$ for all the four terms. This is done by substituting the values 
$[\omega_1(\lambda_1)]^L$ and
$[\omega_1(\lambda_2)]^L$ given by the Bethe Ansatz equations (65) and by 
using
the following two-particle relations 
\bear
\Lambda^{(1)}(\lambda=\lambda_1, \{ \lambda_l \}) 
\vec{\xi} . \vec{\cal{F}} & =& [\bar{b}(\lambda_1,\lambda_2) 
-\bar{a}(\lambda_1, \lambda_2) ]
\vec{\xi} . \vec{\cal{F}} 
\nonumber\\
\Lambda^{(1)}(\lambda=\lambda_2, \{ \lambda_l \}) 
\vec{\xi} . \vec{\cal{F}} & =& [\bar{b}(\lambda_2,\lambda_1) 
-\bar{a}(\lambda_2, \lambda_1) ]
\vec{\xi} . \vec{\cal{F}} 
\ear

After putting  all these simplifications together, one is still left to verify
that the following identity 
\bear
H_1(x,y,z) +
H_2(x,y,z) & = & 
H_3(x,y,z)[\bar{b}(y,z) 
-\bar{a}(y,z)] 
\nonumber\\ 
&& +H_4(x,y,z)[\bar{b}(z,y) 
-\bar{a}(z,y)] 
\ear
is satisfied. At this point we note that there is a way of rewriting
the term  
$H_4(\lambda,\lambda_1,\lambda_2)$  in 
a more 
symmetrical form. This technical point is discussed
in appendix $C$ and proved very useful in carrying out
the cancellation mechanism 
for general multi-particle states. 

Finally, from equations (57-59) we can read directly the wanted terms, and
the two-particle eigenvalue is 
\bear
\Lambda(\lambda,\{\lambda_{i}\}) & = &
[\omega_1(\lambda)]^L 
\prod_{i=1}^{2} i \frac{\alpha_{2}(\lambda_{i},\lambda)}{\alpha_{9}(\lambda_{i},\lambda)} + 
[\omega_3(\lambda)]^L 
\prod_{i=1}^{2} -i\frac{\alpha_{8}(\lambda,\lambda_{i})}{\alpha_{7}(\lambda,\lambda_{i})}  
\nonumber\\
&& -[\omega_2(\lambda)]^L \prod_{i=1}^{2} -i \frac{\alpha_{1}(\lambda,\lambda_{i})}{\alpha_{9}(\lambda,\lambda_{i})} 
\Lambda^{(1)}(\lambda,\{\lambda_{j}\}) 
\ear

Now  we  reached  a point which is typical of nested Bethe Ansatz
problems, i.e. the solution of the two-particle auxiliary
problem is no longer trivial and it is necessary to implement 
a second Bethe Ansatz. We will postpone this discussion until the 
next subsection
in which  we will present the solution of this problem 
for general multi-particle states. 
Although, for an integrable model, it is believed that the two-particle sector
contains the essential features about the general structure of the
eigenvalues and the Bethe Ansatz equations, similar situation for the 
eigenvectors is still less clear. Before considering
this problem, it is wise to look first for an alternative way of starting
with a general Ansatz, since a ``brute force'' analysis proved to be
rather intricate even for the two-particle state. In fact, there is
a symmetry which we have not yet explored. It consists of seeking
for eigenvectors
which are in some way related to each other via permutation of the 
rapidities. This idea goes along the lines the usual pseudomomenta 
symmetrization imposed to coordinate 
Bethe Ansatz wave functions. For example, let
us consider the two-particle vector in which the constraint
$\hat{g}^{(2)}_{0}(\lambda_{1},\lambda_{2})$ has been 
fixed as in equation (55). Then, it is possible to verify that the 
following exchange property  
\EQ
\vec{\Phi}_{2}(\lambda_{1},\lambda_{2}) =
\frac{\alpha_{1}(\lambda_{1},\lambda_{2})}{\alpha_{2}(\lambda_{1},\lambda_{2})}
\vec{\Phi}_{2}(\lambda_{2},\lambda_{1}). \hat{r}(\lambda_{1},\lambda_{2})
\EN
is satisfied. In order to show that, we used 
a remarkable relation between vector 
$\vec{\xi}$, the auxiliary matrix 
$\hat{r}(\lambda,\mu)$ and the Boltzmann weights given by
\EQ
\vec{\xi}. \hat{r}(\lambda,\mu) = 
\frac{\alpha_{10}(\lambda,\mu)\alpha_{7}(\mu,\lambda)\alpha_{2}(\lambda,\mu)}
{\alpha_{7}(\lambda,\mu)\alpha_{10}(\mu,\lambda)\alpha_{1}(\lambda,\mu)} \vec{\xi}
\EN

Alternatively, we can reverse the arguments demanding that the eigenvectors
satisfy the exchange symmetry (69). This procedure gives us a restriction to
function
$\hat{g}^{(2)}_{0}(\lambda_{1},\lambda_{2})$ and it is  
an elegant way of fixing the linear combination from the very beginning. 
Now it is easy to understand the reason why an ``ordering'' factor
had emerged in the ``brute-force'' 
analysis of the two-particle state. For example, the
simplest way to
generate the unwanted terms 
$\vec{B}(\lambda)  
\otimes \vec{\Phi}_1(\lambda_1) $ and 
$[\vec{\xi}.(\vec{B^{*}}(\lambda)\otimes \hat{I})]  
\otimes \vec{\Phi}_1(\lambda_1) $ is  by  using the right-hand side of
equation (69) instead of the left-hand side we used in the whole
``brute force''
analysis. In this 
way we obviously generate only one contribution to
such unwanted terms which carries the ``ordering'' factor explicitly.

In
principle, such symmetrization mechanism can be implemented to any 
multi-particle state, and as we shall see below, it indeed help us to 
handle
the problem of constructing a general  
$n$-particle state Ansatz. We will start our discussion considering
the three-particle state. This state is expected to be 
a composition between the term representing the creation of three holes 
(arbitrary spins) on different sites and the three possible ways of combining 
pairs of holes with a single excitation. Within our algebraic framework the 
Ansatz encoding these features is
\bear
\vec{\Phi}_{3}(\lambda_{1},\lambda_{2},\lambda_{3}) &  =&
\vec{B}(\lambda_{1}) \otimes \vec{B}(\lambda_{2}) \otimes \vec{B}(\lambda_{3}) + [\vec{B}(\lambda_{1}) \otimes  \vec{\xi} F(\lambda_{2}) B(\lambda_{3})]  \hat{g}^{(3)}_{0}(\lambda_{1},\lambda_{2},\lambda_{3}) 
\nonumber\\
&& +
[\vec{\xi} \otimes F(\lambda_{1}) \vec{B}(\lambda_{3}) B(\lambda_{2})]
\hat{g}^{(3)}_{1}(\lambda_{1},\lambda_{2},\lambda_{3}) +
[\vec{\xi} \otimes F(\lambda_{1}) \vec{B}(\lambda_{2}) B(\lambda_{3})]
\hat{g}^{(3)}_{2}(\lambda_{1},\lambda_{2},\lambda_{3})
\nonumber\\
\ear
where the coefficientes 
$\hat{g}^{(3)}_{j}(\lambda_{1},\lambda_{2},\lambda_{3})$ are going to
be determined assuming a priori an exchange property (cf. equation (78))
for the
$\lambda_{1} \leftrightarrow \lambda_{2}$ and 
$\lambda_{2} \leftrightarrow \lambda_{3}$ permutations.  To see how this
works in practice, let us first implement the permutation between the
rapidities $\lambda_2$ and $\lambda_3$. To this end we use the commutation
relation (25) to reorder these rapidities in the permuted three-particle
vector
$\vec{\Phi}_{3}(\lambda_{1},\lambda_{3},\lambda_{2})$. This allows us to
write the following relation
\bear
\frac{\alpha_1(\lambda_2,\lambda_3)}{\alpha_2(\lambda_2,\lambda_3)}
\vec{\Phi}_{3}(\lambda_{1},\lambda_{3},\lambda_{2}).
\hat{r}_{23}(\lambda_2,\lambda_3)  &  =&
\vec{B}(\lambda_{1}) \otimes \vec{B}(\lambda_{2}) 
\otimes \vec{B}(\lambda_{3})  
+i\frac{\alpha_{10}(\lambda_2,\lambda_3)}{\alpha_7(\lambda_2,\lambda_3)}
[\vec{B}(\lambda_{1}) \otimes  \vec{\xi} F(\lambda_{2}) B(\lambda_3)]
\nonumber\\
&& +[\vec{B}(\lambda_1) \otimes \vec{\xi} F(\lambda_{3}) B(\lambda_{2})][ 
-i\frac{\alpha_{10}(\lambda_2,\lambda_3)}{\alpha_7(\lambda_2,\lambda_3)} 
+\frac{\alpha_1(\lambda_2,\lambda_3)}{\alpha_2(\lambda_2,\lambda_3)}
\nonumber\\
&& \times \hat{g}^{(3)}_{0}(\lambda_{1},\lambda_{3},\lambda_{2}).
\hat{r}_{23}(\lambda_2,\lambda_3)  ]
\nonumber \\
&& +[\vec{\xi} \otimes F(\lambda_{1}) \vec{B}(\lambda_{2}) B(\lambda_{3})]
\frac{\alpha_1(\lambda_2,\lambda_3)}{\alpha_2(\lambda_2,\lambda_3)}
\hat{g}^{(3)}_{1}(\lambda_{1},\lambda_{3},\lambda_{2}).
\hat{r}_{23}(\lambda_2,\lambda_3) 
\nonumber\\
&& +[\vec{\xi} \otimes F(\lambda_{1}) \vec{B}(\lambda_{3}) B(\lambda_{2})]
\frac{\alpha_1(\lambda_2,\lambda_3)}{\alpha_2(\lambda_2,\lambda_3)}
\hat{g}^{(3)}_{2}(\lambda_{1},\lambda_{3},\lambda_{2}).
\hat{r}_{23}(\lambda_2,\lambda_3)
\nonumber\\
\ear

Imposing the exchange property to the three-particle state,
i.e. that the right-hand sides 
of equations (71) and (72) are equal, we are able to derive 
constraints  to functions
$\hat{g}^{(3)}_{j}(\lambda_{1},\lambda_{2},\lambda_{3})$. We find that
it is sufficient to have 
\EQ
\hat{g}^{(3)}_{0}(\lambda_{1},\lambda_{2},\lambda_{3}) =
i\frac{\alpha_{10}(\lambda_{2},\lambda_{3})}
{\alpha_{7}(\lambda_{2},\lambda_{3})}
\EN
and 
\EQ
\hat{g}^{(3)}_{2}(\lambda_{1},\lambda_{2},\lambda_{3}) =
\frac{\alpha_1(\lambda_2,\lambda_3)}{\alpha_2(\lambda_2,\lambda_3)}
\hat{g}^{(3)}_{1}(\lambda_{1},\lambda_{3},\lambda_{2}).
\hat{r}_{23}(\lambda_2,\lambda_3)
\EN
where we used the identities
$\hat{r}_{23}(\lambda_2,\lambda_3).
\hat{r}_{23}(\lambda_3,\lambda_2) = \hat{I}$ and $\alpha_1(\lambda,\mu)=
\alpha_2(\mu,\lambda)$. We recall that relation (70) helps us to
cancel out the third term of equation (72). Now it remains to determine
function
$\hat{g}^{(3)}_{1}(\lambda_{1},\lambda_{2},\lambda_{3})$ and this can be done
by using the permutation between the variables $\lambda_1$ and $\lambda_2$.
The technical steps of this computation are more involving, since it is
necessary to use other commutation rules and  some identities between
the Boltzmann weights. The details are presented 
in appendix $D$ and here we quote our result for the remaining
functions
\bear
\hat{g}^{(3)}_{1}(\lambda_{1},\lambda_{2},\lambda_{3}) & = &
i\frac{\alpha_{10}(\lambda_{1},\lambda_{2})}
{\alpha_{7}(\lambda_{1},\lambda_{2})}
i\frac{\alpha_{2}(\lambda_{3},\lambda_{2})}
{\alpha_{9}(\lambda_{3},\lambda_{2})} \nonumber \\
\hat{g}^{(3)}_{2}(\lambda_{1},\lambda_{2},\lambda_{3}) &  = &
i\frac{\alpha_{10}(\lambda_{1},\lambda_{3})}
{\alpha_{7}(\lambda_{1},\lambda_{3})}
i\frac{\alpha_{1}(\lambda_{2},\lambda_{3})}
{\alpha_{9}(\lambda_{2},\lambda_{3})}
\hat{r}_{23}(\lambda_{2},\lambda_{3})
\ear

To make sure we are on the right track, we  have
checked that the three-particle ``easy'' 
unwanted terms  are automatically canceled out provided we fix the constraints
$\hat{g}^{(3)}_{j}(\lambda_{1},\lambda_{2},\lambda_{3})$ as
in equations  (73) and (75). We note that functions 
$\hat{g}^{(3)}_{0}(x,y,z)$ and
$\hat{g}^{(2)}_{0}(y,z)$ are identical, and this allows us to rewrite
the three-particle vector in terms of the following
recurrence relation 
\bear
\vec {\Phi}_{3}(\lambda_{1},\lambda_{2},\lambda_{3}) & = &
\vec {B}(\lambda_{1}) \otimes \vec {\Phi}_{2}(\lambda_{2},\lambda_{3})
\nonumber\\
&& + \sum_{j=2}^3
\left [ \vec{\xi} \otimes F(\lambda_1) 
 \vec {\Phi}_{1}(\lambda_{2},\ldots,\lambda_{j-1},\lambda_{j+1},\ldots,\lambda_{3}) B(\lambda_j) \right ] \hat{g}^{(3)}_{j-1}(\lambda_{1},\lambda_{2},\lambda_{3})
\nonumber\\
\ear

This expression is rather illuminating, because it suggests
that we can write  a general $n$-particle state in terms of
the $(n-1)$-particle and $(n-2)$-particle states via a recurrence relation. 
From our expressions for the two-particle 
and the three-particle states it is not
difficult to guess that the $n$-particle vector should be given by
\bear
\vec {\Phi}_{n}(\lambda_{1},\ldots ,\lambda_{n}) & = &
\vec {B}(\lambda_{1}) \otimes \vec {\Phi}_{n-1}(\lambda_{2},\ldots ,\lambda_{n})
\nonumber \\
&& + \sum_{j=2}^n 
\left [ \vec{\xi} \otimes F(\lambda_1) 
 \vec {\Phi}_{n-2}(\lambda_{2},\ldots,\lambda_{j-1},\lambda_{j+1},\ldots,\lambda_{n}) B(\lambda_j) \right ] \hat{g}^{(n)}_{j-1}(\lambda_{1},\ldots ,\lambda_{n})
\nonumber \\
\ear
where here we  
formally identified $\vec{\Phi}_{0}$ with the unity vector. Our next step
is to implement the
symmetrization scheme for such multi-particle state Ansatz.
The best way to proceed here is to use 
mathematical induction, i.e we assume that the $(n-2)$-particle and the
$(n-1)$-particle states were already symmetrized to infer the constraints
$\hat{g}^{(n)}_{j}(\lambda_{1},\ldots ,\lambda_{n}) $ for the $n$-particle
state. For this purpose we impose that any consecutive permutation between
the rapidities $\lambda_{j-1}$ and $\lambda_j$ ($ j=2, \ldots, n$) satisfies
the following exchange property 
\EQ
\vec {\Phi}_{n}(\lambda_{1},\ldots,\lambda_{j-1},\lambda_{j},\ldots,\lambda_{n}) =
\frac{\alpha_{1}(\lambda_{j-1},\lambda_{j})}{\alpha_{2}(\lambda_{j-1},\lambda_{j})}
\vec{\Phi}_{n}(\lambda_{1},\ldots,\lambda_{j},\lambda_{j-1},\ldots,\lambda_{n}).
\hat{r}_{j-1,j}(\lambda_{j-1},\lambda_{j})
\EN
where the indices under $\hat{r}_{j-1,j}(\lambda_{j-1},\lambda_{j})$ 
emphasize the positions on the $n$-particle space 
$1 \otimes \ldots \otimes j-1 \otimes j \ldots \otimes n$ in which this 
matrix acts non-trivially. 

Now starting with the latest permutation $j=n$ we go ahead comparing the
terms proportional to 
$[ \vec{\xi} \otimes F(\lambda_1) 
\vec {\Phi}_{n-2}(\lambda_{2},\ldots,
\lambda_{j-2},\lambda_{j},\ldots,\lambda_{n}) B(\lambda_{j-1})  ] $ and $
[ \vec{\xi} \otimes F(\lambda_1) 
\vec {\Phi}_{n-2}(\lambda_{2},\ldots,\lambda_{j-1}, $ $
\lambda_{j+1},\ldots,\lambda_{n}) B(\lambda_j)  ] $ in both 
sides of the exchange relation (78). At each step, this
yields a set of  relations between the functions
$\hat{g}^{(n)}_{j}(\lambda_{1},\ldots ,\lambda_{n}) $ which are 
further simplified by using  explicitly both 
the unitarity condition  and the Yang-Baxter equation for the auxiliary
$r$-matrix. Up to $j=3$ we find that such functions satisfy the following
recurrence relation 
\EQ
\hat{g}^{(n)}_{j-1}(\lambda_{1},\ldots,\lambda_{j-1},\lambda_{j},\ldots  ,\lambda_{n})  = 
\frac{\alpha_{1}(\lambda_{j-1},\lambda_{j})}{\alpha_{2}(\lambda_{j-1},\lambda_{j})}
\hat{g}^{(n)}_{j-2}(\lambda_{1},\ldots,\lambda_{j},\lambda_{j-1},\ldots  ,\lambda_{n})
\hat{r}_{j-1,j}(\lambda_{j-1},\lambda_{j})
\EN

Next we implement the symmetrization $\lambda_1 \leftrightarrow \lambda_2$ 
along the lines sketched in appendix $D$ for the three-particle state. In
this case we have to eliminate the term proportional to
$[ \vec{\xi} \otimes F(\lambda_1) 
\vec {\Phi}_{n-2}(\lambda_{3},\ldots,\lambda_{n}) B(\lambda_{2})  ] $
which only occurs in the left-hand side of the exchange relation (78).
This condition helps us to determine the expression for the first
constraint and we have
\EQ
\hat{g}^{(n)}_{1}(\lambda_{1},\ldots ,\lambda_{n})  = 
i\frac{\alpha_{10}(\lambda_{1},\lambda_{2})}
{\alpha_{7}(\lambda_{1},\lambda_{2})}
\prod^{n}_{k=3} i\frac{\alpha_{2}(\lambda_{k},\lambda_{2})}
{\alpha_{9}(\lambda_{k},\lambda_{2})}
\EN

Finally, the set of relations (79) and (80)  are solved recursively and 
we find that the $n$-particle vector is
\bear
\vec {\Phi}_{n}(\lambda_{1},\ldots ,\lambda_{n}) & = &
\vec {B}(\lambda_{1}) \otimes \vec {\Phi}_{n-1}(\lambda_{2}, 
\ldots ,\lambda_{n})
 + 
\sum_{j=2}^n 
i\frac{\alpha_{10}(\lambda_{1},\lambda_{j})}{\alpha_{7}(\lambda_{1},\lambda_{j})}
\prod^{n}_{\stackrel{k=2}{k \neq j}}
i\frac{\alpha_{2}(\lambda_{k},\lambda_{j})}
{\alpha_{9}(\lambda_{k},\lambda_{j})} 
\nonumber \\
&& \times \left [ \vec{\xi} \otimes F(\lambda_1) 
 \vec {\Phi}_{n-2}(\lambda_{2},\ldots,\lambda_{j-1},\lambda_{j+1},\ldots,\lambda_{n}) B(\lambda_j) \right ] 
\nonumber \\
&& \times \prod_{k=2}^{j-1} \frac{\alpha_{1}(\lambda_{k},\lambda_{j})}{\alpha_{2}(\lambda_{k},\lambda_{j})}
\hat{r}_{k,k+1}(\lambda_{k},\lambda_{j})
\ear

At this point it is fair to remark that the recursive way we found for 
the eigenvectors were inspired to some extent on an early work of 
Tarasov on the Izergin-Korepin model \cite{TAR1}. 
Our construction, however, has the important novelty of allowing
a general 
``exclusion statistics''  between the non-commutative and the
commutative creation fields 
and therefore paving the way for further applications and extensions.
Indeed,  the non-trivial way that both
the ``exclusion'' vector  and the auxiliary matrix enters
in the eigenvectors expression (81) makes our formula rather general, being
able to 
accommodate the solution of a wider class of integrable models. This situation
has to be contrasted to that of multi-state 6-vertex
generalizations \cite{RES,BAB}, 
in which the eigenvectors are  easily given by tensoring the creation fields
and there is no explicit dependence of the underlying
algebra. 

Let us now return to the
problem of finding the eigenvalues of the transfer matrix, keeping in
mind the recurrence relation (81) for the eigenvectors. To gain
some insight about this problem we first investigate how the wanted and
unwanted terms are collected for the three-particle state. Besides the
commutation rules for the diagonal fields, we also have to use
our previous results for the two-particle state (cf. (57-59)) wherever
there is the need to carry the diagonal operators through the
vector
$\vec{\Phi}_{2}(\lambda_{2},\lambda_3) $. This recursive way not only helps
us to better simplify the wanted terms but also makes it possible to gather
the unwanted terms in rather closed forms. 
This  analysis is presented in appendix $D$ since it still
involves some extra technicalities.
Having at hand the two-particle and the three-particle data
we can move forward to the analysis of the four-particle state and 
so forth. In general, for $n \leq 3$, the knowledge of the $(n-1)$-particle
and the $(n-2)$-particle results dictates the behaviour of the $n$-particle
state. By using mathematical induction we are able to determine
the general structure for the 
multi-particle states and the final results are  
\bear
B(\lambda) \ket{\Phi_{n}(\lambda_{1},\ldots,\lambda_n)} & = &
[\omega_1(\lambda)]^{L} \prod_{j=1}^{n} 
i\frac{\alpha_{2}(\lambda_{j},\lambda)}{\alpha_{9}(\lambda_{j},\lambda)} 
\ket{\Phi_{n}(\lambda_{1},\ldots,\lambda_n)}
\nonumber \\
&& -\sum_{j=1}^{n}  
[\omega_1(\lambda_{j})]^{L}  \ket{\Psi^{(1)}_{n-1}(\lambda,\lambda_{j}; \{ \lambda_{l} \})}  
\nonumber \\
&& + \sum_{j=2}^{n} \sum_{l=1}^{j-1} H_1(\lambda,\lambda_l,\lambda_j) 
[\omega_1(\lambda_l) \omega_1(\lambda_j)]^{L} 
\ket{\Psi^{(3)}_{n-2}(\lambda,\lambda_j,\lambda_l;\{ \lambda_k \} ) }
\ear
\bear
D(\lambda) \ket{\Phi_{n}(\lambda_{1},\ldots,\lambda_n)} & = &
[\omega_3(\lambda)]^{L} \prod_{j=1}^{n} 
-i\frac{\alpha_{8}(\lambda,\lambda_j)}{\alpha_{7}(\lambda,\lambda_j)} 
\ket{\Phi_{n}(\lambda_{1},\ldots,\lambda_n)}
\nonumber \\
&& -\sum_{j=1}^{n} 
[\omega_2(\lambda_{j})]^{L} 
\Lambda^{(1)}(\lambda=\lambda_j,\{ \lambda_l \})
\ket{\Psi^{(2)}_{n-1}(\lambda,\lambda_{j}; \{ \lambda_{l} \})}  
\nonumber \\
&& + \sum_{j=2}^{n} \sum_{l=1}^{j-1}H_{2}(\lambda,\lambda_l,\lambda_j) 
[\omega_2(\lambda_1) \omega_2(\lambda_2)]^{L} 
\Lambda^{(1)}(\lambda=\lambda_j,\{ \lambda_k \})
\nonumber\\
&& \times \Lambda^{(1)}(\lambda=\lambda_l,\{ \lambda_k \})
\ket{\Psi^{(3)}_{n-2}(\lambda,\lambda_j,\lambda_l;\{ \lambda_k \} ) }
\ear
\bear
\sum_{a=1}^{2} {\hat{A}}_{aa}(\lambda) 
\ket{\Phi_{n}(\lambda_{1},\ldots,\lambda_n)} & = &
[\omega_2(\lambda)]^{L} \prod_{j=1}^{n} 
-i\frac{\alpha_{1}(\lambda,\lambda_j)}{\alpha_{9}(\lambda,\lambda_j)} 
\Lambda^{(1)}(\lambda,\{ \lambda_l \})
\ket{\Phi_{n}(\lambda_{1},\ldots,\lambda_n)}
\nonumber \\
&& -\sum_{j=1}^{n}  
[\omega_2(\lambda_{j})]^{L} 
\Lambda^{(1)}(\lambda=\lambda_j,\{ \lambda_l \})
\ket{\Psi^{(1)}_{n-1}(\lambda,\lambda_{j};\{ \lambda_{l}\} )}
\nonumber \\
&& -\sum_{j=1}^{n} [\omega_1(\lambda_{j})]^{L} 
\ket{\Psi^{(2)}_{n-1}(\lambda,\lambda_{j};\{ \lambda_{l} \} )}
\nonumber \\
&& - \sum_{j=2}^{n} \sum_{l=1}^{j-1} {H}_3(\lambda,\lambda_l,\lambda_j) 
[\bar{a}(\lambda_l,\lambda_j) -\bar{b}(\lambda_l,\lambda_j)] 
[\omega_1(\lambda_l) \omega_2(\lambda_j)]^{L} 
\nonumber\\
&& \times \Lambda^{(1)}(\lambda=\lambda_j,\{ \lambda_k \})
\ket{\Psi^{(3)}_{n-2}(\lambda,\lambda_j,\lambda_l;\{ \lambda_k \} ) }
\nonumber \\
&& -\sum_{j=2}^{n} \sum_{l=1}^{j-1} H_3(\lambda,\lambda_j,\lambda_l)
\frac{\alpha_1(\lambda_l,\lambda_j)}{\alpha_2(\lambda_l,\lambda_l)}
[\omega_1(\lambda_j) \omega_2(\lambda_l)]^{L} 
\nonumber\\
&& \times \Lambda^{(1)}(\lambda=\lambda_l,\{ \lambda_k \})
\ket{\Psi^{(3)}_{n-2}(\lambda,\lambda_j,\lambda_l;\{ \lambda_k \} ) }
\ear

Similarly to what happened to the two-particle and 
the three-particle cases we have
three families
of unwanted terms. As before they are written 
in terms of the creation 
operators and the general expressions 
are 
\bear
\ket{\Psi^{(1)}_{n-1}(\lambda,\lambda_{j};\{ \lambda_{l}\} )} & = &
i\frac{\alpha_{5}(\lambda_{j},\lambda)}
{\alpha_{9}(\lambda_{j},\lambda)} \prod^n_{ \stackrel {k=1}{ k \neq j}} 
i\frac{\alpha_{2}(\lambda_{k},\lambda_j)}{\alpha_{9}(\lambda_{k},\lambda_j)} 
\vec{B}(\lambda) \otimes \vec{\Phi}_{n-1}(\lambda_1, \ldots, 
\check{\lambda}_j, \ldots,\lambda_{n}) 
\nonumber\\
&& \times \hat{O}^{(1)}_j(\lambda_j;\{ \lambda_k \} ) .\vec{\cal{F}}
\ket{0}
\ear
\bear
\ket{\Psi^{(2)}_{n-1}(\lambda,\lambda_{j};\{ \lambda_{l}\} )} & = &
i\frac{\alpha_{10}(\lambda,\lambda_j)}
{\alpha_{7}(\lambda,\lambda_j)} \prod^n_{ \stackrel {k=1}{ k \neq j}} 
i\frac{\alpha_{2}(\lambda_{k},\lambda_j)}{\alpha_{9}(\lambda_{k},\lambda_j)} 
[\vec{\xi}.(\vec{B^{*}}(\lambda)\otimes \hat{I})]  
\otimes \vec{\Phi}_{n-1}(\lambda_1, \ldots, 
\check{\lambda}_j, \ldots,\lambda_{n}) 
\nonumber\\
&& \times \hat{O}^{(1)}_j(\lambda_j;\{ \lambda_k \} ) .\vec{\cal{F}}
\ket{0}
\ear
\bear
\ket{\Psi^{(3)}_{n-2}(\lambda,\lambda_{j}, \lambda_l;\{ \lambda_{k}\} )} & = &
\prod^{n}_{\stackrel {k=1}{ \neq j,l}} i\frac{\alpha_2(\lambda_k,\lambda_j)}
{\alpha_9(\lambda_k,\lambda_j)} 
i\frac{\alpha_2(\lambda_k,\lambda_l)}
{\alpha_9(\lambda_k,\lambda_l)} 
F(\lambda) \vec{\xi} 
\otimes \vec{\Phi}_{n-2}(\lambda_1, \ldots, \check{\lambda}_l, \ldots, 
\check{\lambda}_j, \ldots,\lambda_{n}) 
\nonumber\\
&& \times \hat{O}^{(2)}_{lj}(\lambda_l, \lambda_j;\{ \lambda_k \} ) 
.\vec{\cal{F}}
\ket{0}
\ear
where the symbol $\check{\lambda}_j$ means that the rapidity 
$\lambda_j$ is absent from the set $\{ \lambda_1, \ldots, \lambda_n \} $.
For $n \geq 3$ it was necessary to introduce a
second ``ordering'' factor in order to better represent the third type of
unwanted terms (cf. appendix $D$).  Its
task is  similar to that played by the
first ``ordering'' factor  with the difference that now two rapidities are
reordered. In other words, this second ``ordering'' factor brings 
the rapidities $\lambda_l$ and $\lambda_j$ ($l <j$)
to the first two positions
in the eigenvector formula (81), and a simple calculation shows that 
its expression is
\bear
\hat{O}^{(2)}_{lj}(\lambda_l,\lambda_j;\{ \lambda_k \}) & =&
\prod_{k=1}^{l-1} \frac{\alpha_1(\lambda_k,\lambda_j)}
{\alpha_2(\lambda_k,\lambda_j)} 
\hat{r}_{k+1,k+2}(\lambda_k,\lambda_j) 
\prod_{k=l+1}^{j-1} \frac{\alpha_1(\lambda_k,\lambda_j)}
{\alpha_2(\lambda_k,\lambda_j)} 
\hat{r}_{k,k+1}(\lambda_k,\lambda_j) 
\nonumber\\
&& \times \prod_{k=1}^{l-1} \frac{\alpha_1(\lambda_k,\lambda_l)}
{\alpha_2(\lambda_k,\lambda_l)} 
\hat{r}_{k,k+1}(\lambda_k,\lambda_l) 
\ear

Before discussing the results, we should note that the above 
expressions for multi-particle states indeed reproduce our previous
findings for the two-particle (after considering appendix $C$) and 
the one-particle states. Now, 
from  equations (82-84),  it is direct  to read of
the $n$-particle eigenvalue expression,
namely
\bear
\Lambda(\lambda,\{\lambda_{j}\}) & = &
[\omega_1(\lambda)]^L 
\prod_{j=1}^{n} i \frac{\alpha_{2}(\lambda_{j},\lambda)}{\alpha_{9}(\lambda_{j},\lambda)} + 
[\omega_3(\lambda)]^L 
\prod_{j=1}^{n} -i\frac{\alpha_{8}(\lambda,\lambda_{j})}{\alpha_{7}(\lambda,\lambda_{j})}  
\nonumber\\
&& -[\omega_2(\lambda)]^L \prod_{j=1}^{n} -i \frac{\alpha_{1}(\lambda,\lambda_{j})}{\alpha_{9}(\lambda,\lambda_{j})} 
\Lambda^{(1)}(\lambda,\{\lambda_{l}\}) 
\ear

Following the 
same arguments given for the two-particle state, and in particular the
discussion presented at the end of appendix $C$,  we easily derive that
the unwanted terms vanish provided the rapidities
satisfy the following  Bethe Ansatz equations  
\EQ
\left[ \frac{\omega_1(\lambda_{i})}{\omega_2(\lambda_{i})} \right]^{L} = 
\Lambda^{(1)}(\lambda=\lambda_{i},\{\lambda_{j}\}), ~~ i=1, \ldots, n
\EN

Once again, the final results have been  expressed in terms
of the underlying auxiliary problem, which for a general multi-particle
state is defined by
\EQ
 T^{(1)}(\lambda,\{\lambda_{i}\})_{a_{1} \cdots a_{n}}^{b_{1} \cdots b_{n}}
{\cal{F}}^{b_{n} \cdots b_{1}} = \Lambda^{(1)}(\lambda,\{\lambda_{i}\})  
{\cal{F}}^{a_{n} \cdots a_{1}}
\EN
where the inhomogeneous transfer matrix $T^{(1)}(\lambda,\{\lambda_i\})$ is
\EQ
T^{(1)}(\lambda,\{\lambda_{i}\})_{b_{1} \cdots b_{n}}^{a_{1} \cdots a_{n}} = 
\hat{r}_{b_{1}d_{1}}^{c_{1}a_{1}}(\lambda,\lambda_{1})
\hat{r}_{b_{2}c_{2}}^{d_{1}a_{2}}(\lambda,\lambda_{2}) \ldots
\hat{r}_{b_{n}c_{1}}^{d_{n-1}a_{n}}(\lambda,\lambda_{n})
\EN

As we have commented before these results are direct extensions of those
obtained for the two-particle state. We see that the Bethe Ansatz 
equations and the eigenvalues still depend on an additional
auxiliary eigenvalue problem. In the language of condensed matter  we would say
that so far we managed to solve the ``charge'' degrees of freedom but
still remains the diagonalization of the ``spin'' sector. As we shall see
next the ``spin'' problem can also be solved in terms of the algebraic Bethe
Ansatz approach.

\subsection{The eigenvalues and the nested Bethe Ansatz}

The task of this section is the diagonalization of the auxiliary
transfer matrix $T^{(1)}(\lambda,\{ \lambda_j \})$. For this purpose we have
to set up another Bethe Ansatz which will result in ``nested'' Bethe Ansatz
equations for the rapidities we began with. This problem, however,
is equivalent to the solution of the $6$-vertex model in presence of
inhomogeneities and it has been extensively discussed in the literature 
(see e.g. refs. \cite{KO,RES,BAB}). Therefore we will only sketch the main
steps of the solution for sake of completeness. First we write the
transfer matrix $T^{(1)}(\lambda,\{ \lambda_j \})$ as the trace of
the following monodromy matrix
\EQ
{\cal T}^{(1)}(\lambda,\{\lambda_{j}\}) = 
{\cal L}^{(1)}_{{\cal{A}}^{(1)} n}(\lambda,\lambda_{n})
{\cal L}^{(1)}_{{\cal{A}}^{(1)} n-1}
(\lambda,\lambda_{n-1}) \ldots 
{\cal L}^{(1)}_{{\cal{A}}^{(1)} 1}(\lambda,\lambda_{1})
\EN
where ${\cal{A}}^{(1)}$ is the two-dimensional ``spin'' auxiliary space.
The Lax operator
${\cal L}^{(1)}_{{\cal{A}}^{(1)}j}(\lambda,\lambda_j)$ 
is  related to the auxiliary matrix $\hat{r}(\lambda,\lambda_j)$ 
by a permutation on the 
$C^{2} \times C^{2}$ space and its matrix elements are 
\EQ
{\cal L}^{(1)}_{{\cal{A}}^{(1)}j}(\lambda,\lambda_j) = 
\pmatrix{
1  &0  &0  &0  \cr
0  &\bar{b}(\lambda,\lambda_j)  &\bar{a}(\lambda,\lambda_j)  &0  \cr
0  &\bar{a}(\lambda,\lambda_j)  &\bar{b}(\lambda,\lambda_j)  &0  \cr
0  &0  &0  &1  \cr}
\EN

We now go ahead applying the $ABCD$ algebraic Bethe Ansatz 
framework \cite{STF,FA,FA1} for an inhomogeneous transfer matrix. Writing
the monodromy matrix as
\EQ
{\cal T}^{(1)}(\lambda,\{\lambda_{j}\}) =
\pmatrix{
A^{(1)}(\lambda,\{\lambda_{j}\})  &  B^{(1)}(\lambda,\{\lambda_{j}\})  \cr
C^{(1)}(\lambda,\{\lambda_{j}\})  &  D^{(1)}(\lambda,\{\lambda_{j}\})  \cr}
\EN
and taking as the reference state the vector
\EQ
\ket{0^{(1)}} = \prod_{j=1}^{n} \otimes 
\pmatrix{
1  \cr
0  \cr}_j
\EN
we find 
the following relations
\bear
A^{(1)}(\lambda,\{\lambda_{j}\})\ket{0^{(1)}} & = &\ket{0^{(1)}}
\nonumber\\
D^{(1)}(\lambda,\{\lambda_{j}\})\ket{0^{(1)}} & = &\prod_{j=1}^{n} 
\bar{b}(\lambda,\lambda_{j})\ket{0^{(1)}}
\nonumber \\
C^{(1)}(\lambda,\{\lambda_{j}\})\ket{0^{(1)}} & = & 0
\ear

The field
$ B^{(1)}(\lambda,\{\lambda_{j}\}) $  plays the role of a creation operator
over the reference state. To get its commutation rules we solve the
Yang-Baxter algebra
for the monodromy matrix ${\cal{T}}^{(1)}(\lambda,\{ \lambda_j \})$ using
as intertwiner the 
auxiliary matrix (26).  
This yields the following relations 
\bear
A^{(1)}(\lambda,\{\lambda_{j}\})B^{(1)}(\mu,\{\lambda_{j}\}) & =&
\frac{1}{\bar{b}(\mu,\lambda)}B^{(1)}(\mu,\{\lambda_{j}\})
A^{(1)}(\lambda,\{\lambda_{j}\})
\nonumber \\
&& - \frac{\bar{a}(\mu,\lambda)}{\bar{b}(\mu,\lambda)}B^{(1)}(\lambda,\{\lambda_{j}\})
A^{(1)}(\mu,\{\lambda_{j}\}) 
\nonumber\\
D^{(1)}(\lambda,\{\lambda_{j}\})B^{(1)}(\mu,\{\lambda_{j}\})& =&
\frac{1}{\bar{b}(\lambda,\mu)}B^{(1)}(\mu,\{\lambda_{j}\})
D^{(1)}(\lambda,\{\lambda_{j}\})
\nonumber \\
 && - \frac{\bar{a}(\lambda,\mu)}{\bar{b}
(\lambda,\mu)}B^{(1)}(\lambda,\{\lambda_{j}\})
D^{(1)}(\mu,\{\lambda_{j}\}) 
\nonumber\\
\left [ B^{(1)}(\mu,\{\lambda_{j}\}),B^{(1)}(\lambda,\{\lambda_{j}\})
\right ] &= &0
\ear

Next we have to make an Ansatz for the eigenstates of $T^{(1)}(\lambda,\{
\lambda_j \})$. This is the ``spin'' part of the multi-particle states 
and it is given by the product 
$ \displaystyle \prod_{l=1}^{m} 
B^{(1)}(\mu_{l},\{ \lambda_j \} ) \ket{0^{(1)}}$ whose components are 
precisely identified with the coefficients ${\cal{F}}^{a_n \ldots a_1}$. With
the help of commutation rules (98) we are able to carry on the operators 
$A^{(1)}(\lambda,\{\lambda_{j}\}) + 
D^{(1)}(\lambda,\{\lambda_{j}\})$ through all the creation fields 
$B^{(1)}(\mu_{l},\{ \lambda_j \})  $ 
leading us to the following result for the auxiliary eigenvalue
\EQ
\Lambda^{(1)}(\lambda,\{\lambda_{j}\},\{\mu_{l}\}) =
\prod_{l=1}^{m} \frac{1}{\bar{b}(\mu_{l},\lambda)} +
\prod_{j=1}^{n} \bar{b}(\lambda , \lambda_{j}) \prod_{l=1}^{m} 
\frac{1}{\bar{b}(\lambda,\mu_{l})}
\EN
provided  the numbers $\{\mu_{l}\}$ satisfy the additional 
restriction
\EQ
\prod_{j=1}^{n} \bar{b}(\mu_{l} , \lambda_{j}) = 
- \prod_{k=1}^{m} \frac{\bar{b}(\mu_{l},\mu_{k})}{\bar{b}(\mu_{k},\mu_{l})},~~
l=1,\ldots,m
\EN

Finally, we use the auxiliary eigenvalue
expression to rewrite our previous results for 
the eigenvalues  and Bethe Ansatz equations 
of the ``covering'' vertex model. Substituting the expression (99) in equations
(89,90) and using the second relation of  equation (28) we obtain that the eigenvalue is
\bear
\Lambda(\lambda,\{\lambda_{j}\}, \{ \mu_l \}) & = &
[\omega_1(\lambda)]^L 
\prod_{j=1}^{n} i \frac{\alpha_{2}(\lambda_{j},\lambda)}{\alpha_{9}(\lambda_{j},\lambda)} + 
[\omega_3(\lambda)]^L 
\prod_{j=1}^{n} -i\frac{\alpha_{8}(\lambda,\lambda_{j})}{\alpha_{7}(\lambda,\lambda_{j})}  
\nonumber\\
&& -[\omega_2(\lambda)]^L  \left \{
\prod_{j=1}^{n} -i 
\frac{\alpha_{1}(\lambda,\lambda_{j})}{\alpha_{9}(\lambda,\lambda_{j})} 
\prod_{l=1}^{m}
\frac{1}{\bar{b}(\mu_l,\lambda)} 
+\prod_{j=1}^{n} -i 
\frac{\alpha_{8}(\lambda,\lambda_{j})}{\alpha_{7}(\lambda,\lambda_{j})} 
\prod_{l=1}^{m}
\frac{1}{\bar{b}(\lambda,\mu_l)} 
\right \}
\nonumber \\
\ear
while the Bethe Ansatz equations for the 
rapidities $\{ \lambda_j \} $ becomes
\EQ
\left[ \frac{\omega_1(\lambda_{j})}{\omega_2(\lambda_{j})} \right]^{L} = 
\prod_{l=1}^{m}
\frac{1}{\bar{b}(\mu_l,\lambda_j)} 
\EN

Now we are almost ready to make a comparison with the Lieb's and Wu's results
\cite{LW}. First we introduce a new set 
of variables $z_{\pm}(\lambda_j)$ defined by
\EQ
z_{-}(\lambda_j) =\frac{a(\lambda_j)}{b(\lambda_j)} e^{2h(\lambda_j)}~~~
z_{+}(\lambda_j) =\frac{b(\lambda_j)}{a(\lambda_j)} e^{2h(\lambda_j)}
\EN

Considering this definition and taking into account
the  transformation (32) as well as the identities (29,30), we are able to rewrite the expression for the eigenvalue as
\bear
(-i)^{n} \Lambda(\lambda,\{z_{\pm}(\lambda_{j})\},\{\tilde{\mu}_{l}\}) & = &
[\omega_1(\lambda)]^L \prod_{j=1}^{n} \frac{b(\lambda)}{a(\lambda)}
\left [ \frac{1+z_{-}(\lambda_{j})/z_{+}(\lambda)}{1-z_{-}
(\lambda_{j})/z_{-}(\lambda)} \right ]
\nonumber \\
&& + [\omega_3(\lambda)]^L \prod_{j=1}^{n}  \frac{b(\lambda)}{a(\lambda)}
 \left [ \frac{1+z_{-}(\lambda_{j})z_{-}
(\lambda)}{1-z_{-}(\lambda_{j})z_{+}(\lambda)} \right ]
 \nonumber \\
&& -[\omega_2(\lambda)]^{L} \left \{
\prod_{j=1}^{n}\frac{b(\lambda)}{a(\lambda)} 
\left [ \frac{1+z_{-}(\lambda_{j})/z_{+}(\lambda)}
{1-z_{-}(\lambda_{j})/z_{-}(\lambda)} \right ]
\right.
\nonumber\\
&& \left. \times \prod_{l=1}^{m} \frac{z_{-}(\lambda)-1/z_{-}(\lambda) 
- \tilde{\mu}_{l} + U/2}{z_{-}(\lambda)-1/z_{-}(\lambda)
 - \tilde{\mu}_{l} - U/2}
\right.
\nonumber\\ 
&& \left. +\prod_{j=1}^{n}  \frac{b(\lambda)}{a(\lambda)}
 \left [ \frac{1+z_{-}(\lambda_{j})z_{-}(\lambda)}
{1-z_{-}(\lambda_{j})z_{+}(\lambda)} \right ]
\prod_{l=1}^{n} \frac{1/z_{+}(\lambda)-z_{+}(\lambda)
 - \tilde{\mu}_{l} - U/2}{1/z_{+}(\lambda)-z_{+}(\lambda) 
- \tilde{\mu}_{l} + U/2}  \right \} 
\nonumber\\
\ear
and the nested Bethe Ansatz equations are now  given by
\bear
[z_{-}(\lambda_{j})]^L & = &
\prod_{l=1}^{m}  \frac{z_{-}(\lambda_{j})-1/z_{-}(\lambda_{j}) - \tilde{\mu}_{l} + U/2}{z_{-}(\lambda_{j})-1/z_{-}(\lambda_{j}) - \tilde{\mu}_{l} - U/2},
\nonumber\\
\prod_{j=1}^{n}  \frac{z_{-}(\lambda_{j})-1/z_{-}(\lambda_{j}) - \tilde{\mu}_{l} - U/2}{z_{-}(\lambda_{j})-1/z_{-}(\lambda_{j}) - \tilde{\mu}_{l} + U/2} & =&
-\prod_{k=1}^{m} \frac{\tilde{\mu}_{l}-\tilde{\mu}_{k}+U}{\tilde{\mu}_{l}-\tilde{\mu}_{k}-U},~l=1, \ldots, m 
\ear

From the above expressions we note that function
$\Lambda(\lambda,\{z_{\pm}(\lambda_{j})\},\{\tilde{\mu}_{l}\})$  is 
analytic in $\lambda$. This happens because the condition
of having zero residues  on both
direct $z_{-}(\lambda_j)$ and ``crossed'' 
$z_{+}(\lambda_j)$ channels  
is clearly fulfilled by the nested Bethe Ansatz
equations.
The next step is to expand the logarithm of the eigenvalue 
$\Lambda(\lambda,\{z_{\pm}(\lambda_{j})\},\{\tilde{\mu}_{l}\})$ in powers
of $\lambda$ and up to second order in the expansion we find
\bear
\ln  \left [ \Lambda(\lambda,\{z_{\pm}(\lambda_{j})\},\{\tilde{\mu}_{l}\})
\right ] & = &
\frac{i\pi}{2}n +\sum_{j=1}^{n} \ln[z_{-}(\lambda_j)] +
\nonumber \\
&& +\lambda \left [ \sum_{j=1}^{n} \left [ 
z_{-}(\lambda_j) +1/z_{-}(\lambda_j)  \right ]
+\frac{U}{4}(L-2n) \right ]
\nonumber\\
&& +\lambda^2 \left [ \sum_{j=1}^{n} \left [ z_{-}^2(\lambda_j) 
-1/z_{-}^2(\lambda_j)
- U[z_{-}(\lambda_j) -1/z_{-}(\lambda_j)] \right ] -2L \right ]
\nonumber\\
&&+{\cal{O}}(\lambda^3)
\ear

The ${\cal{O}}(\lambda)$ term parametrizes the spectrum of the Hubbard 
Hamiltonian and to recover the Lieb's and Wu's results we just have to reexpress
the variable
$z_{-}(\lambda_j)$  in terms of the hole momenta $k_j$ by
\EQ
z_{-}(\lambda_j) = e^{i k_j} 
\EN

Considering this relation, the eigenenergies of the Hubbard model are
\EQ
E_n(L) = \frac{U(L-2n)}{4} +\sum_{j=1}^{n} 2 \cos(k_j)
\EN
and the momenta $k_j$ satisfy the following Bethe Ansatz equations 
\bear
e^{iLk_j} & = &
\prod_{l=1}^{m}  \frac{\sin(k_j) - \bar{\mu}_{l} - iU/4}{\sin(k_j) - \bar{\mu}_{l} +i U/4},
\nonumber\\
\prod_{j=1}^{n}  \frac{\sin(k_j) - \bar{\mu}_{l} + i U/4}{\sin(k_j) - \bar{\mu}_{l} - iU/4} & =&
-\prod_{k=1}^{m} \frac{\bar{\mu}_{l}-\bar{\mu}_{k}-iU/2}{\bar{\mu}_{l}-\bar{\mu}_{k} +iU/2},~l=1, \ldots, m 
\ear
where we also used $\tilde{\mu}_l =2i \bar{\mu}_l $ to bring our equations 
in the Lieb's and Wu's form. A careful reader might note that the above 
Bethe Ansatz equations
have an extra minus factor in front of the 
coupling $U$ in comparison to the original
ones. This is because we are using the language of holes instead of particles
and this means that
the integers $n$ and $m$ are the  total number of holes and the
number of holes with spin up, respectively. It is well known that via
a particle-hole transformation the kinetic term of the Hamiltonian gets
an extra minus sign, which changes the sign of factor 
$U/t$ entering in the Bethe Ansatz equations. Similar reasoning
can be carried out for others conserved charges. For example, the first
non-trivial current
commuting with the Hamiltonian \cite{SA1,SA2}
is  
\bear
J & = &\sum_{j=1}^{L} 
 c^{\dagger}_{j \uparrow} c_{j+2 \uparrow} 
-c^{\dagger}_{j+2 \uparrow} c_{j \uparrow} 
\nonumber \\
&&  +U(c^{\dagger}_{j \uparrow} c_{j+1 \uparrow} 
-c^{\dagger}_{j+1 \uparrow} c_{j \uparrow} )(n_{j+1 \downarrow} 
+n_{j \downarrow} -1 ) + [\uparrow \leftrightarrow \downarrow]
\ear
and from equation (106) it follows that the spectrum (modulo a constant) 
of this charge is
\EQ
E_n^{J}(L) = 2i\sum_{j=1}^{n}  \left [ \sin(2 k_j) -U\sin(k_j) \right ]
\EN

We would like to close this section commenting on the construction of
the eigenvectors in the terms of the ``dual'' field $\vec{B}^{*}(\lambda)$.
The  equivalence between the commutation rules for the 
fields $\vec{B}(\lambda)$ and
$\vec{B^{*}}(\lambda)$ allow us to follow straightforwardly
the whole construction of section 4.1 and it is
not difficult
to derive 
formula for the  ``dual'' eigenvectors
$\vec {\Phi^{*}}_{n}(\lambda_{1},\ldots ,\lambda_{n})$. Formally, we
can apply the ``dual'' transformation in expression (81). This leads us to
following ``dual'' recurrence relation
\bear
\vec {\Phi^{*}}_{n}(\lambda_{1},\ldots ,\lambda_{n}) & = &
\vec {B^{*}}(\lambda_{1}) \otimes \vec {\Phi^{*}}_{n-1}(\lambda_{2}, 
\ldots ,\lambda_{n})
+\sum_{j=2}^n 
i\frac{\alpha^{*}_{10}(\lambda_{1},\lambda_{j})}
{\alpha^{*}_{7}(\lambda_{1},\lambda_{j})}
\nonumber\\
&& \prod^{n}_{\stackrel{k=2}{k \neq j}}
i\frac{\alpha^{*}_{2}(\lambda_{k},\lambda_{j})}
{\alpha^{*}_{9}(\lambda_{k},\lambda_{j})} 
\prod_{k=2}^{j-1} \frac{\alpha^{*}_{1}(\lambda_{k},\lambda_{j})}{\alpha^{*}_{2}(\lambda_{k},\lambda_{j})}
\hat{r}^{*}_{k,k+1}(\lambda_{k},\lambda_{j})
\nonumber \\
&& \times \left [ \vec{\xi} \otimes F(\lambda_1) 
 \vec {\Phi^{*}}_{n-2}(\lambda_{2},\ldots,\lambda_{j-1},\lambda_{j+1},\ldots,\lambda_{n}) D(\lambda_j) \right ] 
\ear

We expect that the corresponding eigenvalues 
$\Lambda^{*}(\lambda,\{\lambda_{j}\},\{\tilde{\mu}_{l}\})$ should also
be related to 
$\Lambda(\lambda,\{\lambda_{j}\},\{\tilde{\mu}_{l}\})$ in some way. This is
indeed the case if we shift all
the rapidities around the ``crossing'' point $\pi/2$, and the relation we found
is
\EQ
\Lambda^{*}(\pi/2- \lambda,\{\pi/2- \lambda_{j}\},\{
\pi/2-\tilde{\mu}_{l}\})
=(-1)^{n} \Lambda(\lambda,\{\lambda_{j}\},\{\tilde{\mu}_{l}\})
\EN

With this we complete our analysis of the graded eigenvalue problem and
in the next section we shall discuss some other complementary results which
can be obtained within the $ABCDF$ formalism.

\section{Complementary results}

In this section we shall first consider the solution of the coupled 
spin model with  twisted boundary conditions. This allow us to 
illustrate the difference between the Hubbard and the coupled spin models 
from the viewpoint of their Bethe Ansatz solution. Next we consider
the well known $SU(2)$ symmetries of the Hubbard model \cite{SU2,EKS,GOM}.
We will show 
that the eigenvectors (81) are highest weights of both the $SU(2)$ Lie
algebra of rotations and the $\eta$-paring $SU(2)$ symmetry. Thus we are able
to 
recover the results by Essler, Korepin and Schoutens \cite{EKS} from an
algebraic point of view.

\subsection{Twisted boundary conditions}

We begin recalling that twisted boundary conditions are in general 
associated to certain gauge invariances of the Yang-Baxter algebra. 
The integrability condition (1) is still valid when 
${\cal L}_{{\cal A}i}(\lambda) \rightarrow G_{{\cal A}}{\cal L}_{{\cal A}i}(\lambda)$ provided the gauge matrix $G_{{\cal A}}$ satisfies \cite{DEVI}
\EQ
\left[ R(\lambda,\mu), G_{{\cal A}} \otimes G_{{\cal A}} \right] = 0
\EN

This means that a vertex model defined by the transfer matrix 
$T_{G}(\lambda) = Tr_{{\cal A}} {\cal T}_{G}(\lambda)$ whose monodromy matrix 
is 
\EQ
{\cal T}_{G}(\lambda) = 
G_{{\cal A}}{\cal L}_{{\cal A}L}(\lambda){\cal L}_{{\cal A}L-1}(\lambda)
\ldots {\cal L}_{{\cal A}1}(\lambda)
\EN
still remains integrable. One way of seeing the connection to twisted boundary
conditions is, for example, to derive the quantum Hamiltonian $H_G$ commuting
with the transfer matrix 
$T_{G}(\lambda)$.  To this end we assume that the Lax operator is regular
at some value of the spectral parameter, say 
${\cal L}_{{\cal A}i}(0)=
P_{{\cal A}i}$ where $P_{{\cal{A}}i}$ 
is the $C^{4} \otimes C^{4}$ permutation operator.
Then the  
local quantum Hamiltonian $H_G$ is  \cite{LU,TAR}
\EQ
H_G = T^{-1}_{G}(0) T'_{G}(0)
\EN
where symbol $'$ stands for the derivative on $\lambda$. By using the 
permutation properties
\EQ
\hat{O}_{{\cal A}i} P_{{\cal A}j} = P_{{\cal A}j} \hat{O}_{ji}, ~~
P_{{\cal A}i} \hat{O}_{{\cal A}j} = \hat{O}_{ij} P_{{\cal A}i}, ~~
Tr_{{\cal A}} [G_{{\cal A}} P_{{\cal A} \alpha}] = G_{\alpha}
\EN
and after few algebraic manipulations we derive that (see e.g.  
ref. \cite{BAT})
\EQ
H= \sum_{i=1}^{L-1} h_{i,i+1} + G^{-1}_{L} h_{L,1} G_{L}
\EN
where $h_{ij}=[P{\cal L}'(0)]_{ij}$ and we assumed that $G_{{\cal A}}$ is 
invertible. The last term in the Hamiltonian (118) reflects the presence of 
non-trivial boundary conditions. In the context of the coupled spin model (7), 
it is straightforward to see that we get 
twisted boundary conditions  by
taking the following gauge 
\EQ
G_{{\cal{A}}} = \pmatrix{
e^{-i\phi_{1}/2} & 0 \cr
0 & e^{i\phi_{1}/2} \cr
}
\otimes 
\pmatrix{
e^{-i\phi_{2}/2} & 0 \cr
0 & e^{i\phi_{2}/2} \cr
}
\EN

Clearly, such gauge matrix  fulfill the integrability condition (114).  
In order to diagonalize $T_{G}(\lambda)$ 
we only need to introduce few modifications on the 
formalism developed in the previous sections. It is fundamental that this 
gauge  does 
not spoil the triangular form of the monodromy ${\cal T}_{G}(\lambda)$ when 
it acts on the ferromagnetic reference $\ket{0}$. The diagonal 
operators of ${\cal T}_{G}(\lambda)$, however,  pick up extra phase 
factors and now we have the following relations
\bear
B(\lambda)\ket{0} = e^{-i(\phi_{1}+\phi_{2})/2}[\omega_1(\lambda)]^{L}\ket{0} ~ D(\lambda)\ket{0} = 
e^{i(\phi_{1}+\phi_{2})/2}[\omega_3(\lambda)]^{L}\ket{0}
\nonumber \\
A_{11}(\lambda)\ket{0} = e^{-i(\phi_{1}-\phi_{2})/2}[\omega_2(\lambda)]^{L}\ket{0} ~
A_{22}(\lambda)\ket{0} = e^{i(\phi_{1}-\phi_{2})/2}[\omega_2(\lambda)]^{L}\ket{0}
\nonumber \\
\ear

The next step is to solve the commutation rules in the standard 
Yang-Baxter formalism, since we are considering the coupled spin model. These 
commutation rules have basically the same structure of  those worked out for
the graded case, apart from 
few signs and imaginary factors. We have collected them 
in appendix B and we note that the corresponding $6$-vertex 
auxiliary matrix has now an extra sign in the amplitude $\bar{b}(\lambda,\mu)$.
Therefore, the nested part always gets twisted, emphasizing 
the difference between 
the Hubbard and the coupled spin models for closed boundary conditions. Since
now  the
basic ingredients have been set up we can  follow  closely the
steps of sections 3 and 4.  Here we are interested in the
eigenvalues of the twisted model and now we begin to
summarize our final findings.
Taking into account the relations (120) and the commutation rules B.12-B.22 
we derive that the eigenvalues of transfer matrix $T_{G}(\lambda)$ is
\bear
\Lambda_{G}(\lambda,\{\lambda_{j}\}) &  = &
e^{-i(\phi_{1}+\phi_{2})/2} [\omega_1(\lambda)]^L 
\prod_{j=1}^{n} -\frac{\alpha_{2}(\lambda_{j},\lambda)}{\alpha_{9}(\lambda_{j},\lambda)} + 
e^{i(\phi_{1}+\phi_{2})/2} [\omega_3(\lambda)]^L 
\prod_{j=1}^{n} -\frac{\alpha_{8}(\lambda,\lambda_{j})}{\alpha_{7}(\lambda,\lambda_{j})}  
\nonumber \\
&& +\prod_{j=1}^{n} - \frac{\alpha_{1}(\lambda,\lambda_{j})}{\alpha_{9}(\lambda,\lambda_{j})} 
\Lambda^{(1)}_{G}(\lambda,\{\lambda_{l}\}) 
\ear
where the variables $\{\lambda_{j}\}$ satisfy the Bethe Ansatz 
equations
\EQ
\left[ \frac{\omega_1(\lambda_{j})}{\omega_2(\lambda_{j})} 
\right]^{L} = -(-1)^{n}e^{i(\phi_{1}+\phi_{2})/2} 
\Lambda^{(1)}_{G}(\lambda=\lambda_{j},\{\lambda_{l}\}),~ j=1, \ldots, n
\EN

It turns out that the auxiliary problem  gets 
also an extra  modification besides the sign on amplitude
$\bar{b}(\lambda,\mu)$. The auxiliary problem
absorbs the twisting on the diagonal
fields $ A_{11}(\lambda)$ and $A_{22}(\lambda)$ and now  function 
$\Lambda^{(1)}_{G}(\lambda,\{\lambda_{j}\})$ is  the eigenvalue 
of the following auxiliary transfer matrix
\EQ
{\cal T}^{(1)}_{G}(\lambda,\{\lambda_{i}\}) = 
Tr_{{\cal A}} \left[ G^{(1)}_{{\cal A}}\tilde{{\cal L}}^{(1)}_{{\cal A}n}(\lambda,\lambda_{n})\tilde{{\cal L}}^{(1)}_{{\cal A}n-1}(\lambda,\lambda_{n-1}) \ldots {\tilde{\cal L}}^{(1)}_{{\cal A}1}(\lambda) \right]
\EN
where the Lax operator 
$\tilde{{\cal L}}^{(1)}_{{\cal A}j}(\lambda,\lambda_{j})$ and the matrix
$ {G}^{(1)}_{{\cal A}}$ are given by
\EQ
\tilde{{\cal L}}^{(1)}_{{\cal A}j}(\lambda,\lambda_j) = 
\pmatrix{
1  &0  &0  &0  \cr
0  &-\bar{b}(\lambda,\lambda_j)  &\bar{a}(\lambda,\lambda_j)  &0  \cr
0  &\bar{a}(\lambda,\lambda_j)  &-\bar{b}(\lambda,\lambda_j)  &0  \cr
0  &0  &0  &1  \cr}~~,~~
G^{(1)}_{{\cal A}} = 
\pmatrix{
e^{-i(\phi_{1}-\phi_{2})/2} & 0 \cr
0 & e^{i(\phi_{1}-\phi_{2})/2} \cr}
\EN

The solution of this auxiliary problem is once again standard. Following 
the lines of section 4.2  we find that the auxiliary eigenvalue expression is
\EQ
\Lambda^{(1)}\left(\lambda,\{\lambda_{i}\},\{\mu_{j}\} \right) =
e^{-i(\phi_{1}-\phi_{2})/2}
\prod_{l=1}^{m} \frac{1}{-\bar{b}(\mu_{l},\lambda)} +
e^{i(\phi_{1}-\phi_{2})/2}
\prod_{j=1}^{n}
-\bar{b}(\lambda,\lambda_{j})
\prod_{l=1}^{m} - \frac{1}{\bar{b}(\lambda,\mu_{l})}
\EN

Collecting these results altogether and substituting the 
variables $z_{\pm}(\lambda_j)$ and $\tilde{\mu}_l$ we find that
the eigenvalues of $T_{G}(\lambda)$ can be written as
\bear
\Lambda_{G}(\lambda,\{z_{\pm}(\lambda_{j})\},\{\tilde{\mu}_{l}\}) &  = &
(-1)^{n} e^{-i(\phi_{1}+\phi_{2})/2} 
[\omega_1(\lambda)]^L \prod_{j=1}^{n} \frac{b(\lambda)}{a(\lambda)} \frac{1+z_{-}(\lambda_{j})/z_{+}(\lambda)}{1-z_{-}
(\lambda_{j})/z_{-}(\lambda)} 
\nonumber \\
&&+ e^{i(\phi_{1}+\phi_{2})/2} [\omega_3(\lambda)]^L \prod_{j=1}^{n}  \frac{b(\lambda)}{a(\lambda)} \frac{1+z_{-}(\lambda_{j})z_{-}(\lambda)}{1-z_{-}(\lambda_{j})z_{+}(\lambda)} 
\nonumber \\
&&+(-)^{m}e^{-i(\phi_{1}-\phi_{2})/2}[\omega_2(\lambda)]^{L}
\left \{
\prod_{j=1}^{n} \frac{b(\lambda)}{a(\lambda)} 
\frac{1+z_{-}(\lambda_{j})/z_{+}(\lambda)}{1-z_{-}(\lambda_{j})/z_{-}(\lambda)}
\right. 
\nonumber\\
&& \left. \times \prod_{l=1}^{m} \frac{z_{-}(\lambda)-1/z_{-}(\lambda) 
- \tilde{\mu}_{l} + U/2}{z_{-}(\lambda)-1/z_{-}(\lambda)
 - \tilde{\mu}_{l} - U/2}
+(-1)^{n}e^{i(\phi_{1}-\phi_{2})}
\right. 
\nonumber\\
&& \left. \times \prod_{j=1}^{n}  \frac{b(\lambda)}{a(\lambda)}
\frac{1+z_{-}(\lambda_{j})z_{-}(\lambda)}{1-z_{-}(\lambda_{j})z_{+}(\lambda)}
\prod_{l=1}^{m} \frac{1/z_{+}(\lambda)-z_{+}(\lambda) - \tilde{\mu}_{l}
 - U/2}{1/z_{+}(\lambda)-z_{+}(\lambda) - \tilde{\mu}_{l} + U/2} \right \}
\nonumber \\
\ear
while the nested Bethe Ansatz equation are given by
\bear
 (-1)^{m+n} e^{-i\phi_{2}} 
[z_{-}(\lambda_{k})]^L & = &
-\prod_{j=1}^{m}  \frac{z_{-}(\lambda_{k})-1/z_{-}(\lambda_{k}) - \tilde{\mu}_{j} + U/2}{z_{-}(\lambda_{k})-1/z_{-}(\lambda_{k}) - \tilde{\mu}_{j} - U/2},
~k=1, \ldots, n
\nonumber \\
\prod_{k=1}^{n}  \frac{z_{-}(\lambda_{k})-1/z_{-}(\lambda_{k}) - \tilde{\mu}_{l} - U/2}{z_{-}(\lambda_{k})-1/z_{-}(\lambda_{k}) - \tilde{\mu}_{l} + U/2} & =&
-(-1)^{n} e^{-i(\phi_{1}-\phi_{2})}
\prod_{j=1}^{m} \frac{\tilde{\mu}_{l}-\tilde{\mu}_{j}+U}{\tilde{\mu}_{l}-\tilde{\mu}_{j}-U},~~l=1, \ldots, m 
\nonumber\\
\ear

In order to  get the results for the Hubbard model with twisted boundary
conditions we substitute 
the angles (10) in the above expressions. We should also remember
that we are using the language of holes and therefore the
integers $n$ and $m$ are identified with the total number of holes $N^{h}$
and the number of holes with spin up $N^{h}_{\uparrow}$, respectively.
This cancels extra phase 
factors in the Bethe Ansatz equations (127) and we recover the known set of 
nonlinear equations parametrizing the spectrum of the twisted Hubbard 
model \cite{SS,MFY}. 
Let us close this discussion by mentioning a 
possible application of these twisted Bethe Ansatz results. Consider the 
Hubbard model perturbed by a particle current term (see e.g ref. \cite{CAM})
with periodic
boundary conditions. This model is described by
the Hamiltonian
\EQ
{H_c}(U,\lambda_{c}) =
H(U,\phi_{\uparrow}=0,\phi_{\downarrow}=0) - 
i\lambda_{c}\sum_{i=1}^{L} \sum_{\sigma=\pm} (c^{\dagger}_{i+1\sigma}c_{i\sigma} - c^{\dagger}_{i\sigma} c_{i+1\sigma})
\EN

In the spin language, this perturbation is a Dzyaloshinsky-Moriya 
interaction in the azimuthal direction, playing the role of a ``vertical''
magnetic field. Similar to what happens in the spin case \cite{AWR}, the 
fermionic current perturbation can be 
gauged away by using the canonical transformation \cite{MFY}
\EQ
c_{k\sigma} \rightarrow e^{i\frac{(2k-3)\phi}{2}} c_{k\sigma}~~,~~
\tan(\phi) = \lambda_{c}
\EN
allowing us to derive the relation
\EQ
{H_c}(U,\lambda_{c}) =
\sqrt{1+\lambda^{2}_{c}}   H( \frac{U}{\sqrt{1+\lambda^{2}_{c}}},\phi_{\uparrow}=\phi L,\phi_{\downarrow}=\phi L)
\EN

Thus, the spectrum of ${H_c}(U,\lambda_{c})$ is related to that of the Hubbard model with certain twisted boundary conditions and 
renormalized coupling. Similar reasoning also works if we add a
spin current term $J_s = \displaystyle
-i\lambda_{s}\sum_{i=1}^{L} \sum_{\sigma=\pm} \sigma 
(c^{\dagger}_{i+1\sigma}c_{i\sigma} - 
c^{\dagger}_{i\sigma} c_{i+1\sigma})$ \cite{CAM} 
to the Hubbard Hamiltonian. In 
this case, after performing the transformation 
\EQ
c_{k\uparrow} \rightarrow e^{i\frac{(2k-3)\phi}{2}} c_{k\uparrow}~~,~~ 
c_{k\downarrow} \rightarrow e^{-i\frac{(2k-3)\phi}{2}} c_{k\downarrow} 
\EN
we find the that the Hamiltonian $H_s(U,\lambda_s)$ of the Hubbard model
perturbed by the spin current satisfies
\EQ
{H_s}(U,\lambda_{s}) =
\sqrt{1+\lambda^{2}_{s}}   H( \frac{U}{\sqrt{1+\lambda^{2}_{s}}},\phi_{\uparrow}=\phi L,\phi_{\downarrow}=-\phi L)
\EN

Before closing this section we would like to comment on possible extensions
of the results we have obtained so far. First 
it is possible to diagonalize a two-parameter family of vertex models
whose Lax operator is ${\cal{L}}^{(\theta_0)}(\lambda) =P R(\lambda,\theta_0)$
\cite{SA2,SHWA}. Its Bethe Ansatz solution follows directly from the 
results of this section, since the main change is only concerned
with the action of the fields on the reference state. It turns out that
now the bare pseudomomenta (left-hand side of first equation (127)) depends on
the variable $\theta_0$ as 
$[\frac{\alpha_2(\lambda,\theta_0)}{-\alpha_9(\lambda,\theta_0)}]^{L}$. 
Also, the whole formalism can be extended to treat
the Hubbard model in the presence of chemical potential \cite{GUAN}. 
Finally, for further results on twisted boundary
conditions see for instance ref. \cite{DEG}.

\subsection{$SU(2)$ symmetries}

In this subsection we  
investigate the highest weights properties of the
eigenvectors  constructed in section 4,
with respect to the two $SU(2)$ symmetries of 
the Hubbard model \cite{SU2}. Few years ago,
Essler Korepin and Schoutens \cite{EKS} 
have shown that certain ``regular'' 
states obtained from the coordinate Bethe Ansatz wave function
are highest weight states of both the $SU(2)$ algebra of rotations and 
$\eta$-pairing $SU(2)$ symmetry. The idea here is to explore the algebraic 
machinery we developed in the previous section 
to study this problem from an algebraic  perspective, in close analogy with  
the discussion by Takhtajan and Faddeev\cite{TAFA1} 
for the Heisenberg model. For this purpose we will use the results of 
G\"ohmann and Murakami \cite{GOM} who recently showed that the graded 
monodromy matrix indeed commutes with these two $SU(2)$ Lie algebras. 
More precisely, following the notation of ref. \cite{GOM}
we have
\EQ
[{\cal T}(\lambda),S^{\alpha}]_{Q} = -[{\cal T}(\lambda),{\sum}^{\alpha}]_{A},
~ \alpha=+,-,z
\EN
and
\EQ
[{\cal T}(\lambda),\eta^{\alpha}]_{Q} = -[{\cal T}(\lambda),\tilde{\sum}^{\alpha}]_{A},
~ \alpha=+,-,z
\EN
where the subscripts $Q$ and $A$ emphasize in which space, 
quantum or auxiliary, 
the commutators are taken, respectively. The SU(2) generators of rotations 
$S^{\alpha}$ and those of the $\eta$-pairing symmetry $\eta^{\alpha}$ are 
defined by \cite{GOM}
\EQ
S^{+} = -\sum_{j=1}^{L} c_{j\uparrow}^{\dagger}c_{j\downarrow}~~,~~
S^{-} = -\sum_{j=1}^{L} c_{j\downarrow}^{\dagger}c_{j\uparrow}~~,~~
S^{z} = \sum_{j=1}^{L}(n_{j\uparrow}-n_{j\downarrow})
\EN
and
\EQ
\eta^{+} = \sum_{j=1}^{L} (-1)^{j+1} c_{j\uparrow}^{\dagger}c_{j\downarrow}^{\dagger}~~,~~
\eta^{-} = \sum_{j=1}^{L} (-1)^{j+1} c_{j\downarrow} c_{j\uparrow}~~,~~
\eta^{z} = \sum_{j=1}^{L}(n_{j\uparrow}+n_{j\downarrow}-1)
\EN
while the matrices $\sum^{\alpha}$ and $\tilde{\sum}^{\alpha}$ are \cite{GOM}
\EQ
{\sum}^{+}=\sigma^{+} \otimes \sigma^{-}~~,~~
{\sum}^{-}=\sigma^{-} \otimes \sigma^{+}~~,~~
{\sum}^{z}=\frac{1}{2}(\sigma^{z} \otimes \hat{I} - \hat{I} \otimes \sigma^{z})
\EN
\EQ
\tilde{\sum}^{+}=\sigma^{+} \otimes \sigma^{+}~~,~~
\tilde{\sum}^{-}=\sigma^{-} \otimes \sigma^{-}~~,~~
\tilde{\sum}^{z}=\frac{1}{2}(\sigma^{z} \otimes \hat{I} + \hat{I} \otimes \sigma^{z})
\EN

Let us begin by considering the $\eta$-pairing symmetry.  The identity (134)
enables us to compute the commutators of the 
creation fields $\vec{B}(\lambda)$ and $F(\lambda)$ with the 
$SU(2)$ $\eta$-pairing generators. For the component $\eta^{z}$ we find
\EQ
[\eta^{z},\vec{B}(\lambda)]=-\vec{B}(\lambda)~~,~~
[\eta^{z},F(\lambda)]=-2F(\lambda)
\EN
while for $\eta^{+}$ we have
\EQ
[\eta^{+},\vec{B}(\lambda)]=-\vec{C}^{*}(\lambda)~~,~~
[\eta^{+},F(\lambda)]=B(\lambda)-D(\lambda)
\EN

We see that formula (139) corroborates the physical
interpretation we have proposed for the creation fields 
$\vec{B}(\lambda)$ and $F(\lambda)$, i.e. that
they create a single and a doubly occupied hole on the full
band pseudovacuum. For example, from this equation it is straightforward to
derive 
\EQ
\eta^{z} \ket{ {\Phi}_{n}(\lambda_{1},\cdots ,\lambda_{n})}
= (L-n) \ket{ {\Phi}_{n}(\lambda_{1},\cdots ,\lambda_{n})}
\EN
where we used the property  $\eta^{z}\ket{0}=L\ket{0}$. 

We note that the  above result is valid for arbitrary values of the rapidities.
However, this 
is no longer true when we consider the 
annihilation property  of the raising operator $\eta^{+}$. In what
follows we shall show that
\EQ
\eta^{+} \ket{ {\Phi}_{n}(\lambda_{1},\cdots ,\lambda_{n})}=0
\EN
provided the rapidities $\{\lambda_{j}\}$ satisfy the Bethe Ansatz 
equations derived in section 4.

 To verify the above annihilation property it is
instructive first to study the case of few particles over the reference
state and afterwards use mathematical induction for the general case. 
From equation (136) this is clearly correct for the
reference state. For the one-particle state, by using the first commutator
(140), it is easy to show that
\EQ
\eta^{+} \ket{\Phi_{1}(\lambda_{1})} = 
\eta^{+} \vec{B}(\lambda_{1}) . \vec{{\cal F}} \ket{0} = 
[\eta^{+},\vec{B}(\lambda_{1})] .  \vec{{\cal F}} \ket{0} = 
-\vec{C}^{*}(\lambda_{1}) .  \vec{{\cal F}} \ket{0} = 0
\EN

The Bethe Ansatz restrictions start to emerge in the two-particle 
state analysis. For this state the commutators (140) produce 
\bear
\eta^{+} \ket{\Phi_{2}(\lambda_{1},\lambda_{2})} &=&
\vec{B}(\lambda_{1}) \otimes \eta^{+} \vec{B}(\lambda_{2}) . \vec{{\cal F}} \ket{0} - \vec{C}^{*}(\lambda_{1}) \otimes \vec{B}(\lambda_{2}) . \vec{{\cal F}} \ket{0} 
\nonumber \\
&&+ \frac{i\alpha_{10}(\lambda_{1},\lambda_{2})}{\alpha_{7}(\lambda_{1},\lambda_{2})}[B(\lambda_{1})-D(\lambda_{1})] B(\lambda_{2}) \vec{\xi}.\vec{{\cal F}} \ket{0}
\ear

The first term in the above equation vanishes
by the same arguments used 
in the one-particle state 
analysis. To simplify the second term we use commutation 
rule (B.3) and finally the third term is easily estimated from the diagonal
relation (23). Putting these simplifications together we find
\bear
\eta^{+} \ket{\Phi_{2}(\lambda_{1},\lambda_{2})} & = &
\frac{i\alpha_{10}(\lambda_{1},\lambda_{2})}
{\alpha_{7}(\lambda_{1},\lambda_{2})}
\left [ [w_{1}(\lambda_{1})w_{1}(\lambda_{2})]^{L}- 
[w_{2}(\lambda_{1})w_{2}(\lambda_{2})]^{L} \right ] \vec{\xi}.\vec{\cal{F}} \ket{0}
\nonumber \\
& = & \frac{i\alpha_{10}(\lambda_{1},\lambda_{2})}
{\alpha_{7}(\lambda_{1},\lambda_{2})}
\left[ [w_{1}(\lambda_{1})w_{1}(\lambda_{2})]^{L}
\right.
\nonumber\\
&& \left. - [w_{2}(\lambda_{1})w_{2}(\lambda_{2})]^{L}
\Lambda^{(1)}(\lambda=\lambda_{1},\{\lambda_{l}\})
\Lambda^{(1)}(\lambda=\lambda_{2},\{\lambda_{l}\}) \right] 
\vec{\xi}.\vec{\cal{F}} \ket{0}
\nonumber\\
& = &0
\ear
where in the second  line we used the
following two-particle identity, $\Lambda^{(1)}(\lambda=\lambda_{1},\{\lambda_{l}\})
\Lambda^{(1)}(\lambda=\lambda_{2},\{\lambda_{l}\})=1$. Clearly, 
the term in brackets 
vanishes due to the Bethe Ansatz equations (90).

Next we consider the three-particle state. We shall see that
a general pattern in the analysis begins to
emerge here. After using the  
commutator relations (140) we have  
\bear
\eta^{+} \ket{\Phi_{3}(\lambda_{1},\lambda_{2},\lambda_{3})} & = &
\vec{B}(\lambda_{1}) \otimes \eta^{+} \vec{\Phi}_{2}(\lambda_{2}
,\lambda_{3}) . \vec{{\cal F}} \ket{0} -
\vec{C}^{*}(\lambda_{1}) \otimes \vec{\Phi}_{2}(\lambda_{2},\lambda_{3}) . \vec{{\cal F}} \ket{0}
\nonumber \\
&& +\vec{\xi}[B(\lambda_{1})-D(\lambda_{1})] \otimes \vec{B}(\lambda_{3}) B(\lambda_{2}) g^{(3)}_{1}(\lambda_{1},\lambda_{2},\lambda_{3}) . \vec{{\cal F}} \ket{0}
\nonumber \\
&& +\vec{\xi}[B(\lambda_{1})-D(\lambda_{1})] \otimes \vec{B}(\lambda_{2}) B(\lambda_{3}) g^{(3)}_{2}(\lambda_{1},\lambda_{2},\lambda_{3}) . \vec{{\cal F}} \ket{0}
\ear

The first  term is computable directly  from the first line of equation (145),
after making the 
replacements $\lambda_{1} \rightarrow \lambda_{2}$ 
and $\lambda_{2} \rightarrow \lambda_{3}$. The third and fourth terms are
estimated with the help of commutations rules (35-36).
The simplifications for the second term is more complicated since it involves
the knowledge of an extra commutation rule, besides relation (B.3),
between the fields
$\vec{C}^{*}(\lambda)$ and $F(\mu)$. This relation is 
given by
\bear
\vec{C}^{*}(\lambda)F(\mu) & = & -i\frac{\alpha_{9}(\lambda,\mu)}{\alpha_{7}(\lambda,\mu)} F(u)\vec{C}^{*}(\lambda) + 
i\frac{\alpha_{10}(\lambda,\mu)}{\alpha_{7}(\lambda,\mu)}
\vec{\xi}.[\hat{A}(\lambda) \otimes \vec{B}^{*}(\mu)]
\nonumber \\
&& -\frac{\alpha_{4}(\lambda,\mu)}{\alpha_{7}(\lambda,\mu)} 
\vec{B}(\lambda) D(\mu)
+\frac{\alpha_{5}(\lambda,\mu)}{\alpha_{7}(\lambda,\mu)} \vec{B}(\mu) D(\lambda)
\ear

Collecting all the pieces together is remarkable to see  that many terms 
have opposite
signs and thus they are trivially canceled out. However, there is
a non-trivial simplification yet to be carried out. This is related to the 
terms proportional to $[\vec{\xi}.(\vec{B}^{*}(\lambda_{1}) \otimes \hat{I})]$  
and they vanish thanks to the following identity
\EQ
\frac{\alpha_{10}(x,z)}{\alpha_{7}(x,z)}
\frac{\alpha_{9}(x,y)}{\alpha_{7}(x,y)}+
\frac{\alpha_{5}(y,z)}{\alpha_{9}(y,z)}
\frac{\alpha_{10}(x,y)}{\alpha_{7}(x,y)}+
\frac{\alpha_{10}(x,z)}{\alpha_{7}(x,z)}
\frac{\alpha_{2}(z,y)}{\alpha_{9}(z,y)}=0
\EN

After these simplifications, the remaining terms are only proportional 
to $[\vec{\xi} \otimes \vec{B}(\lambda_{j})]$ and they can be compactly written 
in the following way 
\bear
\eta^{+} \ket{\Phi_{3}(\lambda_{1},\lambda_{2},\lambda_{3})} =
\sum_{j=2}^{3}\sum_{l=1}^{j-1} [\vec{\xi} \otimes 
\vec{\Phi}_1(\lambda_{1},\ldots,\check{\lambda}_{l},\ldots,
\check{\lambda}_{j},\ldots,\lambda_{3})] 
\hat{Q}^{(3)}_{lj}(\lambda_{l},\lambda_{j};\{\lambda_{k}\}) . \vec{{\cal F}} \ket{0}
\ear

The first term 
$\hat{Q}^{(3)}_{12}(\lambda_1,\lambda_2;\{\lambda_k\})$ 
is easily figured out because it has only 
two main contributions coming from the second and the third terms of equation 
(146). The other two are obtained from this term via consecutive permutation 
of rapidities through the exchange property (78). The expressions for 
these coefficients are
\bear
\hat{Q}^{(3)}_{lj}(\lambda_{l},\lambda_{j};\{\lambda_{k}\}) & = &
\left[ [w_{1}(\lambda_{l})w_{1}(\lambda_{j})]^{L}- 
[w_{2}(\lambda_{l})w_{2}(\lambda_{j})]^{L}
\Lambda^{(1)}(\lambda=\lambda_{l},\{\lambda_{k}\})
\Lambda^{(1)}(\lambda=\lambda_{j},\{\lambda_{k}\}) \right] 
\nonumber \\
&& \times 
i\frac{\alpha_{10}(\lambda_{l},\lambda_{j})}{\alpha_{7}(\lambda_{l},\lambda_{j})}
\prod^{3}_{\stackrel {k=1}{k \neq j,l}}
\frac{\alpha_{1}(\lambda_{l},\lambda_{k})}
{i\alpha_{9}(\lambda_{l},\lambda_{k})}\frac{\alpha_{1}(\lambda_{j},\lambda_{k})}
{i\alpha_{9}(\lambda_{j},\lambda_{k})}
\hat{O}_{lj}^{(2)}(\lambda_l,\lambda_j; \{\lambda_k\})
\ear
and they vanish again as a consequence of the Bethe Ansatz equations (90).

Now using mathematical 
 induction it is possible to write the action of the raising operator on 
a general  $n$-particle state as
\bear
\eta^{+} \ket{\Phi_{n}(\lambda_{1},\ldots,\lambda_{n})} =
\sum_{j=2}^{n}\sum_{l=1}^{j-1} [\vec{\xi} \otimes \vec{\Phi}_{n-2}(\lambda_{1},\ldots,\check{\lambda}_{l},\ldots,\check{\lambda}_{j},\ldots,\lambda_{n})] 
\hat{Q}^{(n)}_{lj}(\lambda_{l},\lambda_{j};\{\lambda_{k}\}) . \vec{{\cal F}} \ket{0}
\ear

As before, it is convenient first to compute the simplest coefficient 
$\hat{Q}^{(n)}_{12}(\lambda_{1},\lambda_{2};\{\lambda_{k}\})$ and then 
take advantage of the 
permutation property (78) 
to obtain the remaining ones. For this term we have just 
two contributions coming from 
\EQ
I:= 
\vec{\xi} \otimes B(\lambda_1)
\vec{\Phi}_{n-2}(\lambda_{3}, \dots ,\lambda_{n})B(\lambda_{2})
\hat{g}^{(n)}_{1}(\lambda_{1}, \dots ,\lambda_{n}) . \vec{{\cal F}} \ket{0} 
\EN
and
\EQ
II:=-\vec{C}^{*}(\lambda_{1}) \otimes \vec{\Phi}_{n-1}(\lambda_{2}, \ldots ,\lambda_{n})] . \vec{{\cal F}} \ket{0} 
\EN

We compute the first part by carrying the scalar operator 
$B(\lambda_{1})$ through the vector
$\vec{\Phi}_{n-2}(\lambda_{3}, \ldots ,\lambda_{n})$ 
keeping only the ``wanted terms'' proportional to 
$B(\lambda_{1})$. This is very similar to what we did in appendix 
$D$ and we find
\bear
I:= 
[w_{1}(\lambda_{1})w_{1}(\lambda_{2}) ]^{L}
\frac{i\alpha_{10}(\lambda_{1},\lambda_{2})}
{\alpha_{7}(\lambda_{1},\lambda_{2})}
\prod^n_{\stackrel{k=1}{k \neq 1,2}} \frac{i\alpha_{2}(\lambda_{k},\lambda_{1})}
{\alpha_{9}(\lambda_{k},\lambda_{1})}\frac{i\alpha_{2}(\lambda_{k},\lambda_{2})}
{\alpha_{9}(\lambda_{k},\lambda_{2})}
 [\vec{\xi} \otimes 
\vec{\Phi}_{n-2}(\lambda_{3}, \dots ,\lambda_{n})] . \vec{{\cal F}} \ket{0} 
\ear

The second part is more involving since we have to carry two 
operators of type $\hat{A}(\lambda)$ through vector
$\vec{\Phi}_{n-2}(\lambda_{3}, 
\ldots ,\lambda_{n})$. This means that we have to compute the expression
\EQ
II:=-\frac{i\alpha_{10}(\lambda_{1},\lambda_{2})}
{\alpha_{7}(\lambda_{1},\lambda_{2})}
\xi_{\alpha \beta} \hat{A}_{\alpha b_{1}}(\lambda_{1}) \hat{A}_{\beta b_{2}}(\lambda_{2}) [\vec{\Phi}_{n-2}(\lambda_{3}, \ldots ,\lambda_{n})]_{b_{3} \ldots b_{n}} {\cal F}^{b_{n} \ldots b_{1}} \ket{0} 
\EN
which after some algebra can be compacted back as
\bear
II:=-\frac{i\alpha_{10}(\lambda_{1},\lambda_{2})}
{\alpha_{7}(\lambda_{1},\lambda_{2})}
[w_{2}(\lambda_{1})w_{2}(\lambda_{1}) ]^{L}
\prod^{n}_{\stackrel{k=1}{k \neq 1,2}} \frac{\alpha_{1}(\lambda_{1},\lambda_{k})}
{i\alpha_{9}(\lambda_{1},\lambda_{k})}\frac{\alpha_{1}(\lambda_{2},\lambda_{k})}
{i\alpha_{9}(\lambda_{2},\lambda_{k})} 
\nonumber \\
\times [\vec{\xi} \otimes \vec{\Phi}_{n-2}(\lambda_{3}, \ldots ,\lambda_{n})]_{\alpha_{1} \ldots \alpha_{n}} 
[T^{(1)}(\lambda=\lambda_{1},\{\lambda_l\})
T^{(1)}(\lambda=\lambda_{2},\{ \lambda_l \})]^{b_{1} \ldots b_{n}}_{\alpha_{1} \ldots \alpha_{n}} {\cal F}^{b_{n} \ldots b_{1}} \ket{0} 
\ear
Finally, putting together expressions (154) and (156) and also 
using the 
auxiliary eigenvalue definition (91) we find
\bear
\hat{Q}^{(n)}_{12}(\lambda_{1},\lambda_{2};\{\lambda_{k}\}) &=&
\left[ 
[w_{1}(\lambda_{1})w_{1}(\lambda_{2})]^{L}- [w_{2}(\lambda_{1})w_{2}(\lambda_{2})]^{L}
\Lambda^{(1)}(\lambda=\lambda_{1},\{\lambda_{k}\})
\Lambda^{(1)}(\lambda=\lambda_{2},\{\lambda_{k}\}) \right] 
\nonumber \\
&& \times \frac{i\alpha_{10}(\lambda_{1},\lambda_{2})}
{\alpha_{7}(\lambda_{1},\lambda_{2})}
\prod^{n}_{\stackrel{k=1}{k \neq 1,2}} \frac{\alpha_{1}(\lambda_{1},\lambda_{k})}
{i\alpha_{9}(\lambda_{1},\lambda_{k})}\frac{\alpha_{1}(\lambda_{2},\lambda_{k})}
{i\alpha_{9}(\lambda_{2},\lambda_{k})} 
\ear
which once again vanishes due to the Bethe Ansatz equations. All the other 
coefficients are obtained by permuting the rapidities and by taking into 
account the exchange property (78), and  as a result 
they get an extra multiplicative 
``ordering'' factor $\hat{O}_{lj}^{(2)}(\lambda_l,\lambda_j; \{\lambda_k\})$.
Since the Bethe Ansatz equations are 
invariant under indices relabeling, 
they vanish too.  This completes the proof that
the eigenvectors (81) are highest weight states of the $\eta$-pairing 
symmetry.

Next we turn to examine the highest weight property of the $SU(2)$ algebra of 
rotations.  Now the commutators of the creation fields
with the $SU(2)$ generators  are obtained from equation (133). For the 
component
$S^{z}$  we find 
\EQ
[S^{z},B_{1}(\lambda)]=B_{1}(\lambda)~~,~~
[S^{z},B_{2}(\lambda)]=-B_{2}(\lambda)~~,~~
[S^{z},F(\lambda)]=0
\EN
and for $S^{+}$  we have
\EQ
[S^{+},B_{1}(\lambda)]=0, ~~
[S^{+},B_{2}(\lambda)]=B_{1}(\lambda), ~~
[S^{+},F(\lambda)]=0
\EN

First of all, it is not difficult 
to see that eigenvector (81)  will be hardly 
annihilated 
by the raising operator 
$S^{+}$ unless further restriction are assumed.
To illustrate this fact in a simple example let us consider 
the one-particle state. By using the commutators (159) we find
\EQ
S^{+} \ket{\Phi_{1}(\lambda_{1})} = B_{1}(\lambda_{1}){\cal F}^{2} \ket{0} 
\EN
where we used that 
$S^{+} \ket{0}= 0$. Therefore, to assure the highest weight property for
the one-particle state
we must set ${\cal{F}}^{2} =0$. This is an example of what was called
``regular'' Bethe states in ref.\cite{EKS}, and in general these states
are obtained by projecting out  
the negative sectors of the magnetization operator
$S^{z}$. This later condition 
is easily implemented for the eigenvector (81)
if one uses the commutators (158).

To see how this works in practice let us consider 
the two-particle state. In this case
it is obvious that we have to set ${\cal{F}}^{22}=0$ , and after that we
find
\bear
S^{+} \ket{\Phi_{2}(\lambda_{1},\lambda_{2})}_{regular} & =&
S^{+} \left[
B_{1}(\lambda_{1})B_{1}(\lambda_{2}){\cal F}^{11}+
B_{1}(\lambda_{1})B_{2}(\lambda_{2}){\cal F}^{21}
+B_{2}(\lambda_{1})B_{1}(\lambda_{2}){\cal F}^{12}+
\right.
\nonumber\\
&& \left. +i\frac{\alpha_{10}(\lambda_{1},\lambda_{2})}{\alpha_{7}(\lambda_{1},\lambda_{2})} F(\lambda_{1}) B(\lambda_{2}) \vec{\xi} . \vec{{\cal F}} 
\right] \ket{0}
\nonumber\\
&=& [\sum_{P}{\cal F}^{12}] B_{1}(\lambda_{1}) B_{1}(\lambda_{2}) \ket{0}
\ear
where the sum is over permutations on the indices of the coefficient $
{\cal{F}}^{a_2 a_1}$. In this case it is straightforward to verify
that this sum indeed vanishes by directly solving the auxiliary eigenvalue
problem (56). The deeper reason behind this fact, however, is that the
vanishing  of such sum is precisely related to the highest weight property
of the Bethe wave functions of the $XXX$ Heisenberg model with $two$ sites.
We should recall here that the components of this wave function are identified
with the coefficients
${\cal{F}}^{a_2 a_1}$.
From this discussion, it becomes evident 
that the whole procedure can be applied 
to any multi-particle state. As an example, in table 1 we summarize our
findings up to the four-particle state
\label{ta1}
\btb
\caption{
 The ``regular'' multi-particle states 
properties up to  $n=4$.}
\bc
\bt{|c|c|c|c|} \hline
     &$n$  &$S^{z}\ket{\Phi_{n}(\lambda_1,\ldots,\lambda_n)}_{regular}$  &$S^{+}\ket{\Phi_{n}(\lambda_1,\ldots,\lambda_n)}_{regular}=0$     \\ \hline\hline
&$2$   &$2$  &$none$            \\ \hline
&$2$   &$0$  &$ \displaystyle \sum_{P}{\cal F}^{12}=0$            \\ \hline
&$3$   &$3$  &$none$            \\ \hline
&$3$   &$1$  &$ \displaystyle \sum_{P}{\cal F}^{112}=0$            \\ \hline
&$4$   &$4$  &$none$            \\ \hline
&$4$   &$2$  &$ \displaystyle \sum_{P}{\cal F}^{1112}=0$            \\ \hline
&$4$   &$0$  &$\displaystyle \sum_{P}{\cal F}^{112\bar{2}}=
\displaystyle \sum_{P}{\cal F}^{11\bar{2}2}=\displaystyle \sum_{P}{\cal F}^{1\bar{2}12}= \displaystyle \sum_{P}{\cal F}^{\bar{2}112}=0$            \\ \hline
\et
\ec
\etb

The columns of table 1 refer 
to the particle number, magnetization values 
and the sufficient vanishing condition for $S^{+}$
annihilate the ``regular'' part of eigenvector (81), respectively.  In the sum
the symbol $\bar{a}$ means that the 
$a$-$th$ element is maintain  fixed under permutations. The generalization
to multi-particle state is done by induction and the sufficient
vanishing conditions are made of the many possible permutation over the
coefficients ${\cal{F}}^{a_n \ldots a_1} $ having positve magnetization.
As before,
these conditions are fulfilled as a consequence of 
the highest weight property of the Bethe states of the $XXX$ Heisenberg 
spin chain in a lattice with size $n$. Since this later point
has been well explained 
by Essler, Korepin and Schoutens \cite{EKS}, there is no need 
to proceed with details, and thus we conclude our proof that
$S^{+}\ket{\Phi_{n}(\lambda_{1},\cdots,\lambda_{n})}_{regular}=0$ here.

Finally, we remark that similar properties can be also verified
for the ``dual'' eigenvector. The only difference is
that now the
``regular'' states are defined by projecting out the $positive$ sector
of the magnetization. At this level, the
eigenvector and its ``dual'' becomes complementary eigenstates.

\section{The $ABCDF$ framework for the Bariev model }

The purpose of this section is to illustrate that the
$ABCDF$ framework developed in the previous sections is by no means only
applicable to
the Hubbard model. In order to show that, we consider a 
second interesting 
model of interacting $XY$ chains whose corresponding $R$-matrix also 
does not have the difference property. The model was 
originally formulated by Bariev 
\cite{BAR} and its one-dimensional Hamiltonian is 
\EQ
H = \sum_{i=1}^{L} (\sigma_{i}^{+}\sigma_{i+1}^{-} + \sigma_{i}^{-}\sigma_{i+1}^{+})(1+V\tau_{i+1}^{z}) + 
(\tau_{i}^{+}\tau_{i+1}^{-} + \tau_{i}^{-}\tau_{i+1}^{+})(1+V\sigma_{i}^{z})
\EN
where $V$ is a coupling constant. In the 
language of fermions $V$ plays the role of 
a bond-charge interaction and Hamiltonian (162) resembles the model of hole 
superconductivity proposed by Hirsch \cite{HI}.

In the context of the quantum inverse scattering method this model has 
recently been investigated by 
Zhou \cite{ZHOU} and Shiroishi and 
Wadati \cite{SHWA2} who found two distincts 
covering vertex models for the Bariev 
Hamiltonian. In this section we  apply the $ABCDF$ formalism for the former 
solution\footnote{Part of our results were first announced in ref. \cite{PMB}. See also ref. \cite{ZHOU1}}. 
In this case, the proposed Lax operator was \cite{ZHOU}
\EQ
{\cal L}_{{\cal A}j}^{(B)}(\lambda) = 
{\cal L}_{{\cal A}j}^{(1)}(\lambda){\cal L}_{{\cal A}j}^{(2)}(\lambda)
\EN
where
\EQ
{\cal L}_{{\cal A}j}^{(1)}(\lambda) =
\frac{1}{2}(1+\sigma_{j}^{z}\sigma_{{\cal A}}^{z}) +
\frac{\lambda}{2}(1-\sigma_{j}^{z}\sigma_{{\cal A}}^{z})
exp(\beta \tau^{+}_{{\cal A}}\tau^{-}_{{\cal A}}) +
(\sigma_{j}^{+}\sigma_{{\cal A}}^{-} + \sigma_{j}^{-}\sigma_{{\cal A}}^{+})
\sqrt{1+\lambda^{2} exp(2\beta \tau^{+}_{{\cal A}}\tau^{-}_{{\cal A}})}
\EN
and
\EQ
{\cal L}_{{\cal A}j}^{(2)}(\lambda) =
\frac{1}{2}(1+\tau_{j}^{z}\tau_{{\cal A}}^{z}) +
\frac{\lambda}{2}(1-\tau_{j}^{z}\tau_{{\cal A}}^{z})
exp(\beta \sigma^{+}_{{\cal A}}\sigma^{-}_{{\cal A}})+
(\tau_{j}^{+}\tau_{{\cal A}}^{-} + \tau_{j}^{-}\tau_{{\cal A}}^{+})
\sqrt{1+\lambda^{2} exp(2\beta \sigma^{+}_{{\cal A}}\sigma^{-}_{{\cal A}})}
\EN

The relation between the parameter 
$\beta$ and the coupling constant $V$ 
is determinated by computing the expression 
$ P \frac{d}{d \lambda} {\cal{L}}^{(B)}(\lambda)$ on $\lambda=0$. After
performing the rescaling 
$\lambda \rightarrow \frac{\lambda e^{-\beta/2}}{\cosh(\beta/2)}$ we found
\EQ
\bar{h}=e^{\beta}= \frac{1+V}{1-V}
\EN

	The $R$-matrix solving the Yang-Baxter algebra for this choice of Lax 
operator was also found by Zhou. Its explicit $16 \times 16$ form is \cite{ZHOU}
{\scriptsize
\bear
 R(\lambda,\mu) = 
\pmatrix{
 \rho_{1} &0 &0 &0 &0  &0  &0  &0  &0  &0  &0  &0  &0  &0  &0  &0    \cr
 0 &\rho_{2} &0 &0 &\rho_{3}  &0  &0  &0  &0  &0  &0  &0  &0  &0  &0  &0    \cr
 0 &0 &\rho_{2} &0 &0  &0  &0  &0  &\rho_{3}  &0  &0  &0  &0  &0  &0  &0    \cr
 0 &0 &0 &\rho_{4} &0  &0  &\rho_{5}  &0  &0  &\rho_{6}  &0  &0  &\rho_{9}  &0  &0  &0    \cr
 0 &\rho_{3} &0 &0 &\rho_{2}  &0  
&0  &0  &0  &0  &0  &0  &0  &0  &0  &0    \cr
 0 &0 &0 &0 &0  &\rho_{1}  &0  &0  &0  &0  &0  &0  &0  &0  &0  &0    \cr
 0 &0 &0 &\rho_{12} &0  &0  &\rho_{7}  &0  &0  &\rho_{15}  &0  &0  &\rho_{5}  &0  &0  &0    \cr
 0 &0 &0 &0 &0  &0  &0  &\rho_{8}  &0  &0  &0  &0  &0  &\rho_{11}  &0  &0    \cr
 0 &0 &\rho_{3} &0 &0  &0  &0  &0  &\rho_{2}  &0  &0  &0  &0  &0  &0  &0    \cr
 0 &0 &0 &\rho_{13} &0  &0  &\rho_{15}  &0  &0  &\rho_{10}  &0  &0  &\rho_{6}  &0  &0  &0    \cr
 0 &0 &0 &0 &0  &0  &0  &0  &0  &0  &\rho_{1}  &0  &0  &0  &0  &0    \cr
 0 &0 &0 &0 &0  &0  &0  &0  &0  &0  &0  &\rho_{8}  &0  &0  &\rho_{11}  &0    \cr
 0 &0 &0 &\rho_{14} &0  &0  &\rho_{12}  
&0  &0  &\rho_{13}  &0  &0  
&\rho_{4} &0  &0  &0    \cr
 0 &0 &0 &0 &0  &0  &0  &\rho_{11}  &0  &0  &0  &0  &0  &\rho_{8}  &0  &0    \cr
 0 &0 &0 &0 &0  &0  &0  &0  &0  &0  &0  &\rho_{11}  &0  &0  &\rho_{8}  &0    \cr
 0 &0 &0 &0 &0  &0  &0  &0  &0  &0  &0  &0  &0  &0  &0  &\rho_{1}    \cr}
\nonumber \\
\ear
}
where the fifteen non-null Boltzmann weights $\rho_{j}(\lambda,\mu)$, 
$j=1,\ldots,15$ have been 
collected in appendix $E$. We remark that we have verified 
that this $R$-matrix indeed satisfies the Yang-Baxter equation (6).
	
We note that the structure of such $R$-matrix is very similar to that 
found for the Hubbard model and consequently one could easily guess that the 
$ABCDF$ formalism should work for this embedding as well.  It is not
difficult 
to adapt 
the main steps of section 3 in order  to obtain 
the appropriate commutation rules for 
such classical vertex  analog of the 
Bariev model. The most important commutation rules have been summarized in 
appendix $E$. The interesting feature here is the structure which comes up
for both the 
``exclusion'' 
vector and the auxiliary $r$-matrix. We found that they are given by
\EQ
{\vec{\xi}}^{(B)} = 
\matrix{(
0  &1  &1/\bar{h}  &0 )  \cr},~
\hat{r}^{(B)}(\lambda,\mu) = 
\pmatrix{
1  &0  &0  &0  \cr
0  &a^{(B)}(\lambda,\mu)  &b^{(B)}(\lambda,\mu)  &0  \cr
0  &b^{(B)}(\lambda,\mu)  &a_{1}^{(B)}(\lambda,\mu)  &0  \cr
0  &0  &0  &1  \cr}
\EN
where the weights $a^{(B)}(\lambda,\mu)$ and  $b^{(B)}(\lambda,\mu)$ are
\EQ
a^{(B)}(\lambda,\mu) = \frac{\lambda(1-\bar{h}^{2})}{\lambda-\bar{h}^{2}\mu}
,~~
a_{1}^{(B)}(\lambda,\mu) = \frac{\mu(1-\bar{h}^{2})}{\lambda-\bar{h}^{2}\mu}
,~~
b^{(B)}(\lambda,\mu) = -\frac{\bar{h}(\lambda-\mu)}{\lambda-\bar{h}^{2}\mu}
\EN

From equation (169), it is  easily recognizable that the auxiliary 
$r$-matrix has the structure of an 
asymmetrical and anisotropic $6$-vertex model because the parametrization
leading to the difference property for $\hat{r}^{(B)}(\lambda,\mu)$ is now
standard, namely $\lambda = \exp(i k)$. In this case the hidden symmetry is
of Hecke type because such auxiliary $r$-matrix can be produced as a result
of 
Baxterization of the 
Hecke algebra (see e.g. ref. \cite{WDA}). We recall here that this later 
symmetry was first 
noted by Hikami and Murakami by exploiting the continuum
limit of the  Bariev Hamiltonian \cite{HIMU}.
Interesting enough, we note that
the ``exclusion'' statistics for 
``spins'' degrees of freedom   seems to be of
anyonic type with a  phase $\beta$  
which depends on the strength of the coupling constant $V$ (see equation (166)).
It remains to be seen if this feature will also be manifested in physical
quantities computable by Bethe Ansatz methods such as 
in the low temperature behaviour of the
free energy (conformal limit) and in the scattering of  the elementary
excitations.

Let us now discuss the construction of the eigenvalues and the 
eigenvectors for this classical analog of Bariev model. It turns out that 
such formulation goes fairly parallel to the one already presented in section 4 
and in appendix $D$. For this reason we shall avoid unnecessary repetition, 
and from now on we concentrate our attention only to the 
basic points. We start directly with the 
two-particle state analysis since it has already proved to
contain  sufficient information about  the  main steps entering in
the relevant
computations. Afterwards, 
generalization to multi-particle states  is made following similar 
discussion presented in appendix $D$. Our previous experience with the
Hubbard model
suggests us to begin with a 
symmetrized two-particle vector. As before, the main 
trick is to look at the commutation rule between the two creation fields of 
type $\vec{B}(\lambda)$. From equation (E.17) it is not difficult to guess that 
such vector is 
\EQ
{\vec{\Phi}}^{(B)}_{2}(\lambda_{1},\lambda_{2}) =
\vec{B}(\lambda_{1}) \otimes \vec{B}(\lambda_{2})  
-\frac{\rho_{5}(\lambda_{1},\lambda_{2})}{\rho_{9}(\lambda_{1},\lambda_{2})}
{\vec{\xi}}^{(B)} F(\lambda_{1}) B(\lambda_{2})
\EN
which is indeed the case thanks to the following identity
\EQ
\vec{\xi}^{(B)}.\hat{r}^{(B)}(\lambda,\mu) = 
\frac{\rho_{12}(\lambda,\mu)\rho_{9}(\mu,\lambda)}
{\rho_{9}(\lambda,\mu)\rho_{5}(\mu,\lambda)} {\vec{\xi}}^{(B)}
\EN

We go ahead computing the action of the diagonal fields on the 
two-particle state Ansatz. Here we shall use fully the
permutation property of the eigenvector, specially the simplifications
mentioned at the end of 
appendix $C$. Considering the commutations rules
of appendix $E$ and  following the calculations of section 4, we find that
the expressions for the action of the diagonal fields on the two-particle
state are
\bear
B(\lambda) \ket{\Phi^{(B)}_{2}(\lambda_{1},\lambda_2)} & = &
\prod_{j=1}^{2} 
\frac{\rho_{1}(\lambda_{j},\lambda)}{\rho_{3}(\lambda_{j},\lambda)} 
\ket{\Phi^{(B)}_{2}(\lambda_{1},\lambda_2)}
\nonumber \\
&& -\sum_{j=1}^{2}  
  \ket{\Omega^{(1)}_{1}(\lambda,\lambda_{j}; \{ \lambda_{k} \})}  
\nonumber \\
&& + H^{(B)}_1(\lambda,\lambda_1,\lambda_2)  
\ket{\Omega^{(3)}_{0}(\lambda,\lambda_j,\lambda_l;\{ \lambda_k \} ) }
\ear
\bear
D(\lambda) \ket{\Phi^{(B)}_{2}(\lambda_{1},\lambda_2)} & = &
[\lambda^{2}]^{L} \prod_{j=1}^{2} 
\frac{\rho_{11}(\lambda,\lambda_j)}{\rho_{9}(\lambda,\lambda_j)} 
\ket{\Phi^{(B)}_{2}(\lambda_{1},\lambda_2)}
\nonumber \\
&& -\sum_{j=1}^{2} 
[\bar{h} \lambda_{j}]^{L} 
\Lambda^{(1)}_{(B)}(\lambda=\lambda_j,\{ \lambda_l \})
 \ket{\Omega^{(2)}_{1}(\lambda,\lambda_{j}; \{ \lambda_{k} \})}  
\nonumber \\
&& + H^{(B)}_{2}(\lambda,\lambda_1,\lambda_2) 
[\bar{h}^{2} \lambda_1 \lambda_2]^{L} 
\ket{\Omega^{(3)}_{0}(\lambda,\lambda_j,\lambda_l;\{ \lambda_k \} ) }
\ear
\bear
\sum_{a=1}^{2} {\hat{A}}_{aa}(\lambda) 
\ket{\Phi^{(B)}_{2}(\lambda_{1},\lambda_2)} & = &
[\bar{h}\lambda]^{L} \prod_{j=1}^{2} 
\frac{\rho_{1}(\lambda,\lambda_j)}{\rho_{3}(\lambda,\lambda_j)} 
\Lambda^{(1)}_{(B)}(\lambda,\{ \lambda_l \})
\ket{\Phi^{(B)}_{2}(\lambda_{1},\lambda_2)}
\nonumber \\
&& -\sum_{j=1}^{2}  
[\bar{h}(\lambda_{j})]^{L} 
\Lambda^{(1)}_{(B)}(\lambda=\lambda_j,\{ \lambda_k \})
\ket{\Omega^{(1)}_{1}(\lambda,\lambda_{j};\{ \lambda_{k}\} )}
\nonumber \\
&& -\sum_{j=1}^{2}  
\ket{\Omega^{(2)}_{1}(\lambda,\lambda_{j};\{ \lambda_{k} \} )}
\nonumber \\
&& + H^{(B)}_3(\lambda,\lambda_1,\lambda_2) 
[ \bar{h} \lambda_2]^{L} 
\Lambda^{(1)}_{(B)}(\lambda=\lambda_2), \{ \lambda_k \})
\ket{\Omega^{(3)}_{0}(\lambda,\lambda_j,\lambda_l;\{ \lambda_k \} ) }
\nonumber \\
&& + H^{(B)}_4(\lambda,\lambda_1,\lambda_2)
[\bar{h} \lambda_1 ]^{L} 
\Lambda^{(1)}_{(B)}(\lambda=\lambda_1), \{ \lambda_k \})
\ket{\Omega^{(3)}_{0}(\lambda,\lambda_j,\lambda_l;\{ \lambda_k \} ) }
\nonumber\\
\ear
where we used the relations $B(\lambda) \ket{0}=\ket{0}$, $A_{aa}(\lambda)=
[\bar{h} \lambda]^{L} \ket{0}$ and $D(\lambda) \ket{0}= [\lambda^{2}]^{L} \ket{0}$ 
which are determined by acting the Lax operator on the 
ferromagnetic pseudovacuum. As before, 
$\Lambda^{(1)}_{(B)}(\lambda,\{ \lambda_l \})$ 
is the eigenvalue of the auxiliary problem (56) whose 
$r$-matrix is now $\hat{r}^{(B)}(\lambda,\mu)$. Furthermore, the expressions
for  the unwanted terms  are
\bear
\ket{\Omega^{(1)}_{1}(\lambda,\lambda_j;\{ \lambda_l \} ) } =
\frac{\rho_{2}(\lambda_{j},\lambda)}{\rho_{3}(\lambda_{j},\lambda)}
\prod^2_{\stackrel{k=1}{k \neq j}}
\frac{\rho_{1}(\lambda_{k},\lambda_{j})}{\rho_{3}(\lambda_{k},\lambda_{j})}
\left[ \vec{B}(\lambda) \otimes \vec{B}(\lambda_{k}) \right]
\prod_{k=1}^{j-1} \hat{r}^{(B)}_{k,k+1}(\lambda_{k},\lambda_{j}) .\vec{{\cal F}} \ket{0}
\nonumber \\
\ear
\bear
\ket{\Omega^{(2)}_{1}(\lambda,\lambda_j;\{ \lambda_l \} ) } =
\frac{\rho_{5}(\lambda,\lambda_{j})}{\rho_{9}(\lambda,\lambda_{j})}
\prod^2_{\stackrel{k=1}{k \neq j}} 
\frac{\rho_{1}(\lambda_{j},\lambda_{k})}{\rho_{3}(\lambda_{j},\lambda_{k})}
[ \vec{\xi}^{(B)}.(\vec{B}^{*}(\lambda) \otimes \hat{I})] \otimes \vec{B}(\lambda_{k})
\prod_{k=1}^{j-1} \hat{r}^{(B)}_{k,k+1}(\lambda_{k},\lambda_{j}) .\vec{{\cal F}} \ket{0}
\nonumber \\
\ear
\bear
\ket{\Omega^{(3)}_{0}(\lambda,\lambda_j,\lambda_l;\{ \lambda_k \} ) } =
F(\lambda) \vec{\xi}^{(B)} .\vec{{\cal F}} \ket{0}
\ear

Finally, the functions $H^{(B)}_{l}(x,y,z)$, $l=1,\ldots,4$ 
are given by
\EQ
H^{(B)}_{1}(x,y,z) = \frac{\rho_{5}(y,z)\rho_{4}(y,x)}
{\rho_{9}(y,z)\rho_{9}(y,x)} + 
\frac{\rho_{1}(y,x)\rho_{2}(z,x)\rho_{12}(y,x)}
{\rho_{3}(y,x)\rho_{3}(z,x)\rho_{9}(y,x)}
\EN
\EQ
H^{(B)}_{2}(x,y,z) = \frac{\rho_{4}(x,y)\rho_{5}(y,z)}
{\rho_{9}(x,y)\rho_{9}(y,z)}  -
\frac{\rho_{5}(x,z)\rho_{2}(x,y)}
{\rho_{9}(x,z)\rho_{9}(x,y)}
\EN
\EQ
H^{(B)}_{3}(x,y,z) = [\frac{\rho_{1}(x,y)\rho_{2}(x,z)\rho_{5}(x,y)}
{\rho_{3}(x,y)\rho_{3}(x,z)\rho_{9}(x,y)} - 
\frac{\rho_{5}(x,y)\rho_{2}(x,y)\rho_{2}(y,z)}
{\rho_{9}(x,y)\rho_{3}(x,y)\rho_{3}(y,z)}][a^{(B)}(y,z)+\frac{b^{(B)}(y,z)}{\bar{h}}]
\EN
\EQ
H^{(B)}_{4}(x,y,z) =
\frac{\rho_{1}(x,z)\rho_{2}(x,y)\rho_{5}(x,z)}
{\rho_{3}(x,z)\rho_{3}(x,y)\rho_{9}(x,z)}
-\frac{\rho_{5}(x,z)\rho_{2}(x,z)\rho_{2}(z,y)}
{\rho_{9}(x,z)\rho_{3}(x,z)\rho_{3}(z,y)}
\EN

	In order to cancel out the unwanted terms it is sufficient to impose the
 following Bethe Ansatz restriction to the rapidities
\EQ
[\lambda_{i}\bar{h}]^{-L} = -\Lambda^{(1)}_{B}(\lambda=\lambda_{i},\{\lambda_{j}\}),~~
i=1,2.
\EN
since this condition eliminates automatically the first two kind of unwanted 
terms. Moreover, this helps us to gather the four unwanted terms 
proportional to $F(\lambda) \vec{\xi}^{(B)}.\vec{{\cal F}}$ which are finally 
vanished due to the identity
\EQ
H_{1}^{(B)}(x,y,z)+H_{2}^{(B)}(x,y,z)=H_{3}^{(B)}(x,y,z)+H_{4}^{(B)}(x,y,z)
\EN

To obtain the two-particle eigenvalue we collect  the wanted 
terms and by using the expression for the Boltzmann weights (see appendix $E$) we 
find
\EQ
\Lambda^{(B)}(\lambda,\{\lambda_{j}\}) = 
\prod_{j=1}^{2} \frac{\bar{h}^{-1}+\bar{h}\lambda_{j}\lambda}{\lambda_{j}-\lambda} + 
\lambda^{2L} \prod_{j=1}^{2} \frac{1+\bar{h}^{2}\lambda_{j}\lambda}{\lambda-\bar{h}^{2}\lambda_{j}} 
+ [\bar{h}\lambda]^{L}\prod_{j=1}^{2}
\frac{\bar{h}^{-1}+\bar{h}\lambda_{j}\lambda}{\lambda-\lambda_{j}}
\Lambda^{(1)}_{B}(\lambda,\{\lambda_{j}\})
\EN

The generalization of these results for multi-particle states goes much 
along the lines discussed in appendix $D$. We start constructing a symmetrized 
$n$-particle vector state which satisfies 
\EQ
\vec{\Phi}^{(B)}(\lambda_{1}, \ldots, \lambda_{j},\lambda_{j+1}, \ldots,\lambda_{n}) = 
\vec{\Phi}^{(B)}(\lambda_{1}, \ldots, \lambda_{j+1},\lambda_{j}, \ldots,\lambda_{n}).\hat{r}^{(B)}(\lambda_{j},\lambda_{j+1})
\EN
and after solving these constraints we have
\bear
\vec {\Phi}^{(B)}_{n}(\lambda_{1},\ldots ,\lambda_{n}) = 
\vec {B}(\lambda_{1}) \otimes \vec {\Phi}^{(B)}_{n-1}(\lambda_{2}, 
\ldots ,\lambda_{n})
 - 
\sum_{j=2}^n 
\frac{\rho_{5}(\lambda_{1},\lambda_{j})}{\rho_{9}(\lambda_{1},\lambda_{j})}
\prod^{n}_{\stackrel{k=2}{k \neq j}} 
\frac{\rho_{1}(\lambda_{k},\lambda_{j})}{\rho_{9}(\lambda_{k},\lambda_{j})} 
\nonumber \\
\times \left [ \vec{\xi} \otimes F(\lambda_1) 
 \vec {\Phi}^{(B)}_{n-2}(\lambda_{2},\ldots,\lambda_{j-1},\lambda_{j+1},\ldots,\lambda_{n}) B(\lambda_j) \right ] \prod_{k=2}^{j-1} 
\hat{r}^{(B)}_{k,k+1}(\lambda_{k},\lambda_{j})
\ear

	From the two-particle analysis it is not difficult to see what should 
be the expressions for the multi-particle eigenvalues and Bethe Ansatz 
equations. For example, the auxiliary eigenvalue expression is the same as 
given in equation (99), replacing $\bar{b}(\lambda,\mu)$ by 
$b^{(B)}(\lambda,\mu)$. To make a comparison
with the previous Bethe Ansatz 
results derived by Bariev \cite{BAR} it is convenient to redefine the spectral 
parameter $\lambda$, the rapidities $\{\lambda_{i}\}$ and the nesting variables 
$\{\mu_{j}\}$ \cite{PMB}. Here we set
\EQ
\lambda=e^{ik},~~ \bar{h} \lambda_{j}=e^{ik_{j}},~~\mu_{j}=e^{i\Lambda_{j}}
\EN

In terms of these new rapidities, our final results for the
eigenvalues are 
\bear
\Lambda(k,\{k_{i}\},\{ \Lambda_j \} ) = \prod_{i=1}^{n} \frac{ \cos(k/2 +k_i/2 -i\beta/2)}
{i \sin( k_i/2 -k/2 +i\beta/2)} 
+ \exp{(i2L k)} \prod_{i=1}^{n} \frac{\cos( k_i/2 +k/2 -i\beta/2)}{i \sin(k/2 -k_i/2 +i\beta/2)}
\nonumber \\
+ \exp{[i(k-i\beta)L]} \left \{ \prod_{i=1}^{n} \frac{i \cos(k/2 +k_i/2 -i\beta/2)}{
\sin(k_i/2 -k/2 +i\beta/2)} \prod_{j=1}^{m} -\frac{ \sin(\Lambda_j/2 -k/2 +i\beta)}{\sin(\Lambda_j/2
-k/2)} + 
 \right. \nonumber \\ \left.
\prod_{i=1}^{n} \frac{i \cos(k/2 +k_i/2 -i\beta/2)}{
\sin(k/2 -k_i/2 +i\beta/2)} \prod_{j=1}^{m} 
-\frac{ \sin(k/2 - \Lambda_j/2  +i\beta)}{\sin(k/2 - \Lambda_j/2)}
\right \}
\ear
while the nested Bethe Ansatz equations for the rapidities $\{k_{i}\}$ and 
$\{ \Lambda_{j}\}$ are
\bear
\exp(ik_i L) & = &- (-1)^{n-m} \prod_{j=1}^{m} \frac{ \sin(k_i/2 - \Lambda_j/2 +i\beta/2)}
{ \sin(k_i/2 - \Lambda_j/2 -i\beta/2)},~~ i=1, \ldots, n 
\nonumber\\
(-1)^{n} \prod_{i=1}^{n} \frac{ \sin(\Lambda_j/2 -k_i/2 -i\beta/2)}
{\sin(\Lambda_j/2 -k_i/2 +i\beta/2)} & = &  - \prod_{k=1}^{m} \frac{\sin(\Lambda_j/2 -\Lambda_k/2- i\beta)}
{\sin(\Lambda_j/2 -\Lambda_k/2+ i\beta)},~~ j=1, \ldots, m
\ear

Finally, to obtain the eigenspectrum of the Hamiltonian (162) we 
expand the transfer matrix eigenvalues in power of the spectral parameter. Up 
to second order we have
\bear
ln[\Lambda(\lambda,\{\lambda_{j}\},\{\mu_{j}\}] = 
\sum_{i}^{n} \frac{1}{h\lambda_{i}} + \bar{h} \lambda\sum_{i}^{n} (\bar{h}\lambda_{i}+\frac{1}{\bar{h}\lambda_{i}}) 
\nonumber \\
+\frac{\bar{h}^{2}\lambda^{2}}{2!}
\sum_{i}^{n}[(\frac{1}{\bar{h}\lambda_{i}})^{2}-(\bar{h}\lambda_{i})^{2}]
+ {\cal{O}}(\lambda^3)
\ear

Considering the ${\cal{O}}(\lambda)$ term of the above
equation and remembering to perform the rescaling $\lambda \rightarrow 
\frac{\lambda e^{-\beta/2}}{\cosh(\beta/2)}$ we conclude that 
the eigenenergies of the Hamiltonian (162) are
\EQ
E_{n} = 2(1+V) \sum_{i=1}^{n} cos(k_{i})
\EN

We conclude remarking that this model can also be solved with  twisted boundary conditions
following precisely the same steps presented in  section 5.1 .

\section{Conclusions }

The main purpose of this paper was to apply the quantum inverse scattering
program for the one-dimensional Hubbard model. We succeeded in developing
a framework which allowed us to present  an algebraic
formulation for the Bethe states of the transfer matrix of the
classical ``covering'' Hubbard model 
proposed earlier by  Shastry  \cite{SA1,SA2}. A hidden 6-vertex symmetry
has been revealed, and it played a fundamental role in the solution
of the transfer matrix eigenvalue problem.
We have found the eigenvalues of the transfer
matrix and showed that its eigenstates are highest weights 
states of both the rotational and the $\eta$-paring $SU(2)$ symmetries. This
later result corroborates the original proof given by Essler, Korepin
and Schoutens \cite{EKS} in terms of coordinate wave functions. We
have also discussed the algebraic solution of models with twisted boundary
conditions and applied the results to the Hubbard model perturbed by
charge and spin currents.

The framework developed in this paper, the $ABCDF$ formalism, 
is indeed suitable to solve
a broad class of integrable systems. As an example, we solved, in section 6,
the classical analog of the Bariev model by this method. There are also
other models that fit in the $ABCDF$ framework, 
such as the trigonometric 
vertex models based on the
$B_n$, $C_n$ , $D_n$ ,
$A_{2n}^{2}$ and $A_{2n-1}^{2}$ algebras as well as 
certain related supersymmetric
models \cite{MP1}. Interesting enough, the former models 
almost exhaust the  Jimbo's and Bazhanov's list of $U_q(G)$ R-matrices \cite{JIBA}, and
only the $D_{n+1}^{2}$ model appears to be not solvable 
within our framework. Anyhow, these examples  
suggest us that
the $ABCDF$ formalism is capable of solving
integrable models having one less trivial
conserved quantum number when compared to 
the $A_n$ multi-state $6$-vertex models with an equivalent Hilbert space.

Finally, the possibility of bringing a variety of 
models under one unifying approach not
only highlight the qualities of the quantum inverse scattering
program but also allows us to better 
understand the relevant properties entering  their 
Bethe Ansatz solution. This  also motives
us to look for further extensions which could shape our knowledge
towards  a possible classification of integrable models 
from an algebraic point of view. An interesting example seems to be
the $D_{n+1}^{2}$ vertex model, which we plan to investigate in a future
work.

\section*{Acknowledgements}
This  
work was support by  FOM (Fundamental Onderzoek der Materie) 
and Fapesp ( Funda\c c\~ao
de Amparo \`a Pesquisa do Estado de S. Paulo). The work of M.J. Martins
was partially done in the frame of Associate Membership programme
of the ICTP, Trieste, Italy.
P.B. Ramos thanks
B. Nienhuis for 
the participation in the Altenberg
summer school and the hospitality of Instituut voor Theoretische Fysica, Amsterdam.

\centerline{\bf Appendix A : Boltzmann weights of the Shastry model }
\setcounter{equation}{0}
\renewcommand{\theequation}{A.\arabic{equation}}

We start this appendix by presenting the ten non-null Boltzmann 
weights of Shastry's $R$-matrix (15). They are given by 
\EQ
\alpha_{1}(\lambda,\mu)= 
\left\{e^{[h(\mu)-h(\lambda)]}a(\lambda)a(\mu)+e^{-[h(\mu)-h(\lambda)]}b(\lambda)b(\mu) \right \} \alpha_{5}(\lambda,\mu)
\EN
\EQ
\alpha_{2}(\lambda,\mu)= 
\left \{ e^{-[h(\mu)-h(\lambda)]}a(\lambda)a(\mu)+e^{[h(\mu)-h(\lambda)]}b(\lambda)b(\mu) \right \} \alpha_{5}(\lambda,\mu)
\EN
\EQ
\alpha_{3}(\lambda,\mu)=
\frac{ e^{[h(\mu)+h(\lambda)]}a(\lambda)b(\mu)
+e^{-[h(\mu)+h(\lambda)]}b(\lambda)a(\mu) }
{a(\lambda)b(\lambda)+a(\mu)b(\mu)} \left \{
\frac{\cosh[h(\mu)-h(\lambda)]}{\cosh[h(\mu)+h(\lambda)]} 
\right \} \alpha_{5}(\lambda,\mu)
\EN
\EQ
\alpha_{4}(\lambda,\mu)=
\frac{e^{-[h(\mu)+h(\lambda)]}a(\lambda)b(\mu)+e^{[h(\mu)+h(\lambda)]}b(\lambda)a(\mu)}
{a(\lambda)b(\lambda)+a(\mu)b(\mu)}
\left \{ \frac{\cosh(h(\mu)-h(\lambda))}{\cosh(h(\mu)+h(\lambda))} 
\right \} \alpha_{5}(\lambda,\mu)
\EN
\EQ
\alpha_{6}(\lambda,\mu)= \left \{
\frac{ e^{[h(\mu)+h(\lambda)]}a(\lambda)b(\mu)-e^{-[h(\mu)+h(\lambda)]}b(\lambda)a(\mu) }
{a(\lambda)b(\lambda)+a(\mu)b(\mu)} \right \} [b^{2}(\mu)-b^{2}(\lambda)]
\frac{\cosh[h(\mu)-h(\lambda)]}{\cosh[h(\mu)+h(\lambda)]} 
\alpha_{5}(\lambda,\mu)
\EN
\EQ
\alpha_{7}(\lambda,\mu)= \left \{
\frac{-e^{-[h(\mu)+h(\lambda)]}a(\lambda)b(\mu)+e^{[h(\mu)+h(\lambda)]}b(\lambda)a(\mu)}
{a(\lambda)b(\lambda)+a(\mu)b(\mu)}
\right \} [b^{2}(\mu)-b^{2}(\lambda)]
\frac{\cosh[h(\mu)-h(\lambda)]}{\cosh[h(\mu)+h(\lambda)]} 
\alpha_{5}(\lambda,\mu)
\EN
\EQ
\alpha_{8}(\lambda,\mu)= 
\left \{ e^{[h(\mu)-h(\lambda)]}a(\lambda)b(\mu)-
e^{-[h(\mu)-h(\lambda)]}b(\lambda)a(\mu) \right \} \alpha_{5}(\lambda,\mu)
\EN
\EQ
\alpha_{9}(\lambda,\mu)= 
\left \{ -e^{-[h(\mu)-h(\lambda)]}a(\lambda)b(\mu)+
e^{[h(\mu)-h(\lambda)]}b(\lambda)a(\mu) \right \} 
\alpha_{5}(\lambda,\mu)
\EN
\EQ
\alpha_{10}(\lambda,\mu)=
\frac{b^{2}(\mu)-b^{2}(\lambda)}{a(\lambda)b(\lambda)+a(\mu)b(\mu)}
\left \{ \frac{\cosh[h(\mu)-h(\lambda)]}{\cosh[h(\mu)+h(\lambda)]} 
\right \} \alpha_{5}(\lambda,\mu)
\EN
where the weight $\alpha_{5}(\lambda,\mu)$ has been used as a 
normalization. We recall that functions $a(\lambda)$ and $b(\lambda)$ 
satisfy the free-fermion condition $a^{2}(\lambda)+b^{2}(\lambda)=1$, and 
in this paper we shall use the
parametrization is $a(\lambda)=\cos(\lambda)$ 
and $b(\lambda)=\sin(\lambda)$. There are certain useful identities satisfied 
by these weights we have used to simplify commutation rules and the 
multi-particle problem. These relations are given by \cite{WA}
\EQ
\alpha_3(\lambda,\mu)= \alpha_1(\lambda,\mu) + \alpha_6(\lambda,\mu)~~
\alpha_4(\lambda,\mu) + \alpha_7(\lambda,\mu)= \alpha_2(\lambda,\mu)
\EN
\EQ
\alpha_2(\lambda,\mu) \alpha_1(\lambda,\mu) -\alpha_9(\lambda,\mu) 
\alpha_8(\lambda,\mu)= \alpha_4(\lambda,\mu) \alpha_3(\lambda,\mu) -
\alpha_{10}^2(\lambda,\mu)= \alpha_5^2( \lambda, \mu)
\EN
\EQ
\alpha_2(\lambda,\mu) \alpha_3(\lambda,\mu) + \alpha_4(\lambda,\mu) 
\alpha_1(\lambda, \mu) = 2 \alpha_5^2(\lambda,\mu)
\EN

\centerline{\bf Appendix B : Extra commutation rules }
\setcounter{equation}{0}
\renewcommand{\theequation}{B.\arabic{equation}}

This appendix is devoted to complement the commutation relations 
presented in 
the main text. For instance, there are some additional commutation 
rules which are important for the complete solution of the two-particle 
state problem. These are relations between the fields $\vec{B}(\lambda)$, 
$\vec{B}^{*}(\lambda)$, $\vec{C}(\lambda)$ and $\vec{C}^{*}(\lambda)$ given 
by
\EQ
C_{a}(\lambda)B_{b}(\mu) =
-\frac{\alpha_{8}(\lambda,\mu)}{\alpha_{9}(\lambda,\mu)} B_{b}(\mu)C_{a}(\lambda) 
+i\frac{\alpha_{5}(\lambda,\mu)}{\alpha_{9}(\lambda,\mu)}
[B(\mu)A_{ab}(\lambda)-B(\lambda)A_{ab}(\mu)]
\EN
\EQ
B_{a}^{*}(\lambda)B_{b}(\mu) =
-\frac{\alpha_{8}(\lambda,\mu)}{\alpha_{9}(\lambda,\mu)} B_{b}(\mu)
B^{*}_{a}(\lambda) 
+i\frac{\alpha_{5}(\lambda,\mu)}{\alpha_{9}(\lambda,\mu)}
[F(\mu)A_{ab}(\lambda)-F(\lambda)A_{ab}(\mu)]
\EN
\bear
C^{*}_{a}(\lambda)B_{b}(\mu) &=&
\frac{\alpha_{3}(\lambda,\mu)}{\alpha_{7}(\lambda,\mu)} B_{a}(\mu)C^{*}_{b}(\lambda)
-\frac{\alpha_{4}(\lambda,\mu)}{\alpha_{7}(\lambda,\mu)}
B_{a}(\lambda)C^{*}_{b}(\mu)
-\frac{\alpha_{6}(\lambda,\mu)}{\alpha_{7}(\lambda,\mu)}
B_{b}(\mu)C^{*}_{a}(\lambda)
\nonumber\\
&& +i\frac{\alpha_{10}(\lambda,\mu)}{\alpha_{7}(\lambda,\mu)} 
\xi_{lm}A_{la}(\lambda)A_{mb}(\mu) 
+i\frac{\alpha_{10}(\lambda,\mu)}{\alpha_{7}(\lambda,\mu)} 
\xi_{ab}[F(\mu)C(\lambda)-B(\mu)D(\lambda)]
\nonumber\\
\ear

In particular, the commutation rule $(B.3)$ is  of considerable importance
in the proof that 
the eigenvectors constructed in 
section 4 are highest weights states of the $SU(2)$ $\eta$-
pairing symmetry (see section 5.2). In order to understand the role of the 
creation field $\vec{B}^{*}(\lambda)$ it is indispensable  to 
derive  its 
commutations relations with the other relevant fields. Between 
$\vec{B}^{*}(\lambda)$ and the diagonal operators we have
\bear
\hat{A}(\lambda) \otimes \vec{B}^{*}(\mu) &= &
-i\frac{\alpha_{1}(\mu,\lambda)}{\alpha_{8}(\mu,\lambda)}
 \hat{r}(\mu,\lambda).
[\vec{B}^{*}(\mu) \otimes 
\hat{A}(\lambda) ]
+i \frac{\alpha_{5}(\mu,\lambda)}{\alpha_{8}(\mu,\lambda)} 
\vec{B^{*}}(\lambda) \otimes \hat{A}(\mu)   
\nonumber \\
&& -i\frac{\alpha_{10}(\mu,\lambda)}{\alpha_{7}(\mu,\lambda)} 
\vec{\xi}^{t} \otimes[\vec{B}(\lambda)D(\mu) 
+i\frac{\alpha_{5}(\mu,\lambda)}{\alpha_{8}(\mu,\lambda)}F(\lambda)\vec{C}^{*}(\mu)  
-i \frac{\alpha_{2}(\mu,\lambda)}{\alpha_{8}(\mu,\lambda)}F(\mu)\vec{C}^{*}(\lambda)]  
\nonumber\\
\ear
\EQ
D(\lambda)\vec{B}^{*}(\mu) = 
i\frac{\alpha_{2}(\lambda,\mu)}{\alpha_{8}(\lambda,\mu)} \vec{B}^{*}(\mu)D(\lambda) -i 
\frac{\alpha_{5}(\lambda,\mu)}{\alpha_{8}(\lambda,\mu)} \vec{B}^{*}(\lambda)D(\mu)
\EN
\bear
B(\lambda)\vec{B}^{*}(\mu) & = &
-i\frac{\alpha_{9}(\mu,\lambda)}{\alpha_{7}(\mu,\lambda)} \vec{B}^{*}(\mu)B(\lambda) 
+ \frac{\alpha_{5}(\mu,\lambda)}{\alpha_{7}(\mu,\lambda)} F(u)\vec{C}(\lambda)
\nonumber \\ 
&& - \frac{\alpha_{4}(\mu,\lambda)}{\alpha_{7}(\mu,\lambda)} F(\lambda)\vec{C}(\mu)
- i \frac{\alpha_{10}(\mu,\lambda)}{\alpha_{7}(\mu,\lambda)} 
[ \vec{B}(\lambda) \otimes \hat{A}(\mu)] . \vec{\xi}^{t}
\ear
while with itself and with the scalar operator $F(\lambda)$ we have
\EQ
\vec{B}^{*}(\lambda) \otimes \vec{B}^{*}(\mu) = \frac{\alpha_{1}(\mu,\lambda)}{\alpha_{2}(\mu,\lambda)}
\hat{r}(\mu,\lambda).
[ \vec{B}^{*}(\mu) \otimes \vec{B}^{*}(\lambda) ] 
+i\frac{\alpha_{10}(\mu,\lambda)}{\alpha_{7}(\mu,\lambda)}  
\{ F(\lambda)D(\mu) - F(\mu)D(\lambda) \} \vec{\xi}^{t}
\EN

\EQ
F(\lambda) \vec{B}^{*}(\mu) = \frac{\alpha_{5}(\mu,\lambda)}{\alpha_{2}(\mu,\lambda)} F(\mu) \vec{B}^{*}(\lambda) -i\frac{\alpha_{9}(\mu,\lambda)}{\alpha_{2}(\mu,\lambda)} \vec{B}^{*}(\mu) F(\lambda)
\EN
\EQ
\vec{B}^{*}(\lambda) F(\mu) = \frac{\alpha_{5}(\mu,\lambda)}{\alpha_{2}(\mu,\lambda)} \vec{B}^{*}(\mu) F(\lambda)  -i \frac{\alpha_{8}(\mu,\lambda)}{\alpha_{2}(\mu,\lambda)} F(\mu) \vec{B}^{*}(\lambda)
\EN

	Lastly, the commutation rules with the annihilation fields $\vec{C}(\lambda)$ and $\vec{C}^{*}(\lambda)$ are 
\EQ
C_{a}^{*}(\lambda)B_{b}^{*}(\mu) =
-\frac{\alpha_{9}(\mu,\lambda)}{\alpha_{8}(\mu,\lambda)} B_{b}^{*}(\mu)C_{a}^{*}(\lambda) 
-i\frac{\alpha_{5}(\mu,\lambda)}{\alpha_{8}(\mu,\lambda)}
[D(\mu)A_{ba}(\lambda)-D(\lambda)A_{ba}(\mu)]
\EN
\bear
C_{a}(\lambda)B^{*}_{b}(\mu) & = &
\frac{\alpha_{3}(\mu,\lambda)}{\alpha_{7}(\mu,\lambda)} B^{*}_{a}(\mu)C_{b}(\lambda)
-\frac{\alpha_{4}(\mu,\lambda)}{\alpha_{7}(\mu,\lambda)}
B^{*}_{a}(\lambda)C_{b}(\mu)
-\frac{\alpha_{6}(\mu,\lambda)}{\alpha_{7}(\mu,\lambda)}
B^{*}_{b}(\mu)C_{a}(\lambda)
\nonumber\\
&&-i\frac{\alpha_{10}(\mu,\lambda)}{\alpha_{7}(\mu,\lambda)} 
\xi_{lm}A_{al}(\lambda)A_{bm}(\mu) 
-i\frac{\alpha_{10}(\mu,\lambda)}{\alpha_{7}(\mu,\lambda)} 
\xi_{ab}[F(\mu)C(\lambda)-D(\mu)B(\lambda)]
\nonumber\\
\ear

The best way of seeing that these later commutations relations are 
connected to those for the field $\vec{B}(\lambda)$ is to read the equations
in terms of their components. For instance, we note that commutation
rule (B.2) is self-dual under the ``dual'' transformation described in section
3. Several other relations have similar property as well.

We close this appendix by presenting the expressions for the 
fundamental commutation rules when we solve the standard Yang-Baxter algebra 
(3). These relations lack the presence of the imaginary factors ``i'' 
and certain 
extra signs when compared to their graded counterparts. Below we list the 
most important relations for the creation fields $\vec{B}(\lambda)$ and 
$F(\lambda)$
\bear
\hat{A}(\lambda) \otimes \vec{B}(\mu) & = &
-\frac{\alpha_{1}(\lambda,\mu)}{\alpha_{9}(\lambda,\mu)}[\vec{B}(\mu) \otimes 
\hat{A}(\lambda) ]. \hat{r}_{TW}(\lambda,\mu)
+\frac{\alpha_{5}(\lambda,\mu)}{\alpha_{9}(\lambda,\mu)} \vec{B}(\lambda) \otimes \hat{A}(\mu)   
\nonumber \\
&& -\frac{\alpha_{10}(\lambda,\mu)}{\alpha_{7}(\lambda,\mu)} [\vec{B^{*}}(\lambda)B(\mu) 
-\frac{\alpha_{5}(\lambda,\mu)}{\alpha_{9}(\lambda,\mu)}F(\lambda)\vec{C}(\mu)  
+\frac{\alpha_{2}(\lambda,\mu)}{\alpha_{9}(\lambda,\mu)}F(\mu)\vec{C}(\lambda)] 
\otimes \vec{\xi}_{TW}
\nonumber\\
\ear
\EQ
B(\lambda)\vec{B}(\mu) = 
-\frac{\alpha_{2}(\mu,\lambda)}{\alpha_{9}(\mu,\lambda)} \vec{B}(\mu)B(\lambda) +\frac{\alpha_{5}(\mu,\lambda)}{\alpha_{9}(\mu,\lambda)} \vec{B}(\lambda)B(\mu),
\EN
\bear
D(\lambda)\vec{B}(\mu) & =& 
-\frac{\alpha_{8}(\lambda,\mu)}{\alpha_{7}(\lambda,\mu)} \vec{B}(\mu)D(\lambda) 
-\frac{\alpha_{5}(\lambda,\mu)}{\alpha_{7}(\lambda,\mu)} F(u)\vec{C^{*}}(\lambda)
\nonumber \\ 
&& +\frac{\alpha_{4}(\lambda,\mu)}{\alpha_{7}(\lambda,\mu)} F(\lambda)\vec{C^{*}}(\mu)
+\frac{\alpha_{10}(\lambda,\mu)}{\alpha_{7}(\lambda,\mu)} 
\vec{\xi}_{TW}. [ \vec{B^{*}}(\lambda) \otimes \hat{A}(\mu)] 
\ear
\bear
\hat{A}_{ab}(\lambda)F(\mu) & = &
[1 + \frac{\alpha^{2}_{5}(\lambda,\mu)}{\alpha_{9}(\lambda,\mu)\alpha_{8}(\lambda,\mu)}] F(\mu)\hat{A}_{ab}(\lambda)
- \frac{\alpha^{2}_{5}(\lambda,\mu)}{\alpha_{9}(\lambda,\mu)\alpha_{8}(\lambda,\mu)} F(\lambda) \hat{A}_{ab}(\mu)
\nonumber \\
&&+ \frac{\alpha_{5}(\lambda,\mu)}{\alpha_{9}(\lambda,\mu)} 
[\vec{B}(\lambda) \otimes \vec{B^{*}}(\mu)]_{ba} 
+\frac{\alpha_{5}(\lambda,\mu)}{\alpha_{8}(\lambda,\mu)} 
[\vec{B^{*}}(\lambda) \otimes \vec{B}(\mu)]_{ab} 
\ear
\EQ
B(\lambda)F(\mu) = 
-\frac{\alpha_{2}(\mu,\lambda)}{\alpha_{7}(\mu,\lambda)} F(\mu)B(\lambda) + 
\frac{\alpha_{4}(\mu,\lambda)}{\alpha_{7}(\mu,\lambda)} F(\lambda)B(\mu)
+\frac{\alpha_{10}(\mu,\lambda)}{\alpha_{7}(\mu,\lambda)} 
\{ \vec{B}(\lambda) \otimes \vec{B}(\mu) \}
.\vec{\xi}_{TW}^{t} 
\EN
\EQ
D(\lambda)F(\mu) = 
-\frac{\alpha_{2}(\lambda,\mu)}{\alpha_{7}(\lambda,\mu)} F(\mu)D(\lambda) +
\frac{\alpha_{4}(\lambda,\mu)}{\alpha_{7}(\lambda,\mu)} F(\lambda)D(\mu)
+\frac{\alpha_{10}(\lambda,\mu)}{\alpha_{7}(\lambda,\mu)} \vec{\xi}_{TW} . 
\{ \vec{B^{*}}(\lambda) \otimes \vec{B^{*}}(\mu) \}
\EN
where $\vec{\xi}_{TW}$ and $\hat{r}(\lambda,\mu)_{TW}$ are given by
\EQ
{\vec \xi}_{TW} = 
\matrix{(
0  &1  &1  &0 )  \cr}~~~
\hat{r}_{TW}(\lambda,\mu) = 
\pmatrix{
1  &0  &0  &0  \cr
0  &\bar{a}(\lambda,\mu)  &-\bar{b}(\lambda,\mu)  &0  \cr
0  &-\bar{b}(\lambda,\mu)  &\bar{a}(\lambda,\mu)  &0  \cr
0  &0  &0  &1  \cr}
\EN

	Furthermore, the relations closing the commutation rules between the 
creation operators $\vec{B}(\lambda)$ and $F(\lambda)$ are
\EQ
\vec{B}(\lambda) \otimes \vec{B}(\mu) = \frac{\alpha_{1}(\lambda,\mu)}{\alpha_{2}(\lambda,\mu)}
[ \vec{B}(\mu) \otimes \vec{B}(\lambda) ] .\hat{r}_{TW}(\lambda,\mu)
\nonumber \\
-\frac{\alpha_{10}(\lambda,\mu)}{\alpha_{7}(\lambda,\mu)}  
\{ F(\lambda)B(\mu) - F(\mu)B(\lambda) \} \vec{\xi}_{TW}
\EN
\EQ
\left[ F(\lambda), F(\mu) \right] = 0
\EN
\EQ
F(\lambda) \vec{B}(\mu) = \frac{\alpha_{5}(\lambda,\mu)}{\alpha_{2}(\lambda,\mu)} F(\mu) \vec{B}(\lambda) +\frac{\alpha_{8}(\lambda,\mu)}{\alpha_{2}(\lambda,\mu)} \vec{B}(\mu) F(\lambda)
\EN
\EQ
\vec{B}(\lambda) F(\mu) = \frac{\alpha_{5}(\lambda,\mu)}{\alpha_{2}(\lambda,\mu)} \vec{B}(\mu) F(\lambda)  -\frac{\alpha_{9}(\lambda,\mu)}{\alpha_{2}(\lambda,\mu)} F(\mu) \vec{B}(\lambda)
\EN

\centerline{\bf Appendix C : The two-particle state}
\setcounter{equation}{0}
\renewcommand{\theequation}{C.\arabic{equation}}

In this appendix we provide details about the technical points entering 
the analysis of the two-particle eigenvalue problem. We begin the discussion
by first considering the 
wanted terms. We recall that
the amplitudes proportional to the first part of the two-particle
eigenstate are easily estimated as a product of the first 
right-hand side terms of the commutation rules (34-36). 
For the second part, however, there are more 
contributions since the action of diagonal operators on the first
part $\vec{B}(\lambda_1) \otimes \vec{B}(\lambda_2) . \vec{ \cal{F}} \ket{0}$
produce at least one extra term proportional to the second part $F(\lambda_1)
\vec{\xi} . \vec{ \cal{F}} \ket{0}$ as well. It turns out, however, that
these contributions miraculously factorize in the same product forms we 
have obtained 
for the first part of the eigenstate. This happens thanks
to remarkable identities
between the Boltzmann weights we begin listing below. For the field
$B(\lambda)$ there are two contributions and they factorize as
\EQ
\frac{\alpha_{2}(y,x)}{\alpha_{7}(y,x)} -
\frac{\alpha_{2}(y,x)}{\alpha_{9}(y,x)}
\frac{\alpha_{5}(z,x)}{\alpha_{9}(z,x)}
\frac{\alpha_{10}(y,x)}{\alpha_{7}(y,x)}
\frac{\alpha_{7}(y,z)}{\alpha_{10}(y,z)}=-\frac{\alpha_{2}(y,x)}{\alpha_{9}(y,x)}
\frac{\alpha_{2}(z,x)}{\alpha_{9}(z,x)}
\EN

Analogously, for the field $D(\lambda)$ we have
\EQ
\frac{\alpha_{2}(x,y)}{\alpha_{7}(x,y)} -
\frac{\alpha_{5}(x,y)}{\alpha_{7}(x,y)}
\frac{\alpha_{10}(x,z)}{\alpha_{7}(x,z)}
\frac{\alpha_{7}(y,z)}{\alpha_{10}(y,z)}=-\frac{\alpha_{8}(x,y)}{\alpha_{7}(x,y)}
\frac{\alpha_{8}(x,z)}{\alpha_{7}(x,z)}
\EN

For the diagonal field $ \displaystyle \sum_{a=1}^{2} 
\hat{A}_{aa}(\lambda)$ 
we have three contributions, where two of them are generated by the first
part of the eigenstate. The identity 
that brings these terms together and  also gives rise
to the auxiliary eigenvalue function is
\bear
2\left[1+\frac{\alpha_{5}^{2}(x,y)}{\alpha_{8}(x,y) \alpha_{9}(x,y)} \right]
& - &\frac{\alpha_{1}(x,y)}{\alpha_{9}(x,y)}
\frac{\alpha_{10}(x,z)}{\alpha_{7}(x,z)}
\frac{\alpha_{5}(x,y)}{\alpha_{8}(x,y)}
\frac{\alpha_{7}(y,z)}{\alpha_{10}(y,z)}
\left[1+\bar{a}(x,y) \right] 
\nonumber\\
&& +\frac{\alpha_{10}(x,y)}{\alpha_{7}(x,y)}
\frac{\alpha_{2}(x,y)}{\alpha_{9}(x,y)}
\frac{\alpha_{5}(x,z)}{\alpha_{9}(x,z)}
\frac{\alpha_{7}(y,z)}{\alpha_{10}(y,z)}
\nonumber \\
& = & -\frac{\alpha_{1}(x,y)}{\alpha_{9}(x,y)}
\frac{\alpha_{1}(x,z)}{\alpha_{9}(x,z)}
\left [ \bar{b}(x,y)+\bar{b}(x,z)-\bar{a}(x,y)\bar{a}(x,z) \right ] 
\nonumber\\
\ear

Next we turn to the analysis of the 
unwanted terms proportional to  $\vec{B}(\lambda) \otimes \vec{B}(\lambda_j)$ 
and $[\vec{\xi} \otimes 
(\vec{B^{*}}(\lambda) \otimes \hat{I})]\otimes \vec{B}(\lambda_j)$. The terms
with $\lambda_j=\lambda_2$ are straightforwardly read from the
commutation rules (34-36) because  only single contributions occur for
each diagonal field. However, for
$\lambda_j =\lambda_1$, the situation is more complicated 
because it involves many different contributions whose
origin is due to the fact that the 
rapidity $\lambda_1$ is wrongly ordered when compared with 
$\lambda_2$. Nevertheless, one expects that 
there should be a better way of recasting these terms since
the Bethe Ansatz equations are usually independent of indices 
relabeling. Indeed, it turns out that these many
contributions can be compactly written by introducing the ``ordering''
factor $O_j^{(1)}(\lambda,\lambda_j; \{ \lambda_k \})$. As before,
in order to factorize these contributions to a single term, we had
to use extra identities 
between the
Boltzmann weights. 
For example, for the field 
$B(\lambda)$ they are
\EQ
\frac{\alpha_{1}(y,x)}{\alpha_{9}(y,x)}
\frac{\alpha_{5}(z,x)}{\alpha_{9}(z,x)} \bar{a}(y,x)-
\frac{\alpha_{10}(y,x)}{\alpha_{7}(y,x)}
\frac{\alpha_{10}(y,z)}{\alpha_{7}(y,z)}  - 
\frac{\alpha_{5}(y,x)}{\alpha_{9}(y,x)}
\frac{\alpha_{5}(z,y)}{\alpha_{9}(z,y)} =
\frac{\alpha_{5}(z,x)}{\alpha_{9}(z,x)}
\frac{\alpha_{1}(y,z)}{\alpha_{9}(y,z)}
\bar{a}(y,z)
\nonumber\\
\EN
and
\EQ
\frac{\alpha_{10}(y,x)}{\alpha_{7}(y,x)}
\frac{\alpha_{10}(y,z)}{\alpha_{7}(y,z)}+
\frac{\alpha_{1}(y,x)}{\alpha_{9}(y,x)}
\frac{\alpha_{5}(z,x)}{\alpha_{9}(z,x)}\bar{b}(y,x)=
\frac{\alpha_{5}(z,x)}{\alpha_{9}(z,x)}
\frac{\alpha_{1}(y,z)}{\alpha_{9}(y,z)}
\bar{b}(y,z)
\EN
where the left-hand side of the above equations represent the 
contributions coming from the ``brute force'' calculations while the
right-hand side exhibits the ``ordering'' factor explicitly.

Similar simplifications can be carried out for the fields $\displaystyle
\sum_{a=1}^{2} \hat{A}_{aa}(\lambda)$ and $D(\lambda)$, but
we skip further details since
there is a much simpler way to understand the origin of such
``ordering'' factor. As it has been explained in section 4, this factor can be
easily derived with the help of the  
exchange property (69). Anyhow, the  coincidence between
the ``brute-force'' computations and the  symmetrization results
gives  us confidence to go ahead using the
symmetrization procedure 
for multi-particle states.

Finally, we  show how the third type of unwanted terms 
generated by the
diagonal field 
$ \displaystyle \sum_{a=1}^2 \hat{A}_{aa}(\lambda)$ 
can be further
simplified.  First it is convenient to rewrite the term proportional
to $[\omega_1(\lambda_1) \omega_2(\lambda_2)]^{L} F(\lambda) \vec{\xi}.
\vec{{\cal{F}}}$ in a way that the auxiliary eigenvalue function 
appears explicitly.
For this purpose we use the  second identity (66),
and rewrite the contribution to the above mentioned unwanted term as
\bear
[w_{1}(\lambda_{1})w_{2}(\lambda_{2})]^{L}
\Lambda^{(1)}(\lambda=\lambda_{2}, \{ \lambda_{l} \}) 
H_{3}(\lambda,\lambda_{1},\lambda_{2}) 
[\bar{b}(\lambda_{1},\lambda_{2})-\bar{a}(\lambda_{1},\lambda_{2})]
F(\lambda) \vec{\xi}. \vec{{\cal F} }
\ear

Next we take advantage of the  symmetrization property of the two-particle
eigenstate and evaluate the contribution proportional
to $[\omega_1(\lambda_2) \omega_2(\lambda_1)]^{L} F(\lambda) \vec{\xi}.
\vec{{\cal{F}}}$ as follows. The idea is to begin with the right-hand side of
equation (69), which remarkably gives 
us precisely the extra $r$-matrix necessary to produce the
auxiliary eigenvalue at $\lambda=\lambda_1$. Obviously, the amplitude 
contributing to this term is proportional to function
obviously 
$H_{3}(\lambda,\lambda_{2},\lambda_{1})$  multiplied by the
extra factor $\frac{\alpha_1(\lambda_1,\lambda_2)}
{\alpha_2(\lambda_1,\lambda_2)}$ coming from the exchange relation.
Putting these information together we are able to rewrite
the second contribution as
\bear
-[w_{1}(\lambda_{2})w_{2}(\lambda_{1})]^{L}
\Lambda^{(1)}(\lambda=\lambda_{1}, \{ \lambda_{l} \}) 
H_{3}(\lambda,\lambda_{2},\lambda_{1}) 
\frac{\alpha_{1}(\lambda_{1},\lambda_{2})}{\alpha_{2}(\lambda_{1},\lambda_{2})}
F(\lambda) \vec{\xi}. \vec{{\cal F} }
\ear

These manipulations make the cancellation of the third type of
unwanted terms more transparent, since it allows us to use the Bethe Ansatz
equations in a more direct way. Indeed, using the Bethe Ansatz equations (65)
in the terms (C.6) and (C.7) and adding them to those coming
from the fields $B(\lambda)$ and $D(\lambda)$, we find that
the unwanted terms proportional to $ F(\lambda) \vec{\xi} .\vec{{\cal{F}}}$ 
are cancelled out thanks to the following identity
\EQ
H_{1}(x,y,z) + H_{2}(x,y,z) =
H_{3}(x,y,z) 
[\bar{b}(y,z)-\bar{a}(y,z)] -
H_{3}(x,z,y)
\frac{\alpha_{1}(y,z)}{\alpha_{2}(y,z)}
\EN

This gives us another 
opportunity to verify the symmetrization scheme. Comparing
(C.8) and (67) we conclude that the following identity
\EQ
H_{3}(x,z,y)
\frac{\alpha_{1}(y,z)}{\alpha_{2}(y,z)}=
H_{4}(x,y,z)
[\bar{a}(z,y)-\bar{b}(z,y)]
\EN
is indeed  satisfied.

We remark that the 
above technicalities are of enormous help  when we consider 
generalization to multi-particle states.  In next appendix we shall discuss
this fact for the three-particle state.

\centerline{\bf Appendix D : The three-particle state }
\setcounter{equation}{0}
\renewcommand{\theequation}{D.\arabic{equation}}

We shall start this appendix showing how the 
permutation symmetry 
$\lambda_{1} \leftrightarrow \lambda_{2}$ is implemented for the
three-particle state. As before, our strategy 
consists in reordering the rapidities $\lambda_1$ and $\lambda_2$ with
the help of the commutation rule (25). This allows us to write the
Ansatz (76) as
\bear
\vec{\Phi}_{3}(\lambda_{1},\lambda_{2},\lambda_{3}) &=&
\frac{\alpha_1(\lambda_1,\lambda_2)}{\alpha_2(\lambda_1,\lambda_2)}
\vec{B}(\lambda_{2}) \otimes \vec{B}(\lambda_{1}) 
\otimes \vec{B}(\lambda_{3})  
.\hat{r}_{12}(\lambda_1,\lambda_2)  
-i\frac{\alpha_{10}(\lambda_1,\lambda_2)}{\alpha_7(\lambda_1,\lambda_2)}
[F(\lambda_{1})  B(\lambda_{2}) \vec{\xi} \otimes \vec{B}(\lambda_3)]
\nonumber\\
&& i\frac{\alpha_{10}(\lambda_1,\lambda_2)}{\alpha_7(\lambda_1,\lambda_2)}
[F(\lambda_{2})  B(\lambda_{1}) \vec{\xi} \otimes \vec{B}(\lambda_3)]
+i\frac{\alpha_{10}(\lambda_2,\lambda_3)}{\alpha_7(\lambda_2,\lambda_3)} 
[\vec{B}(\lambda_1) \otimes \vec{\xi} F(\lambda_{2}) B(\lambda_{3})]
\nonumber \\
&& +[\vec{\xi} \otimes F(\lambda_{1}) \vec{B}(\lambda_{3}) B(\lambda_{2})]
\hat{g}^{(3)}_{1}(\lambda_{1},\lambda_{2},\lambda_{3})
+[\vec{\xi} \otimes F(\lambda_{1}) \vec{B}(\lambda_{2}) B(\lambda_{3})]
\hat{g}^{(3)}_{2}(\lambda_{1},\lambda_{2},\lambda_{3})
\nonumber\\
\ear

Next we use the commutation rule (35) to simplify the second and the
third parts of the above equation, carring 
the scalar
field $B(\lambda_j)$ ($j=1,2$) through the 
creation operator $\vec{B}(\lambda_3)$. This procedure not only helps us
to eliminate the fifth term of equation (D.1) but also prompts the
appearance of a desirable term proportional to
$[\vec{\xi} \otimes F(\lambda_{2}) \vec{B}(\lambda_{3}) B(\lambda_{1})]$.
Now, imposing the exchange property (78) for the rapidities
$\lambda_1$ and $\lambda_2$ we find the following necessary condition
\EQ
\hat{g}^{(3)}_{1}(\lambda_{2},\lambda_{1},\lambda_{3}) 
\frac{\alpha_7(\lambda_2,\lambda_1)}{\alpha_{10}(\lambda_2,\lambda_1)} =
-\frac{\alpha_2(\lambda_3,\lambda_1)}{\alpha_{9}(\lambda_3,\lambda_1)} 
\EN

This relation together with the previous restrictions found in section
4, cf. equations (73-74),  are able 
to determine unambiguously the constraints for the three-particle state. 
The next step is to show the consistency of the whole procedure, i.e. that
the equality between  the remaining terms  
are indeed satisfied.  By using the
commutation rules (41-42) we derive two consistency conditions, given by
\bear
&&[\vec{B}(\lambda_{2}) \otimes \vec{\xi}]F(\lambda_{1})B(\lambda_{3})
\left [
\frac{\alpha_{10}(\lambda_{2},\lambda_{3})}{\alpha_{7}(\lambda_{2},\lambda_{3})}
\frac{\alpha_{5}(\lambda_{1},\lambda_{2})}{\alpha_{2}(\lambda_{1},\lambda_{2})}
-
\frac{\alpha_{1}(\lambda_{1},\lambda_{2})}{\alpha_{2}(\lambda_{1},\lambda_{2})}
\frac{\alpha_{10}(\lambda_{1},\lambda_{3})}{\alpha_{7}(\lambda_{1},\lambda_{3})}\hat{r}_{12}(\lambda_{1},\lambda_{2})
\right ]
\nonumber \\
=
&&[\vec{\xi} \otimes \vec{B}(\lambda_{2})]F(\lambda_{1})B(\lambda_{3})
\left [
-\frac{\alpha_{10}(\lambda_{1},\lambda_{2})}{\alpha_{7}(\lambda_{1},\lambda_{2})}
\frac{\alpha_{5}(\lambda_{3},\lambda_{2})}{\alpha_{9}(\lambda_{3},\lambda_{2})}
\frac{\alpha_{8}(\lambda_{1},\lambda_{2})}{\alpha_{2}(\lambda_{1},\lambda_{2})}+
\frac{\alpha_{8}(\lambda_{1},\lambda_{2})}{\alpha_{2}(\lambda_{1},\lambda_{2})}
\hat{g}^{(3)}_{2}(\lambda_{1},\lambda_{2},\lambda_{3})
\right ]
\nonumber \\
\ear
and
\bear
&&F(\lambda_{2})[\vec{\xi} \otimes \vec{B}(\lambda_{1})]B(\lambda_{3})
\left [
\frac{\alpha_{10}(\lambda_{1},\lambda_{2})}{\alpha_{7}(\lambda_{1},\lambda_{2})}
\frac{\alpha_{5}(\lambda_{3},\lambda_{1})}{\alpha_{9}(\lambda_{3},\lambda_{1})}
+
\frac{\alpha_{5}(\lambda_{1},\lambda_{2})}{\alpha_{2}(\lambda_{1},\lambda_{2})}
\hat{g}^{(3)}_{2}(\lambda_{1},\lambda_{2},\lambda_{3})
\right.
\nonumber \\
&& \left. -\frac{\alpha_{10}(\lambda_{1},\lambda_{2})}{\alpha_{7}(\lambda_{1},\lambda_{2})}
\frac{\alpha_{5}(\lambda_{3},\lambda_{2})}{\alpha_{9}(\lambda_{3},\lambda_{2})}
\frac{\alpha_{5}(\lambda_{1},\lambda_{2})}{\alpha_{2}(\lambda_{1},\lambda_{2})}
-
\frac{\alpha_{1}(\lambda_{1},\lambda_{2})}{\alpha_{2}(\lambda_{1},\lambda_{2})}
\hat{g}^{(3)}_{2}(\lambda_{2},\lambda_{1},\lambda_{3}).
\hat{r}_{12}(\lambda_{1},\lambda_{2})
\right ]
\nonumber \\
&& = -F(\lambda_{2})[\vec{B}(\lambda_{1}) \otimes \vec{\xi}]B(\lambda_{3})
\frac{\alpha_{10}(\lambda_{2},\lambda_{3})}{\alpha_{7}(\lambda_{2},\lambda_{3})}
\frac{\alpha_{9}(\lambda_{1},\lambda_{2})}{\alpha_{2}(\lambda_{1},\lambda_{2})}
\ear

In order to disentangle the above expressions we need the help of certain
useful identities between the ``exclusion'' vector and the
auxiliary $r$-matrix. More precisely, they are
\EQ
[\vec{\xi} \otimes \vec{B}(y)] \hat{r}_{12}(\lambda,\mu) =
[\bar{a}(\lambda,\mu)-\bar{b}(\lambda,\mu)] [\vec{\xi} \otimes \vec{B}(y)] 
\EN
\EQ
[\vec{\xi} \otimes \vec{B}(y)] \hat{r}_{23}(\lambda,\mu) =
[\vec{\xi} \otimes \vec{B}(y)]+\bar{b}(\lambda,\mu)[\vec{B}(y) \otimes \vec{\xi}]
\EN
\EQ
[\vec{B}(y) \otimes \vec{\xi}] \hat{r}_{23}(\lambda,\mu) =
[\bar{a}(\lambda,\mu)-\bar{b}(\lambda,\mu)] [\vec{B}(y) \otimes \vec{\xi}] 
\EN
\EQ
[\vec{B}(y) \otimes \vec{\xi}]\hat{r}_{12}(\lambda,\mu) =
[\vec{B}(y) \otimes \vec{\xi}]+\bar{b}(\lambda,\mu)[\vec{\xi} \otimes \vec{B}(y)]
\EN

Inserting the identities (D.5-D.8) into equations (D.3-D.4)
we end up with four identities among the Boltzmann weights which have been
verified  by using $Mathematica^{TM}$. With this we complete the symmetrization 
analysis for the three-particle state.   

We now turn to the analysis of the eigenvalue problem for the
three-particle state. Let us begin by investigating the action of the scalar
field $B(\lambda)$ on the state (76). The first step  consists to
carry the field $B(\lambda)$ through the creation fields $\vec{B}(\lambda_1)$
and $F(\lambda_1)$ by using the commutation rules (35) and (41-42). Afterwards,
we use directly the known results for the two-particle
state, cf. (57), in order to  turn one more time 
the scalar fields $B(\lambda)$ and $B(\lambda_1)$
over the two-particle state 
$\ket{\Phi_{2}(\lambda_{2},\lambda_{3})} $.  As a third step, we
need to reorder creation fields such as $\vec{B}(\lambda_1)$ and
$\vec{B}(\lambda)$ with the help of commutation rule (25) as well
as keep on carrying the scalar field $B(\lambda)$ until it
reaches the vacuum. After this long but straightforward computations 
we find  the following result
\bear
B(\lambda) \ket{\Phi_{3}(\lambda_{1},\lambda_{2},\lambda_{3})} & = &
[w_{1}(\lambda)]^{L} \prod_{j=1}^{3}
i\frac{\alpha_{2}(\lambda_{j},\lambda)}{\alpha_{1}(\lambda_{j},\lambda)}
\vec{B}(\lambda_1) \otimes 
 \vec{\Phi}_{2}(\lambda_{2},\lambda_{3})  
. \vec{\cal F} \ket{0}
\nonumber\\
&&+[w_{1}(\lambda)]^{L} 
i\frac{\alpha_{2}(\lambda_{3},\lambda)}{\alpha_{9}(\lambda_{3},\lambda)}
L[\lambda,\lambda_1,\lambda_2] 
\vec{\xi} \otimes F(\lambda_{1}) 
\vec{B}(\lambda_{3}) B(\lambda_2)
\hat{g}^{(3)}_1(\lambda_1,\lambda_2,\lambda_3)
. \vec{\cal F} \ket{0}
\nonumber\\
&&+[w_{1}(\lambda)]^{L} 
i\frac{\alpha_{2}(\lambda_{2},\lambda)}{\alpha_{9}(\lambda_{2},\lambda)}
L[\lambda,\lambda_1,\lambda_3] 
\vec{\xi} \otimes F(\lambda_{1}) 
\vec{B}(\lambda_{2}) B(\lambda_3)
\hat{g}^{(3)}_2(\lambda_1,\lambda_2,\lambda_3)
. \vec{\cal F} \ket{0}
\nonumber\\
&& + \mathrm{ unwanted~ terms}
\ear
where function $L[x,y,z]$ is precisely the left-hand side of 
identity (C.1) we have worked out for the two-particle state.
This allows us to factorize the amplitudes for the second and the
third terms of the above equation, and as result all the three wanted
terms have a common amplitude as it should be. To what concerns
the unwanted terms, our
computation shows that
they can be gathered in two basic families. More specificly, 
they are proportional to
\EQ
[w_{1}(\lambda_j)]^{L}
\vec{B}(\lambda) \otimes \vec{\Phi}_{2}(\lambda_{l},\lambda_{k})
\EN
and 
\EQ
[w_{1}(\lambda_j)
w_{1}(\lambda_l)]^{L}
\vec{\xi} \otimes F(\lambda) 
\vec{\Phi}_{1}(\lambda_{k})
\EN

The first term in the family (D.10), say $\lambda_j=\lambda_1$, $\lambda_l=\lambda_2$ and
$\lambda_k =\lambda_3$, is originated from the first part of the
three-particle state when we turn the scalar field $B(\lambda)$
through $\vec{B}(\lambda_1)$. Keeping the second term of the
commutation rule (35), and by using the two-particle 
results (57) to carry $B(\lambda_1)$ through 
$\ket{\Phi_{2}(\lambda_{2},\lambda_{3})} $, we find that its
amplitude is
\EQ
-i\frac{\alpha_{5}(\lambda_{1},\lambda)}{\alpha_{9}(\lambda_{1},\lambda)}
\prod_{k=2}^{3}
i\frac{\alpha_{2}(\lambda_{k},\lambda_1)}{\alpha_{9}(\lambda_{k},\lambda_1)}
\EN

We estimate the amplitudes of the remaining terms in the family
(D.10) by taking
into account the exchange property (78), in  much the same way we
did 
for the two-particle state. This means
that the amplitudes  are going to be multiplied by the first ``ordering''
factors 
$\hat{O}^{(1)}_{j}(\lambda_{j};\{\lambda_{k}\})$, and three possible 
unwanted terms $j=1,2,3$  can be compactly written as
\EQ
-[w_{1}(\lambda_{j})]^{L} 
i\frac{\alpha_{5}(\lambda_{j},\lambda)}{\alpha_{9}(\lambda_{j},\lambda)}
\prod_{ \stackrel {k=1}{i \neq j}}^{3} 
i\frac{\alpha_{2}(\lambda_{k},\lambda_{j})}
{\alpha_{9}(\lambda_{k},\lambda_{j})} 
\vec{B}(\lambda) 
\otimes \vec{\Phi}_{2}(\lambda_{1},\dots,\check{\lambda}_{j},\dots,\lambda_{3}) 
\times
\hat{O}^{(1)}_{j}(\lambda_{j};\{\lambda_{k}\}). \vec{\cal F} \ket{0}
\EN

The contributions to the second family of unwanted terms come from
all the pieces composing the three-particle state. It turns out that
for $k=2,3$ their amplitudes can be computed in a very similar
way we did for the second and third parts of the wanted terms, respectively.
The main difference is that now 
we have to keep track of terms proportional to $F(\lambda)$
rather than  $F(\lambda_1)$. We find that the amplitudes for these
unwanted terms are
\bear
[w_{1}(\lambda_{1})
w_{1}(\lambda_{j})]^{L} 
H_{1}(\lambda,\lambda_{1},\lambda_{j})
\prod_{ \stackrel {k=1}{k \neq 1,j}}^{3} 
i\frac{\alpha_{2}(\lambda_{k},\lambda_{1})}{\alpha_{9}(\lambda_{k},\lambda_{1})}  
i\frac{\alpha_{2}(\lambda_{k},\lambda_{j})}
{\alpha_{9}(\lambda_{k},\lambda_{j})}
&&[F(\lambda) \vec{\xi} \otimes 
 \vec{\Phi}_{1}(\lambda_{2},\dots,\check{\lambda}_{j},\dots,\lambda_{3})]
\nonumber\\
&& \times \hat{O}^{(2)}_{1j}(\lambda_{1},\lambda_{j};\{\lambda_{k}\}). \vec{\cal F} \ket{0} 
\ear
where 
$\hat{O}^{(2)}_{1j}(\lambda_{1},\lambda_{j};\{\lambda_{k}\}) $,$j=1,2$, is
the second type of ``ordering'' factor which has been already defined
in the main text, see equation (88). It should be emphasized that we have
derived the above  factor from a ``brute force '' analysis, and
similar to what happened to the two-particle state, this gives us the
clue to proceed in order to better estimate the remaining unwanted terms 
appearing  in this family. We easily recognize that this
factor is related to the operation of bringing two rapidities in the
first two positions of the eigenvector. Keeping this in mind, we see that
all the  contributions to the second family of unwanted terms can be
written by
\bear
[w_{1}(\lambda_{l})
w_{1}(\lambda_{j})]^{L} 
H_{1}(\lambda,\lambda_{l},\lambda_{j})
\prod_{ \stackrel{k=1}{k \neq l,j}}^{3} 
i\frac{\alpha_{2}(\lambda_{k},\lambda_{l})}{\alpha_{9}(\lambda_{k},\lambda_{l})}  
i\frac{\alpha_{2}(\lambda_{k},\lambda_{j})}{\alpha_{9}(\lambda_{k},\lambda_{j})}
&&[F(\lambda) \vec{\xi} \otimes  
\vec{\Phi}_{1}(\lambda_{1},\ldots,\check{\lambda}_{l},\ldots,
\check{\lambda}_j, \ldots, \lambda_{3})]
\nonumber\\
&& \times \hat{O}^{(2)}_{lj}(\lambda_{l},\lambda_{j};\{\lambda_{k}\}). \vec{\cal F} \ket{0}
\ear

Collecting the expressions 
(D.9),(D.13) and (D.15) we find that the action of the
scalar field $B(\lambda)$ on the three-particle state is described
by the formula (82) with $n=3$. Similar reasoning can be repeated
for the fields 
$D(\lambda)$ and $\displaystyle \sum_{a=1}^{2} \hat{A}_{aa}(\lambda) $, and
only when we are estimating the third type of unwanted terms new technicalities
emerge. 
In what follows we present the details
of these computations in the simplest case, i.e. the situation  where no
``ordering'' factors are needed.  Generalization for the remaining 
terms is along the lines of formula (D.15).
For the field $D(\lambda)$ we find that
such amplitude is
\EQ
H_2(\lambda,\lambda_1,\lambda_2) F(\lambda) \xi_{bc} 
\hat{A}_{b b_{1}}(\lambda_{1}) \hat{A}_{c b_{2}}(\lambda_{2}) 
B_{b_3}(\lambda_3)  {\cal{F}}^{b_{3} b_{2} b_{1}} \ket{0} 
\EN
Now, carrying the operators 
$\hat{A}_{b b_{1}}(\lambda_{1})$ and $ \hat{A}_{c b_{2}}(\lambda_{2})$
through $B_{b_3}(\lambda_3)$ with the help of commutation rule (34) we find
\EQ
[w_{2}(\lambda_{1})w_{2}(\lambda_{2})]^{L}  H_2(\lambda,\lambda_1,\lambda_2)
\prod_{k=1}^{2} \frac{\alpha_{1}(\lambda_{k},\lambda_{3})}
{i\alpha_{9}(\lambda_{k},\lambda_{3})} \xi_{bc}
F(\lambda) B_{\gamma}(\lambda_3) 
\hat{r}_{\alpha c}^{b_{2}b_{3}}(\lambda_2,\lambda_3)
\hat{r}_{\gamma b}^{b_{1}\alpha}(\lambda_1,\lambda_3)
  {\cal{F}}^{b_{3} b_{2} b_{1}} \ket{0} 
\EN
which is further simplified by using the following identity
\EQ
 T^{(1)}(\lambda=\lambda_2,\{\lambda_{l}\})_{c_{1} c_2 c_{3}}^{b_{1} b_2 b_{3}}
 T^{(1)}(\lambda=\lambda_1,\{\lambda_{l}\})_{b c \gamma}^{c_{1} c_2 c_{3}}
=
\hat{r}_{\alpha c}^{b_{2}b_{3}}(\lambda_2,\lambda_3)
\hat{r}_{\gamma b}^{b_{1}\alpha}(\lambda_1,\lambda_3)
\EN

Inserting (D.18) into (D.17) we finally obtain
\EQ
[w_{2}(\lambda_{1})w_{2}(\lambda_{2})]^{L}  H_2(\lambda,\lambda_1,\lambda_2)
\Lambda^{(1)}(\lambda=\lambda_1,\{ \lambda_l \})
\Lambda^{(1)}(\lambda=\lambda_2,\{ \lambda_l \})
\nonumber\\
\prod_{k=1}^{2} \frac{\alpha_1(\lambda_k,\lambda_3)}
{i\alpha_9(\lambda_k,\lambda_3)} 
F(\lambda) \vec{\xi} 
\otimes \vec{B}(\lambda_3)
\EN

For the field $\displaystyle \sum_{a=1}^{2} \hat{A}_{aa}(\lambda) $
we find that one of the contributions is
\EQ
H_3(\lambda,\lambda_1,\lambda_2) \xi_{a b_1} 
F(\lambda) B(\lambda_1)
\hat{A}_{a b_{2}}(\lambda_{2})  
B_{b_3}(\lambda_3)  {\cal{F}}^{b_{3} b_{2} b_{1}} \ket{0} 
\EN     
and when we carry $B(\lambda_1)$ and 
$\hat{A}_{a b_{2}}(\lambda_{1})  $ through $B_{b_3}(\lambda_3)$ we have
\EQ
[w_{1}(\lambda_{1})w_{2}(\lambda_{2})]^{L}  H_3(\lambda,\lambda_1,\lambda_2)
\prod_{k=1}^{2} \frac{\alpha_1(\lambda_k,\lambda_3)}
{i\alpha_9(\lambda_k,\lambda_3)} 
\xi_{a b_1}
F(\lambda) 
\hat{r}_{d a}^{b_{2}b_{3}}(\lambda_2,\lambda_3) B_d(\lambda_3)
  {\cal{F}}^{b_{3} b_{2} b_{1}} \ket{0} 
\EN

Next using the following identity
\EQ
\xi_{a b_1}
\hat{r}_{d a}^{b_{2}b_{3}}(\lambda_2,\lambda_3) B_d(\lambda_3)
 {\cal{F}}^{b_{3} b_{2} b_{1}}  
=
\xi_{\gamma \delta} 
\hat{r}_{\gamma \delta}^{\alpha \beta }(\lambda_1,\lambda_2) 
T^{(1)}(\lambda=\lambda_2,\{\lambda_{l}\})_{\alpha \beta d }^{b_{1} b_2 b_{3}}
B_d(\lambda_3)
 {\cal{F}}^{b_{3} b_{2} b_{1}}  
\EN
we finally find
\EQ
[w_{1}(\lambda_{1})w_{2}(\lambda_{2})]^{L} 
H_3(\lambda,\lambda_1,\lambda_2)[ \bar{a}(\lambda_1 ,\lambda_2) -\bar{b}(\lambda_1 ,\lambda_2)] 
\Lambda^{(1)}(\lambda=\lambda_2,\{ \lambda_l \})
\prod_{k=1}^{2} \frac{\alpha_1(\lambda_k,\lambda_3)}
{i \alpha_9(\lambda_k,\lambda_3)} 
F(\lambda) \vec{\xi} 
\otimes \vec{B}(\lambda_3)
\EN

Lastly, the second contribution coming from the field
$\displaystyle \sum_{a=1}^{2} \hat{A}_{aa}(\lambda) $ is estimated by using
the same trick explained in the previous appendix for the two-particle
state. We further remark that the technical points explained in appendices $C$
and $D$ are valid for many other models such as the Bariev $XY$ chain and
those solved in ref. \cite{MP1}.

\centerline{\bf Appendix E : The Bariev model}
\setcounter{equation}{0}
\renewcommand{\theequation}{E.\arabic{equation}}

We start this appendix by 
presenting the Boltzmann weights $\rho_{j}(\lambda,\mu) j=1,\ldots,15$ found by Zhou \cite{ZHOU}. Normalizing them 
by $\rho_{1}(\lambda,\mu)$ we have
\EQ
\rho_{2}(\lambda,\mu)= \frac{\sqrt{(1+\bar{h}^{2}\lambda^{2})(1+\bar{h}^{2}\mu^{2})}}{(1+\bar{h}^{2}\lambda\mu )}
\EN
\EQ
\rho_{3}(\lambda,\mu)= \frac{\bar{h}(\lambda-\mu)}{1+\bar{h}^{2}\lambda\mu}~~,~~
\rho_{4}(\lambda,\mu)= \rho_{2}(\lambda,\mu)\rho_{2}(\lambda/\bar{h},\mu/\bar{h})
\EN
\EQ
\rho_{5}(\lambda,\mu)= \rho_{3}(\lambda,\mu) \frac{\sqrt{(1+\bar{h}^{2}u^{2})(1+\lambda^{2})}}{(1+\lambda\mu)}~~,~~
\rho_{6}(\lambda,\mu)= \frac{\rho_{5}(\lambda,\mu)}{\bar{h}}
\EN
\EQ
\rho_{7}(\lambda,\mu)= \frac{1}{1+\bar{h}^{2}\lambda\mu}+
\frac{\bar{h}^{2}(\lambda^2+\mu^2+\lambda^2 \mu^2-\lambda\mu)}
{(1+\lambda\mu)(1+\bar{h}^{2}\lambda\mu)}~~,~~
\rho_{8}(\lambda,\mu)=\rho_{2}(\lambda/\bar{h},\mu/\bar{h})
\EN
\EQ
\rho_{9}(\lambda,\mu)= \frac{\rho_{3}(\lambda,\mu)(\lambda-\bar{h}^{2}\mu)}{\bar{h}(1+\lambda\mu)}
\EN
\EQ
\rho_{10}(\lambda,\mu)=
\frac{\bar{h}^{2}\lambda\mu}{1+\bar{h}^{2}\lambda\mu}+\frac{1+\lambda^{2}+\mu^{2}-\lambda\mu}
{(1+\lambda\mu)(1+\bar{h}^{2}\lambda\mu)}~~,~~
\rho_{11}(\lambda,\mu)=\rho_{3}(\lambda/\bar{h},\mu/\bar{h})
\EN
\EQ
\rho_{12}(\lambda,\mu)= -\rho_{5}(\mu,\lambda)~~,~~
\rho_{13}(\lambda,\mu)= \frac{\rho_{12}(\lambda,\mu)}{\bar{h}}
\EN
\EQ
\rho_{14}(\lambda,\mu)= \rho_{9}(\mu,\lambda)~~,~~
\rho_{15}(\lambda,\mu)= \rho_{3}(\lambda,\mu)\rho_{11}(\lambda,\mu)
\EN

We remark that a rescaling $\lambda \rightarrow \frac{\lambda}{\sqrt{\bar{h}}}$ 
and $\mu \rightarrow \frac{\mu}{\sqrt{\bar{h}}}$ brings these weights in a more 
symmetrical form and this is useful, for example, to check 
the Yang-Baxter equation (6). Besides that 
there are some few identities between those 
weights that were useful in the calculations of section 6. They are 
\EQ
\rho_{15}(\lambda,\mu)[ \rho_{9}(\lambda,\mu)+ \rho_1(\lambda,\mu) ] = 
\rho_{5}(\lambda,\mu)
\rho_{6}(\lambda,\mu),~~
\rho_{6}(\lambda,\mu) \rho_{1}(\lambda,\mu)+ \rho_{5}(\lambda,\mu)  
\rho_{15}(\lambda,\mu) = \rho_{6}(\lambda,\mu)
\rho_{7}(\lambda,\mu)
\EN
\EQ
\rho_{12}(\lambda,\mu)[ \rho_{9}(\lambda,\mu)+ \rho_{1}(\lambda,\mu) ] = \rho_{5}(\lambda,\mu)
\rho_{4}(\lambda,\mu),~~
\rho_{5}(\lambda,\mu) \rho_1(\lambda,\mu)+ \rho_{15}(\lambda,\mu)  
\rho_{6}(\lambda,\mu) = \rho_{5}(\lambda,\mu)
\rho_{10}(\lambda,\mu)
\EN

Next we present the commutation relations we found by solving 
the standard Yang-Baxter algebra (3). Between the diagonal fields and the 
creation operators we have
\bear
\hat{A}(\lambda) \otimes \vec{B}(\mu) &  = &
\frac{\rho_{1}(\lambda,\mu)}{\rho_{3}(\lambda,\mu)}[\vec{B}(\mu) \otimes \hat{A}(\lambda) ]. 
{\hat{r}}^{(B)}(\lambda,\mu)
- \frac{\rho_{2}(\lambda,\mu)}{\rho_{3}(\lambda,\mu)} \vec{B}(\lambda) \otimes \hat{A}(\mu)+   
\nonumber \\
&& \frac{\rho_{5}(\lambda,\mu)}{\rho_{9}(\lambda,\mu)} \left[
 \vec{B^{*}}(\lambda)B(\mu) 
+\frac{\rho_{2}(\lambda,\mu)}{\rho_{3}(\lambda,\mu)}F(\lambda)\vec{C}(\mu)  
-\frac{\rho_{12}(\lambda,\mu)\rho_{1}(\lambda,\mu)}{\rho_{3}(\lambda,\mu)\rho_{5}(\lambda,\mu)}F(\mu)\vec{C}(\lambda) \right ] 
\otimes \vec{\xi}^{(B)}
\nonumber\\
\ear
\EQ
B(\lambda)\vec{B}(\mu) = 
\frac{\rho_{1}(\mu,\lambda)}{\rho_{3}(\mu,\lambda)} \vec{B}(\mu)B(\lambda) - 
\frac{\rho_{2}(\mu,\lambda)}{\rho_{3}(\mu,\lambda)} \vec{B}(\lambda)B(\mu),
\EN
\bear
D(\lambda)\vec{B}(\mu) & = &
\frac{\rho_{11}(\lambda,\mu)}{\rho_{9}(\lambda,\mu)} \vec{B}(\mu)D(\lambda) 
+ \frac{\rho_{8}(\lambda,\mu)}{\rho_{9}(\lambda,\mu)} F(\mu)\vec{C^{*}}(\lambda)
 \nonumber \\ 
&& - \frac{\rho_{4}(\lambda,\mu)}{\rho_{9}(\lambda,\mu)} F(\lambda)\vec{C^{*}}(\mu)
- \frac{\rho_{5}(\lambda,\mu)}{\rho_{9}(\lambda,\mu)} 
\vec{\xi}^{(B)} . \{ \vec{B^{*}}(\lambda) \otimes \hat{A}(\mu) \}
\ear
\bear
\hat{A}_{ab}(\lambda)F(\mu) & = &
\frac{\rho_{11}(\lambda,\mu)}{\rho_{3}(\lambda,\mu)}[1 - \frac{\rho_{8}^{2}(\lambda,\mu)}{\rho_{11}^{2}(\lambda,\mu)}] F(\mu)\hat{A}_{ab}(\lambda)
+ \frac{\rho_{2}(\lambda,\mu)\rho_{8}(\lambda,\mu)}{\rho_{3}(\lambda,\mu)\rho_{11}(\lambda,\mu)} F(\lambda) \hat{A}_{ab}(\mu)
\nonumber \\
&& - \frac{\rho_{2}(\lambda,\mu)}{\rho_{3}(\lambda,\mu)} 
 [\vec{B}(\lambda) \otimes \vec{B^{*}}(\mu)]_{ba} 
+ \frac{\rho_{8}(\lambda,\mu)}{\rho_{11}(\lambda,\mu)} 
[\vec{B^{*}}(\lambda) \otimes \vec{B}(\mu)]_{ab} 
\ear
\EQ
B(\lambda)F(\mu) = 
\frac{\rho_{1}(\mu,\lambda)}{\rho_{9}(\mu,\lambda)} F(\mu)B(\lambda) - 
\frac{\rho_{4}(\mu,\lambda)}{\rho_{9}(\mu,\lambda)} F(\lambda)B(\mu)
- \frac{\rho_{5}(\mu,\lambda)}{\rho_{9}(\mu,\lambda)} 
  \{ \vec{B}(\lambda) \otimes \vec{B}(\mu) \}. [{\vec{\xi}^{(B)}}]^{t} 
\EN
\EQ
D(\lambda)F(\mu) = 
\frac{\rho_{1}(\lambda,\mu)}{\rho_{9}(\lambda,\mu)} F(\mu)D(\lambda) - 
\frac{\rho_{4}(\lambda,\mu)}{\rho_{9}(\lambda,\mu)} F(\lambda)D(\mu)
- \frac{\rho_{5}(\lambda,\mu)}{\rho_{9}(\lambda,\mu)} \vec{\xi}^{(B)} . 
\{ \vec{B^{*}}(\lambda) \otimes \vec{B^{*}}(\mu) \}
\EN

Between the creation fields we have
\EQ
\vec{B}(\lambda) \otimes \vec{B}(\mu) =
[ \vec{B}(\mu) \otimes \vec{B}(\lambda) ]. \hat{r}^{(B)}(\lambda,\mu)
+ \left[ \frac{\rho_{5}(\lambda,\mu)}{\rho_{9}(\lambda,\mu)}  
 F(\lambda)B(\mu) - \frac{\rho_{12}(\lambda,\mu)}
 {\rho_{9}(\lambda, \mu)}F(\mu)B(\lambda) \right] {\vec{\xi}}^{(B)}
\EN
\EQ
\left[ F(\lambda), F(\mu) \right] = 0
\EN
\EQ
F(\lambda) \vec{B}(\mu) = \frac{\rho_{8}(\lambda,\mu)}{\rho_{1}(\lambda,\mu)} F(\mu) \vec{B}(\lambda) + 
 \frac{\rho_{11}(\lambda,\mu)}{\rho_{1}(\lambda,\mu)} \vec{B}(\mu) F(\lambda)
\EN
\EQ
\vec{B}(\lambda) F(\mu) = \frac{\rho_{8}(\lambda,\mu)}{\rho_{1}(\lambda,\mu)} \vec{B}(\mu) F(\lambda) + 
 \frac{\rho_{11}(\lambda,\mu)}{\rho_{1}(\lambda,\mu)} F(\mu) \vec{B}(\lambda)
\EN

The extra relations for the analysis of the two-particle state 
are
\EQ
C_{a}(\lambda)B_{b}(\mu) = 
B_{b}(\mu)C_{a}(\lambda) - \frac{\rho_{2}(\lambda,\mu)}
{\rho_{3}(\lambda,\mu)}[B(\lambda)A_{ab}(\mu) - B(\mu)A_{ab}(\lambda)]
\EN
\EQ
B^{*}_{a}(\lambda)B_{b}(\mu) =
\frac{\rho_{11}(\lambda,\mu)}
{\rho_{3}(\lambda,\mu)}
B_{b}(\mu)B^{*}_{a}(\lambda) 
-\frac{\rho_{2}(\lambda,\mu)}{\rho_{3}(\lambda,\mu)}F(\lambda)A_{ab}(\mu)  
+\frac{\rho_{8}(\lambda,\mu)}{\rho_{3}(\lambda,\mu)}F(\mu)A_{ab}(\lambda)
\EN
\EQ
C^{*}_{a}(\lambda)B_{b}(\mu) \ket{0} = 
\xi^{(B)}_{ab} \frac{\rho_{5}(\lambda,\mu)}{\rho_{9}(\lambda,\mu)}
\left[ B(\mu)D(\lambda) - A_{aa}(\lambda)A_{bb}(\mu) \right]  \ket{0}
\EN


\begin{thebibliography}{}
\bibitem{STF} E.K. Sklyanin, L.A. Takhtadzhan and L.D. Faddeev, {\em Theor.Math.Fiz. 40 (1979) 194}
\bibitem{FA} L.D. Faddeev, {\em  Integrable models 
in $(1+1)$ dimensions  , Les
Houches (1982) p. 561, Elsevier }
\bibitem{FA1} L.A. Takhtajan and L.D. Faddeev, {\em Rus.Math.Sur.34 (1979) 11 };
L.A. Takhtajan, {\em Lectures Notes in Physics vol. 242 
eds. B.S. Shastry, S.S. Jha and V. Singh, Springer-Verlag, p. 175 }, 
P.P. Kulish and E.K. Sklyanin, {\em Lecture Notes in Physics vol. 151, eds. J.
Hietarinta and C. Montonen, Springer-Verlag, p. 62 }
\bibitem{TAK} H.B. Thacker, {\em Rev.Mod.Phys. 53 (1981) 253}
\bibitem{KO} V.E. Korepin, G. Izergin and N.M. Bogoliubov, {\em Quantum Inverse Scattering
Method, Correlation Functions and Algebraic Bethe Ansatz, 1992, (Cambridge University
Press) }
\bibitem{LU} M. L\"uscher, {\em Nucl.Phys. B 117 (1976) 475}
\bibitem{TAR} V.O. Tarasov, L.A. Takhtadzhan and L.D. Faddeev, {\em Theor.Math.Fiz. 57 (1983) 163}
\bibitem{SH} L.I. Shift, {\em Quantum Mechanics, MacGraw-Hill Book Company, Inc. (1955)}
\bibitem{LW} E.H. Lieb and F.Y. Wu,{\em Phys.Rev.Lett.20 (1968) 1445}
\bibitem{YY} C.N.Yang, {\em Phys.Rev.Lett. 19 (1967) 1312}; 
M. Gaudin, {\em Phys.Lett. A 24 (1967) 55 }
\bibitem{SA} B.S. Shastry, {\em Phys.Rev.Lett.56 (1986) 1529}
\bibitem{SA1} B.S. Shastry, {\em Phys.Rev.Lett.56 (1986) 2453}
\bibitem{SA2} B.S. Shastry, {\em J.Stat.Phys.30 (1988) 57}
\bibitem{BA} R.Z. Bariev, {\em Theor.Math.Fiz. 82 (1990) 313}
\bibitem{WA} M. Wadati,E. Olmedilla and Y. Akutsu, {\em J.Phys.Soc.Japan 36 (1987) 340}; E. Olmedilla, M. Wadati and Y. Akutsu, {\em J.Phys.Soc.Japan 36 (1987) 2298}; E. Olmedilla and M. Wadati, {\em Phys.Rev.Lett. 60 (1988) 1595}
\bibitem{PM} P.B. Ramos and M.J. Martins, {\em J.Phys.A: Math.Gen. 30 (1997) 
L195}
\bibitem{KOM}  M. Suzuki, {\em Phys.Rev.B. 31 (1985) 2957}; J.Suzuki, Y.Akutsu
and M. Wadati, {\em J.Phys.Soc.Japan 59 (1990) 2667}; T. Koma, {\em 
Prog.Theor.Phys. 83 (1990) 1445 }
\bibitem{KUP} A. Kl\"umper, {\em Ann.Physik 1 (1992) 540; Z.Phys.B 91
(1993) 507}; A. Kl\"umper and R.Z. Bariev, {\em Nucl. Phys.B 458 (1995) 625};
G. J\"utner, A. Kl\"umper and J. Suzuki, {\em Nucl.Phys.B 487 (1997) 656 }
\bibitem{DEV} C. Destri and H.J.de Vega, {\em Phys.Rev.Lett. 69 (1992) 2313; 
Nucl.Phys.B 438 (1995) 413}
\bibitem{ZOT} X. Zotos, P. Naet and P. Prelov, {\em Phys.Rev.B 55 (1997) 11029}
\bibitem{POI} G. Montanbaux, D. Poiblanc, J. Bellisard and C. Sire, {\em 
Phys.Rev.Lett. 70 (1993) 497}; M. Distasio and X. Zotos, {\em Phys.Rev.Lett. 74 
(1995) 2050}
\bibitem{PM2} P.B. Ramos and M.J. Martins, {\em Nucl.Phys.B 474 (1996) 678}
\bibitem{MP1} M.J. Martins and P.B. Ramos, {\em Nucl.Phys.B 500 (1997) 579}
\bibitem{BAR} R.Z. Bariev, {\em J.Phys.A:Math.Gen. 24 (1991) L549 }
\bibitem{DEVI} H.J.de Vega, {\em Nucl.Phys.B 240 (1984) 495}
\bibitem{INV} M.P. Grabowski and P. Mathieu, {\em Ann.Phys. 243 (1995)};
H. Grosse, {\em Lett.Math.Phys. 18 (1989) 151}
\bibitem{SHWA} M. Shiroishi and M. Wadati, {\em J.Phys.Soc.Japan 64 (1995) 57}
\bibitem{SKU} P.P. Kulish and E.K. Sklyanin, {\em J.Soviet.Math. 19 (1982) 1596}
\bibitem{KUN} A. Kundu, {\em `Quantum Integrable Systems: Construction, Solution,
Algebraic Aspect', hep-th 96/12046}
\bibitem{RES} P.P. Kulish and N.Yu. Reshetikhin, {\em Sov.Phys. JETP 53 (1981) 108}; {\em J.Phys.A: Math.Gen. 16 (1983) L591}
\bibitem{BAB} O. Babelon, H.J.de Vega and C.M. Viallet, {\em Nucl.Phys.B 200 (1982) 266}
\bibitem{UK} D.B. Uglov and V. E.Korepin, {\em Phys.Lett.A 190 (1994) 238}
\bibitem{GOMU1} S. Murakami and F. G\"ohmann, {\em Phys.Lett.A 227 (1997) 216}
\bibitem{TAR1} V.O. Tarasov, {\em Theor.Math.Phys. 76 (1988) 793}
\bibitem{SU2} J.O. Heilmann and E.H. Lieb, {\em Ann.NY.Acad.Sci. 172 
(1971) 584}; C.N. Yang, {\em Phys.Rev.Lett. 63 (1989) 2144} ; M. Pernici, {\em
Europhys.Lett. 12 (1990) 75 }
\bibitem{EKS} F.H.L. Essler, V.E. Korepin and K. Schoutens, {\em Nucl.Phys.B 
372 (1992) 559; Phys.Rev.Lett. 67 (1991) 3848}
\bibitem{GOM} F.G\"ohmann and S.Murakami, {\em J.Phys.A: Math.Gen. 30 (1997) 5269 }
\bibitem{BAT} C.M. Yung and M.T. Batchelor, {\em Nucl.Phys.B 446 (1995) 461}
\bibitem{SS} B.S. Shastry and B. Sutherland, {\em Phys.Rev.Lett. 65 (1990) 243}
\bibitem{MFY} M.J. Martins and R.M. Fye, {\em J.Stat.Phys. 64 (1991) 271}
\bibitem{CAM} J.M.P. Carmelo and P. Horsch, {\em Phys.Rev.Lett. 
68 (1992) 871 }; J.M.P. Carmelo, P.Horsch and A.A. Ovchinnikov, {\em
Phys.Rev.B 46 (1992) 14728 }
\bibitem{AWR} F.C. Alcaraz and W.F. Wreszinski, {\em J.Stat.Phys. 58 (1990) 45}
\bibitem{GUAN} Xi-Wen Guan and S.D. Yang, 
{\em ``
Algebraic Bethe Ansatz for 
one-dimensional Hubbard model with chemical potential'',Jilin preprint (1997)}
\bibitem{DEG} R. Yue and T. Deguchi, {\em J.Phys.A: Math.Gen. 30 (1997) 849};
M. Shiroishi and M. Wadati, {\em ``Integrable boundary conditions 
for the one-dimensional Hubbard model'', J.Phys.Soc.Japan. (1997), to appear}
\bibitem{TAFA1} L.A. Takhtajan and L.D. Faddeev, {\em J.Sov.Math. 24 (1984)
241}
\bibitem{HI} J.E. Hirsch, {\em Physica C 158 (1989) 326}
\bibitem{ZHOU} Huan-Qiang Zhou, {\em Phys.Lett.A 221 (1996) 104}
\bibitem{PMB} M.J. Martins and P.B. Ramos, {\em J.Phys.Math.Gen.30 (1997) L465}
\bibitem{HIMU} K. Hikami and S. Murakami, {\em Phys.Lett.A 221 (1996) 109 }
\bibitem{SHWA2} M. Shiroischi and M. Wadati, {\em J.Phys.A:Math.Gen. 30 (1997) 1115}
\bibitem{ZHOU1} H.Q. Zhou, {\em J.Phys.A:Math.Gen.30 (1997) L423 }
\bibitem{WDA} M. Wadati, T. Deguchi and Y. Akutsu, {\em Phys.Rep. 180 (1987) 247, and references
there in}
\bibitem{JIBA} M. Jimbo, {\em Commun.Math.Phys. 102 (1986) 537 }; V.V. 
Bazhanov, {\em Phys.Lett.B 159 (1985) 321} 
\end{thebibliography}
\end{document}